\documentclass[a4paper,fleqn,usenatbib]{mnras}

\usepackage{latexsym}
\usepackage{amsmath}
\usepackage{amssymb}
\usepackage{fnpos}
\usepackage{hyperref}
\usepackage{tabularx}
\usepackage{xspace}
\usepackage{enumerate}
\usepackage{scalerel}
\usepackage{graphicx}
\usepackage{url}
\usepackage{caption}
\usepackage{longtable}
\usepackage{lscape}
\usepackage[normalem]{ulem} 
\usepackage[dvipsnames]{xcolor} 
\usepackage{wasysym} 
\definecolor{darkgreen}{rgb}{0.0,0.55,0.0}
\definecolor{darkblue}{rgb}{0.0,0.0,0.5}

\newcommand{\eal}[2]{\ifmmode{\mathrm{#1\,#2}}\else{#1\textsc{$\,$\lowercase{#2}}}\fi\xspace}
\newcommand{\feal}[2]{\ifmmode{\mathrm{#1\,#2}}\else{[#1\textsc{$\,$\lowercase{#2}}]}\fi\xspace}
\newcommand{\hfeal}[2]{\ifmmode{\mathrm{#1\,#2}}\else{#1\textsc{$\,$\lowercase{#2}}]}\fi\xspace}

\newcommand{\maps}{Meteoritics \& Planetary Science}

\newcommand{\orcid}[1]{$^{\rm \href{https://orcid.org/#1}{\includegraphics[height=0.6em]{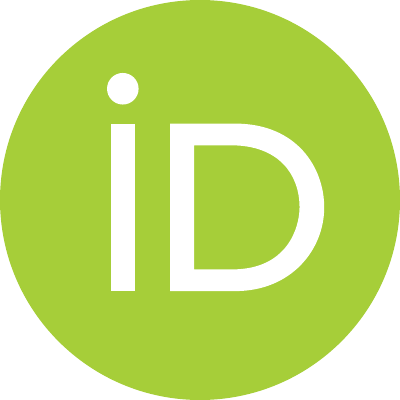}}}$}

\def\jaavso{JAAVSO}

\title[Dust Formation in Novae]{Revisiting the Classics: On the Statistics of Dust Formation in Novae}
\author[Chong et al.]{\parbox{\textwidth}{Atticus Chong\orcid{0009-0003-2656-6811}$^{1}$\thanks{E-mail: \href{mailto:chongatt@msu.edu}{chongatt@msu.edu}, \href{mailto:oh0sitm@gmail.com}{oh0sitm@gmail.com}}, 
Elias Aydi\orcid{0000-0001-8525-3442}$^{2,1}$\thanks{E-mail: \href{mailto:eaydi@ttu.edu}{eaydi@ttu.edu}}, 
Peter Craig\orcid{0000-0002-3673-0668}$^{1}$,
Laura Chomiuk\orcid{0000-0002-8400-3705}$^{1}$, 
Ashley Stone$^{3}$,
Jay Strader\orcid{0000-0002-1468-9668}$^{1}$,
Adam Kawash\orcid{0000-0003-0071-1622}$^{1}$, 
Kirill V. Sokolovsky\orcid{0000-0001-5991-6863}$^{4}$, and 
Fred Walter$^{5}$}
\vspace{0.4cm}\\
\parbox{\textwidth}{
$^{1}$Center for Data Intensive and Time Domain Astronomy, Department of Physics and Astronomy, Michigan State University, East Lansing, MI 48824, USA\\
$^{2}$Department of Physics \& Astronomy, Texas Tech University, Box 41051, Lubbock, TX, 79409-1051, USA\\
$^{3}$Department of Physics \& Astronomy, West Virginia University, White Hall, Morgantown, 26501-6315, WV, USA\\
$^{4}$Department of Astronomy, University of Illinois at Urbana-Champaign, 1002 W. Green Street, Urbana, IL 61801, USA\\
$^{5}$Department of Physics \& Astronomy, Stony Brook University, Stony Brook, NY 11794-2100 \\
}}

\pubyear{2025}

\begin{document}
\label{firstpage}
\pagerange{\pageref{firstpage}--\pageref{lastpage}}
\maketitle

\begin{abstract}

 While nova eruptions produce some of the most common and dramatic dust formation episodes among astrophysical transients, the demographics of dust-forming novae remain poorly understood. Here, we present a statistical study of dust formation in 40 novae with high-quality optical/IR light curves, quantitatively distinguishing dust-forming from non-dust-forming novae while exploring the properties of the dust events. We find that 50--70 \% of novae produce dust, significantly higher than previous estimates. Dust-forming novae can be separated from those that do not show dust formation by using the largest redward ($V-K$) colour change from peak visible brightness; ($V-J$) or ($V-H$) offer useful but less sensitive constraints. This makes optical+IR photometry a powerful tool to quantify dust formation in novae. We find that novae detected in GeV $\gamma$-rays by \emph{Fermi}-LAT appear to form dust more often than novae not detected by \emph{Fermi}, implying a possible connection between $\gamma$-ray producing shocks and dust production. We also find that novae that evolve very quickly ($t_2 < 10$ days) are much less likely to form dust, in agreement with previous findings. We confirm a correlation between $t_2$ and the time of the onset of dust formation (which occurs $\sim$1 week--3 months after maximum light), but conclude that it is primarily an observational artifact driven by dust formation determining when a nova drops 2 mag below peak. The significant fraction of novae that form dust make them ideal laboratories in our Galactic backyard to tackle the puzzle of dust formation around explosive transients.
\end{abstract}



\section{Introduction}

A classical nova eruption is a transient event driven by a thermonuclear runaway on 
the surface of an accreting white dwarf in a close binary system 
(see \citealt{Bode_etal_2008,Della_Valle_Izzo_2020,Chomiuk_etal_2021} for recent reviews). 
The thermonuclear runaway leads to the ejection of at least part of the accreted envelope 
($10^{-7}-10^{-3}$~M$_{\odot}$) with velocities ranging between 200 and 10,000\,km\,s$^{-1}$ 
\citep{Payne-Gaposchkin_1957,Gallagher_etal_1978,Yaron_etal_2005,Aydi_etal_2020b}. 
25--50 novae eruptions are expected each year in our Milky Way Galaxy, 
with around half of them actually being discovered by optical and infrared (IR) surveys \citep{De_etal_2021,Kawash_etal_2022, Zuckerman_etal_2023}. 
Nova eruptions can recur on a given white dwarf, 
sometimes with recurrence times as short as a year \citep[e.g.,][]{Darnley_Henze_2020}, 
but usually with recurrence times longer than human timescales, setting the
observational distinction between `classical' novae (with a single observed eruption)
and `recurrent' novae (with multiple eruptions in recorded history;
\citealt{2010ApJS..187..275S,2021gacv.workE..44D}).

After the nova, nuclear burning can be sustained on the white dwarf surface for days\,--\,years, which manifests as a super-soft X-ray source \citep{Schwarz_etal_2011, Wolf_etal_2013,Page_etal_2020}. Emission from the nuclear-burning white dwarf photo-ionizes the nova ejecta, and the ionization front moves outward in the ejecta with time \citep[e.g.][]{Beck_etal_1990, Williams_1990}; once ionized, the ejecta are expected to be sustained at a temperature around $10^4$ K \citep{Cunningham_etal_2015}. In addition, shocks within the ejecta can further heat portions of the ejecta up to X-ray emitting temperatures \citep{Mukai_etal_2008,Metzger_etal_2015, Aydi_etal_2020a,Gordon_etal_2021}. 

Despite these warm-to-hot temperatures expected and observed in nova ejecta, somehow nova eruptions find a way to form dust. A dust formation event in a nova light curve typically presents as a dip in the optical light curve and/or a rise in the IR brightness, as the dust absorbs optical radiation and then re-emits it mostly in the IR \citep{Gehrz_2008,Strope_etal_2010}.
The `dust dip' in the optical light curve is usually followed by a recovery, caused by the dispersion or destruction of the dust, to a level fainter than that at the start of the dip \citep{Strope_etal_2010}. This evolution of the visible and IR brightness is demonstrated in data from as early as 1976 (e.g., NQ Vul; \citealt{Ney_Hatfield_1978}).

These signatures of dust formation in novae have been observed for over 50 years \citep{Geisel_etal_1970}, and have been reviewed by \citet{Gehrz_1988, Gehrz_2002, Gehrz_2008, Gehrz_etal_1998,  Evans_1990, Evans_1997, Evans_2001, Evans_Rawlings_2008, Evans_etal_2012}, and \citet{Banerjee_Ashok_2012}. 
Dust formation usually occurs between one and five months post-eruption \citep{Williams_etal_2013}, with some rare exceptions where dust forms on timescales that are significantly shorter (e.g., V745 Sco; \citealt{Banerjee_etal_2023}) or longer (e.g., the unique He nova V445 Pup; \citealt{Woudt_etal_2009}). The dust grains in novae are usually carbon-rich (and indeed, dust formation episodes are most commonly hosted by carbon-oxygen white dwarfs), but in some cases they are silicate-rich \citep{Gehrz_etal_1998}. Typical dust masses produced in novae reach $10^{-8}- 10^{-6}$ M$_{\odot}$, comprising
0.001--10~per~cent of the ejecta mass \citep{Gehrz_etal_1998, Gehrz_2008}, but not all novae form dust. While it is clear that much of the newly formed dust is subsequently destroyed, primarily from irradiation by the supersoft X-ray source \citep{Evans_etal_2017, Gehrz_etal_2018}, in at least some novae there is evidence that some dust can survive to late times \citep[e.g.][]{Jose_Hernanz_2007,Evans_etal_2010, Banerjee_etal_2018}.

The formation of dust in explosive transients remains a major outstanding mystery, and novae are no exception. It is unclear exactly where and how dust forms in the nova ejecta, given that they are warm, rapidly expanding, dropping in density, and irradiated by energetic emission. Early models of dust formation pointed out that nova nucleation sites must be neutral and shielded from the hard emission of the supersoft X-ray source \citep{Gallagher_1977, Rawlings_1988, Rawlings_Williams_1989}. 
There has long been evidence for dense structures or clumps in nova ejecta \citep[e.g.][]{Gallagher_Anderson_1976, Williams_1994,Obrien_Bode_2008}, and it has been hypothesized that dust forms in these over-densities \citep{Rawlings_1988}. 
In light of the recent discovery of GeV $\gamma$-rays and strong energetic shocks associated with nova eruptions \citep{Abdo_etal_2010,Ackermann_etal_2014,Metzger_etal_2014,Metzger_etal_2015}, a novel origin of nova dust has been suggested: that dust forms in radiative shock fronts internal to the nova ejecta \citep{Derdzinski_etal_2017}. This theory predicts that the rapidly-cooling, high-density, shocked gas could create the ideal environment for dust to condensate, while shielding it from the irradiation of the central white dwarf \citep{Derdzinski_etal_2017}. This theory is supported by recent observations of radio synchrotron flares---associated with particle acceleration in shocks---contemporaneous with dust dips in the optical
\citep{Chomiuk_etal_2021b,Babul_etal_2022}. Shocks are seen as the primary sites of dust formation in massive-star colliding-wind binaries \citep{Usov_1991, Lau_etal_2022}, bolstering the case that the same shocks that produce $\gamma$-rays in novae may also form dust.

Their nearby distances and $\sim$Eddington luminosities make novae the brightest IR transients regularly detected in the sky, and dust formation keeps them bright at IR wavelengths, often for months. This makes novae one of the best laboratories to observe and explore dust formation around explosive transients. As we will see in this paper, novae regularly reach $K$-magnitudes $< 8$, and the brightest novae saturated our IR detector on the SMARTS 1.3m telescope ($K < 5$). Wide-field time-domain IR surveys are currently coming of age, with $Y$, $J$, and $H$-band surveys like Gattini-IR \citep{De_etal_2020, De_etal_2021}, WINTER \citep{Lourie_etal_2020}, and PRIME \citep{Yama_etal_2023, Hamada_etal_2024}. These surveys will be routinely detecting novae and producing high-quality IR light curves. However, most wide-field
surveys are only observing in a single IR band, and it is currently unclear how well one can determine if a nova formed dust from these
single-band data alone, or in combination with optical monitoring. This paper seeks to clarify what we can expect from these time-domain surveys, and how we can use them to tell if a nova has formed dust. 

With a new hypothesis that dust is formed in nova shocks, and a new generation of IR time-domain surveys, we here revisit the pioneering studies discussed above with a more modern, larger dataset in order to (1) provide an up-to-date estimate of the percentage of novae that produce dust, (2) explore the properties of the dust forming novae, and (3) present a simple, easy-to-measure diagnostic of dust formation in novae, with the aim of motivating future observations and surveys in the near/mid-IR wavelengths. The observations and sample selection is presented in Section~\ref{sec_obs}. Our diagnostic of whether a nova has formed dust is presented in Section~\ref{sec_diagnostic}. In Section~\ref{sec_disc}, we estimate what fraction of novae form dust and test for correlations between dust properties and other characteristics of the novae. We conclude in Section~\ref{sec_conc}.

\section{Observations}
\label{sec_obs}
\subsection{Sample of Nova Light Curves}
The optical and IR light curves analysed in this paper are from the \textit{Stony Brook / SMARTS Atlas of (mostly) Southern Novae}\footnote{\url{https://www.astro.sunysb.edu/fwalter/SMARTS/NovaAtlas/atlas.html}} \citep{Walter_etal_2012}. The Atlas consists of photometric and spectroscopic data of 93 novae in the Galaxy and the Magellanic clouds that erupted during $\sim$2003--2019. These have measurements of varying cadence using the Johnson-Kron-Cousins $B, V, R_C,$ and $I_C$ filters and the CIT/CTIO $J, H,$ and $K$ filters. Most of the photometry was obtained using the ANDICAM dual-channel imager on the 1.3\,m telescope, supplemented by some higher cadence data obtained with the SMARTS 0.9\,m and 1.0\,m telescopes \citep{Walter_etal_2012}.
Data were reduced in a consistent manner using custom pipelines; details of the photometric observations and data reduction can be found in \citet{Walter_etal_2012}. 
Since the uncertainties in the SMARTS photometry balloon to large values for magnitudes $\gtrsim 20$ mag, we take this as our faint detection limit, and adopt a bright saturation limit of 5 mag. We have also ignored any measurements with error $\geq 0.5$ mag.

Of the 93 novae with photometry from the SMARTS Atlas, 79 have at least 20 measurements in each of the $V$-band and the $K$-band (the bands we focus on here for dust diagnostics; see Section~\ref{sec_vk}), which we consider the minimum number of measurements for a nova to be included in our analysis. We then eliminated novae that were severely lacking in data cadence (some novae have large time gaps in their light curves) and novae that were not useful to analyse due to consistently high measurement errors. The literature indicates that dust events usually ($\geq 90$~per~cent) start within the first 150 days of eruption and can last for as little as 40 days \citep{Strope_etal_2010}. All novae eliminated had a time gap of at least 50 days before day 150, during which a dust event could in theory have occurred and been missed. 
There are two exceptions: Nova SMC 2016 and V1311 Sco, which we both eliminate.
The IR photometry of N SMC 2016 has large uncertainties, and there is a large gap in time coverage from day $\sim$120--220 during which a dust event could have occurred. V1311 Sco lacks SMARTS measurements within 27 days of eruption, during which time this rapidly evolving nova drops by $\sim$7.5 mag in the $V$ band (according to AAVSO data), which is most of the ``action" of the nova eruption. We deem too much of this nova eruption was missed by the SMARTS photometry to be included in our sample.

\begin{figure*}
    \centering \includegraphics[width=1.0\linewidth]{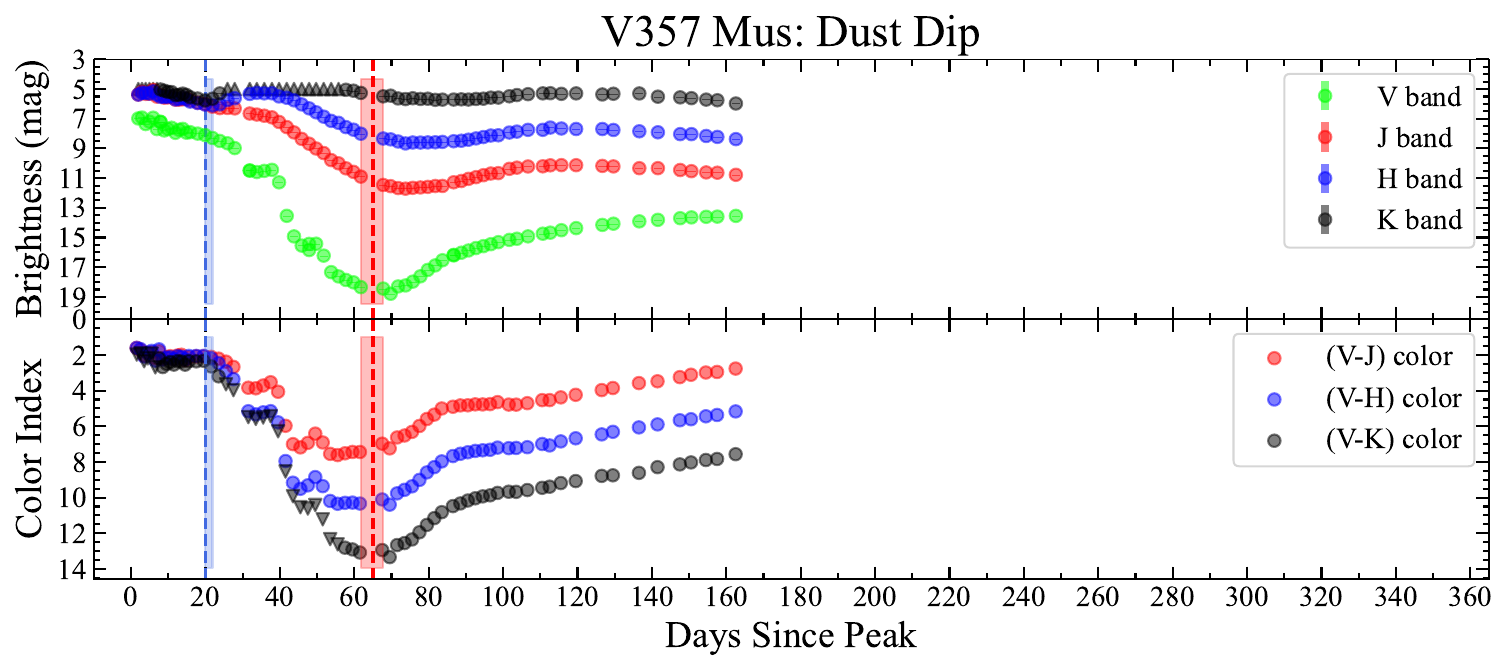}
    \caption{Example $V$, $J$, $H$, and $K$ light curves, and corresponding colour curves of nova V357~Mus, which we classify as `Dust Dip'. Its $V$-band light curve dramatically declines and bottoms out around day 65, while the $K$-band light curve remains relatively constant and even increases a bit during the $V$-band dust dip. $J$- and $H$-bands show behaviour intermediate to that of $V$- and $K$-bands. The blue vertical bar denotes the earliest feature attributable to the dust event (in this case, a rise in the $K$-band light curve), while the red vertical bar marks the largest redward excursion of the ($V-K$) colour (i.e., the `bottom' of the dust dip). The width of the bars is set by the cadence of the photometry at these times.}
    \label{fig:v357mus}
\end{figure*} 

In some cases, the $K$-band light curve is saturated for most of the nova eruption, including during the dust dip. For such novae, we try to complement the SMARTS data with available data in the literature. We use IR data for nova V339~Del from \citet{Burlak+15} and for nova V5668~Sgr from \citet{Banerjee+16} and \citet{Gehrz_etal_2018}. SMARTS/Andicam data were re-reduced for nova V906~Car by \citet{Wee+20} and confirmed as unsaturated --- we use their photometry here.

After excluding novae with poor data cadence and/or quality, we analysed 40 novae that had sufficient early observations and temporal coverage to assess dust formation.
We are only considering the first year of eruption, 
but we expect 
the majority of dust forming novae to form dust over this time frame (e.g. \citealt{Strope_etal_2010}). 
According to compilations by \citet{Strope_etal_2010} and \citet{Williams_etal_2013}, V445~Pup is the nova with the latest dust dip, and it began showing evidence of dust formation around day 218, with a light curve minimum around day 600 \citep{Woudt_etal_2009}; it is also a unique He nova with no H emission detected, which may contribute to its unique dust properties \citep{Ashok_etal_2003,Banerjee_etal_2023b}. If an unexpected change in nova color  was observed at late times, well beyond a year, the association with a dust formation event in the nova ejecta would likely be ambiguous. Therefore, we restrict our study to dust formation events in the first year of the eruption.

We present our sample of novae in Table \ref{tab:sample}, with the time of optical maximum $t_{\mathrm{peak}}$ and the time to decline by two magnitudes from light curve peak, $t_2$. `Class' lists our classification of the nova's dust formation properties (see Section~\ref{sec_cat}), and we also list whether the nova was detected in GeV $\gamma$-rays by the Large Area Telescope \citep[LAT;][]{2009ApJ...697.1071A} on the \textit{Fermi Gamma-ray Space Telescope} (of course this is only feasible for novae that erupted after the launch of \textit{Fermi}).
`t$_{\mathrm{dust, start}}$' refers to the earliest time in the light curve, in days (relative to t$_{\mathrm{peak}}$), that we distinguish a feature in the light curve that is caused by dust formation. `$\Delta t_{\mathrm{dust}}$' is the time from t$_{\mathrm{dust, start}}$ to the `bottom' of the dust event, which is immediately after the ($V-K$) colour appears to have finished the large redward shift caused by dust formation. 
`$\Delta V_{\mathrm{dust}}$' is the change in the $V$-band magnitude from t$_{\mathrm{dust, start}}$ to the bottom of the dust event.
The first reference(s) given under the `Ref.' column (to the left of the slashes) is the source(s) where $t_{\mathrm{peak}}$ and $t_2$ are measured. In cases where the optical peak of the eruption light curve was not well-captured (i.e. there is a gap $>$1 week between pre-eruption non-detection and eruption discovery), we only place an upper limit on $t_2$ (as most optical light curves of nova eruptions become more shallow as they evolve, barring dips caused by dust events). The second reference (between slashes in the `Ref.' column) refers to the work/method used in classifying the dust formation properties. The third reference (to the right of the slashes in the `Ref.' column) is the citation to the $\gamma$-ray detection or upper limit.

\begingroup
\renewcommand{\arraystretch}{1.25}
\begin{table*}\caption{Dust classifications for the novae included in our sample, along with the time of optical peak, t$_2$, if they are detected by \textit{Fermi}-LAT as a $\gamma$-ray source, and the properties of their dust formation feature (should it exist).} \label{tab:sample}
    \begin{tabular}{|c c c c c c c c c|}
    \hline
    Name & t$_{\mathrm{peak}}$ & t$_2$ & Class & $\gamma$-ray detected? & t$_{\mathrm{dust, start}}$ & $\Delta t_{\mathrm{dust}}$ & $\Delta V_{\mathrm{dust}}$ & Ref.\\
     & (MJD) &(days) & & & (days) & (days) & (mag) & \\
    \hline
 V475 Sct & 52883.8$^{+0.5}_{-0.5}$ & 22$^{+1}_{-1}$ & Dust Dip   & -- & 40.0$^{+0.2}_{-0.7}$ & 76.0$^{+1.2}_{-2.9}$ & 6.0 & 1,25/2/--\\ 
 V574 Pup & 53331.7$^{+0.1}_{-0.6}$ & 6.6$^{+0.2}_{-1.0}$ & None   & -- & -- & -- & -- & 3/2/--\\ 
 N LMC 2005 & 53710.1$^{+1.0}_{-1.0}$ & 70.4$^{+2.2}_{-2.2}$ & Dust Dip & -- & 73.0$^{+1.0}_{-0.1}$ & 110.0$^{+2.1}_{-0.1}$ & 1.7 & 4/2/-- \\ 
 RS Oph (2006) & 53778.9$^{+0.1}_{-0.1}$ & 7.9$^{+0.1}_{-0.1}$ & None  & Y$^a$ & -- & -- & -- & 5/6/7\\
 N LMC 2009a & 54864.5$^{+2.5}_{-2.5}$ & 7.0$^{+2.7}_{-2.7}$ & Unsure & -- & -- & -- & -- & 8/2/--\\
 N LMC 2009b & -- & -- & Dust Dip  & -- & 50.0$^{+17.5}_{-39.0}$$^b$ & 90.0$^{+25.4}_{-39.0}$$^b$ & 3.9$^b$ & --/2/--\\ 
 V5584 Sgr & 55138.4$^{+1.4}_{-2.0}$ & 40.7$^{+142.1}_{-27.3}$ & Dust/Unsure$^c$ & N & -- & -- & -- & 9/2/10\\ 
 V5588 Sgr & 55658.1$^{+1.0}_{-1.0}$ & 46.0$^{+1.4}_{-1.0}$ & None  & N & -- & -- & -- & 9/2/10\\ 
 V1312 Sco & 55714.3$^{+0.2}_{-3.3}$ & 12.9$^{+1.3}_{-3.4}$ & None  & N & -- & -- & -- & 9/2/10\\ 
 PR Lup & 55787.4$^{+0.7}_{-0.2}$ & 14.7$^{+0.7}_{-0.6}$ & Dust/Unsure$^c$ & N & -- & -- & -- & 9/2/10 \\ 
 V834 Car & 55987.4$^{+0.1}_{-0.2}$ & 20.1$^{+0.1}_{-0.2}$ & None  & N & -- & -- & -- & 8/2/10\\ 
 N LMC 2012a & 56006.4$^{+6.0}_{-6.0}$ & $<$2.0 & None & -- & -- & -- & -- & 11/2,11/--\\ 
 V1368 Cen & 56008.5$^{+1.8}_{-1.8}$ & 16.1$^{+1.8}_{-1.8}$ & Dust Dip  & N & 36.0$^{+4.7}_{-2.3}$ & 70.0$^{+6.6}_{-3.4}$ & 2.1 & 12/2/10\\ 
 V2676 Oph & 56021.5$^{+1.9}_{-1.0}$ & 83.8$^{+2.1}_{-1.4}$ & Dust Dip  & N & 85.0$^{+2.8}_{-1.2}$ & 100.0$^{+5.5}_{-1.2}$ & 7.4 & 9/2/10\\ 
 V1428 Cen & 56024.8$^{+0.5}_{-0.4}$ & 10.8$^{+0.9}_{-0.5}$ & IR Excess & N & 20.0$^{+4.4}_{-4.7}$ & 55.0$^{+5.6}_{-6.0}$ & 5.0 & 9/2/10\\ 
 V5589 Sgr & 56039.5$^{+0.1}_{-1.6}$ & 2.3$^{+0.6}_{-1.6}$ & None   & N & -- & -- & -- & 9/2/10\\ 
 V2677 Oph & 56068.3$^{+0.2}_{-2.6}$ & 6.7$^{+0.3}_{-2.7}$ & None   & N & -- & -- & -- & 9/2/10\\   
 V1324 Sco & 56098.0$^{+0.4}_{-0.3}$ & 24.2$^{+0.4}_{-0.3}$ & Dust Dip  & Y & 19.0$^{+0.2}_{-2.7}$ & 47.0$^{+0.2}_{-5.5}$ & 8.9 & 9/2/13\\ 
 V5591 Sgr & 56105.3$^{+0.2}_{-0.7}$ & 2.3$^{+0.3}_{-0.7}$ & None   & N & -- & -- & -- & 9/2/10\\ 
 V5592 Sgr & 56115.5$^{+4.3}_{-4.3}$ & $<$16.8 & Dust Dip   & N & 15.0$^{+15.2}_{-1.7}$ & 44.0$^{+15.2}_{-13.9}$ & 6.4 & 12,14/2/10\\ 
 V1533 Sco & 56447.5$^{+0.7}_{-3.8}$ & 7.9$^{+3.1}_{-3.9}$ & IR Excess  & N & 14.0$^{+1.9}_{-1.2}$ & 26.0$^{+2.6}_{-1.7}$ & 1.5 & 9/2/10 \\  
 V339 Del & 56520.7$^{+0.1}_{-0.5}$ & 11.3$^{+0.4}_{-0.5}$ & Dust Dip   & Y & 27.0$^{+2.4}_{-7.6}$ & 55.0$^{+2.7}_{-7.6}$ & 2.5 & 9,26/2/13\\ 
 N LMC 2013 & 56577.3$^{+4.0}_{-4.0}$ & 52$^{+4.9}_{-4.1}$ & Dust Dip   & -- & 48.0$^{+2.9}_{-0.1}$ & 110.0$^{+4.0}_{-4.2}$ & 2.6 & 15,16/2/--\\  
 V1369 Cen & 56640.7$^{+2.0}_{-1.9}$ & 37.7$^{+2.4}_{-4.1}$ & IR Excess  & Y & 30.0$^{+6.7}_{-11.3}$ & 78.0$^{+6.7}_{-11.3}$ & 2.7 & 9/2/17 \\ 
 
 V745 Sco (2014) & 56694.8$^{+1}_{-0.1}$ & 2.5$^{+1}_{-0.5}$ & IR Excess   & N$^d$ & 5.5$^{+0.1}_{-0.9}$$^e$ & 13.0$^{+0.6}_{-1.0}$$^e$ & 3.3$^e$ & 18/23/10\\ 
 V5666 Sgr & 56705.4$^{+1.0}_{-1.0}$  & 68.5$^{+1.1}_{-1.1}$  & Unsure  & N & -- & -- & -- & 12/2/10 \\ 
 V1534 Sco & 56744.5$^{+1}_{-1}$ & 5.6$^{+1.4}_{-1.4}$ & None   & N & -- & -- & -- & 19/2,24/10\\ 
 V1535 Sco & 57064.3$^{+1}_{-1}$ & 9.2$^{+1.4}_{-1.4}$ & None  & N$^d$ & -- & -- & -- & 19/2/10\\  
 V5667 Sgr & 57068.7$^{+2.0}_{-0.9}$ & 64.0$^{+2.0}_{-1.3}$ & Unsure   & N & -- & -- & -- & 9/2/10\\ 
 V5668 Sgr & 57102.4$^{+0.3}_{-0.3}$ & 74.7$^{+3.3}_{-3.8}$ & Dust Dip  & Y & 75.0$^{+6.0}_{-4.1}$ & 105.0$^{+6.7}_{-4.2}$ & 6.7 & 9/2/17 \\ 
 V2944 Oph & 57125.6$^{+1.0}_{-1.0}$ & 32.5$^{+1.1}_{-1.2}$ & Unsure & N & -- & -- & -- & 20/2/10  \\ 
 V3661 Oph & 57459.8$^{+1}_{-1}$ & 3.9$^{+1.4}_{-1.4}$ & Unsure   & -- & -- & -- & -- & 19/2/--\\ 
 V3662 Oph & 57873.2$^{+3}_{-1}$ & 37.0$^{+6.7}_{-5.7}$ & Dust Dip   & -- & 45.0$^{+4.0}_{-0.1}$ & 60.0$^{+4.5}_{-0.1}$ & 4.0 & 9/2/--\\ 
 V1405 Cen & 57891.5$^{+2.0}_{-1.0}$ & 108.5$^{+5.4}_{-14.1}$ & None  & -- & -- & -- & -- & 12/2/--\\ 
 V612 Sct & 57964.0$^{+0.9}_{-1.0}$ & 132.2$^{+78.2}_{-12.5}$ & Unsure   & N & -- & -- & -- & 9/2/9\\ 
 V357 Mus & 58134.4$^{+1.1}_{-11.1}$ & 22.9$^{+1.6}_{-11.1}$ & Dust Dip  & Y & 20.0$^{+1.9}_{-0.1}$ & 65.0$^{+3.3}_{-3.3}$ & 10.3 & 9/2/21 \\ 
 V1661 Sco & 58136.8$^{+1}_{-1}$ & 11.2$^{+3.2}_{-4.2}$ & Dust Dip   & -- & 13.0$^{+2.6}_{-13.0}$ & 55.0$^{+3.0}_{-13.1}$ & 6.2 & 9/2/--\\ 
 FM Cir & 58146.3$^{+1.0}_{-0.9}$ & 133.7$^{+1.0}_{-2.2}$ & Unsure  & N & -- & -- & -- & 9/2/9\\ 
 V1662 Sco & 58159.3$^{+1}_{-1}$ & 14.5$^{+1.1}_{-0.7}$ & IR Excess  & -- & 14.0$^{+0.6}_{-1.5}$ & 47.0$^{+1.6}_{-1.6}$ & 4.3 & 9/2/9 \\ 
 V906 Car & 58205.5$^{+1.0}_{-0.9}$ & 43.7$^{+1.3}_{-1.1}$ & IR Excess  & Y & 20.0$^{+0.6}_{-0.4}$ & 75.0$^{+2.6}_{-1.6}$ & 2.9 & 9/2/22 \\ 
    \hline    
    \end{tabular}
References (light curve parameters/ dust classification/ \emph{Fermi}/LAT detection): 1=\citet{Morgan_etal_2005}; 2= This work; determined using $V$ and $K$ band light curves; 3=\citet{Samus+04}; 4= \citet{Liller+07}; 5=\citet{Hounsell_etal_2010}; 6=\citet{Evans_etal_2007}; 7=\citet{Cheung2022}; 8=\citet{Bode_etal_2016}; 9= Craig et al.\ 2025, in preparation;  10=\citet{Franckowiak_etal_2018}; 11= \citet{Schwarz_etal_2015}; 12= Measured here using AAVSO and ASAS-SN light curves; 13=\citet{Ackermann_etal_2014}; 14=\citet{Nishimura_etal_2012}; 15=\citet{Wyrzykowski_etal_2013}; 16= Measured from OGLE I Band light curve \citet{Mroz2016}; 17=\citet{Cheung_etal_2016}; 18= Molina et al.\ 2024, accepted; 19=\citet{Munari_etal_2017}; 20=\citet{Srivastava_etal_2016};  21=\citet{Gordon_etal_2021}; 22=\citet{Aydi_etal_2020a}; 23=\citet{Banerjee_etal_2023}; 24=\citet{Joshi_etal_2015}; 25=\citet{Chocol_etal_2006}; 26=\citet{Gehrz_etal_2015}
\\
$^a$ While the 2006 eruption of the recurrent nova RS Oph was not observed with \emph{Fermi}/LAT, the 2021 eruption was, and was detected.\\
$^b$ Values are relative to the time of discovery for N LMC 2009b (MJD 54956.0; \citealt{Liller_Monard_2009}) due to a lack of data around light curve peak. 
$^c$ While we are unsure if V5584 Sgr and PR Lup are `Dust Dip' or `IR Excess' novae, it is clear that they formed dust.\\
$^d$ Marginal, sub-threshold hints of detection are seen in \emph{Fermi}/LAT \citep{Franckowiak_etal_2018}.\\
$^e$ V745~Sco continues to redden after the dust event concludes, due to the system's red giant companion star; for this nova, $\Delta t_{\rm dust}$ is the time from t$_{\rm dust,start}$ to when the small bump in the $K$-band caused by the dust event has ended \citep{Banerjee_etal_2023}.
\end{table*}
\endgroup

\subsection{Categorizing Novae by Dust Properties}
\label{sec_cat}
Each of the 40 novae in the sample were manually labelled as one of: `Dust Dip', `IR Excess', `None', and `Unsure', based on inspection of their $V$- and $K$-band light curves. In the near-IR, we focus on the $K$-band (rather than $J/H$) because it provides a longer wavelength baseline and a higher potential contrast with the $V$-band. This behaviour is illustrated in Figure~\ref{fig:v357mus}, where we plot $V$-, $J$-, $H$-, and $K$-band light curves for the nova V357~Mus. This nova shows a dramatic dust dip in the $V$-band, bottoming out around day 65, while the $K$-band plateaus and even brightens a bit during this time. The $J$- and $H$-bands show `in between' behaviour, where the light curves do not decline as much as the $V$-band during the dust dip, but do show some fading that bottoms out around day 75, likely driven by dust formation. 

The `Dust Dip' label refers to novae that have a brightening/plateau in the $K$-band, correlating with a relatively sharp decrease in $V$-band brightness followed by either a recovery or a plateau. V357~Mus is an example of a `Dust Dip' nova (Figure~\ref{fig:v357mus}). We note that not all of the novae in the `Dust Dip' category display features that are easily distinguished as dust when looking at the $V$-band light curve alone. For some, it is only distinguishable when also analysing the increase in $K$-band brightness.

\begin{figure}
    \centering
    \includegraphics[width=1.0\linewidth]{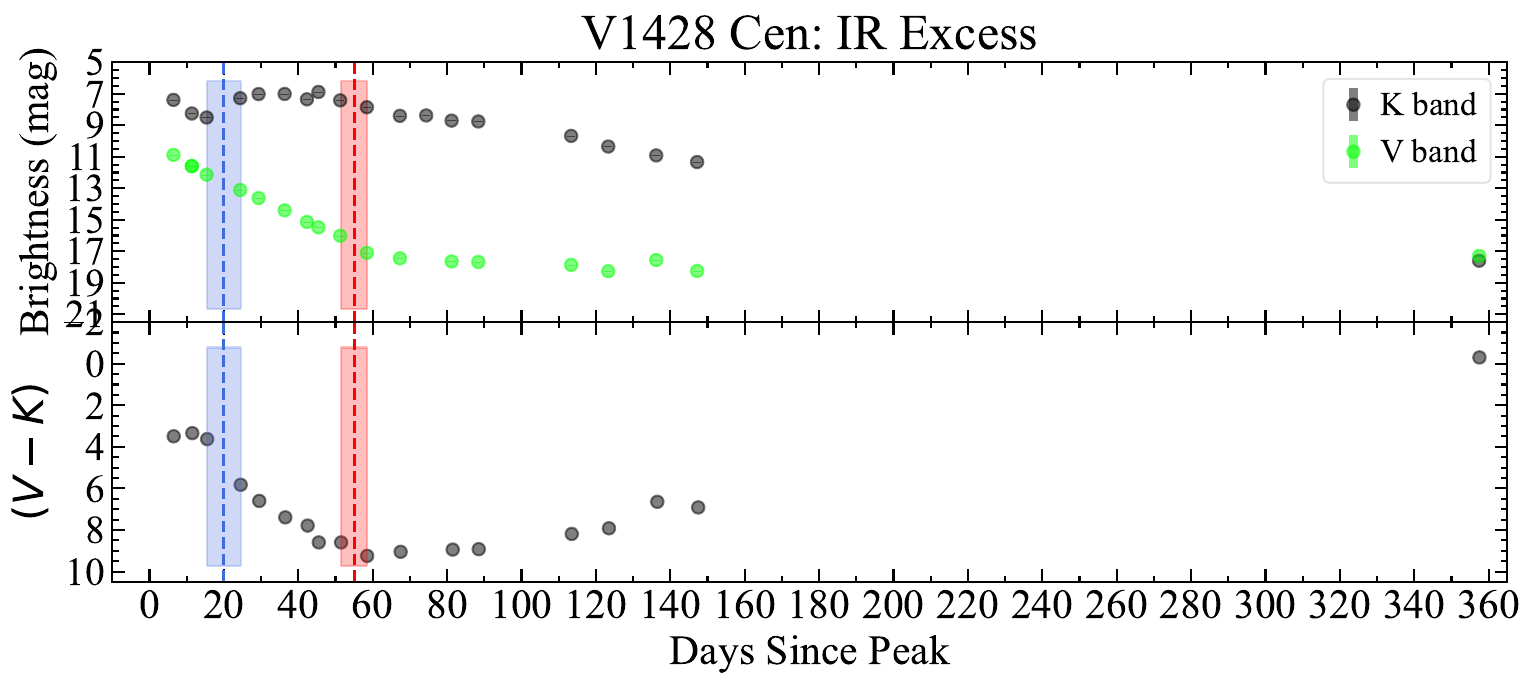}
    \caption{Example $V$ and $K$ light curves, and ($V-K$) colour curve, of a nova that has been classified as `IR Excess': V1428~Cen. Its $V$-band light curves shows a smooth decline while the $K$-band light curve increases in flux around day 20. The ($V-K$) colour curve shows a clear reddening over the course of this dust formation episode. As in Figure~\ref{fig:v357mus}, the blue bar marks the start of the dust dip and the red bar marks the reddest point in the ($V-K$) curve.}
    \label{fig:v1428cen}
\end{figure}

`IR Excess' refers to novae that have a significant brightening in the IR flux at a time when it would be reasonable for a dust event to occur, but this increase in the IR is not reflected by a significant dimming in the optical light, typically expected from dust absorption (see Figure~\ref{fig:v1428cen} for an example). The main difference between the `Dust dip' novae and  `IR Excess' is likely due to the location of dust formation relative to our line of sight: `Dust Dip' events are referred to as \textit{optically thick dust shells} by \citet{Gehrz_1988}, marking dust that is formed in the ejecta along the line of sight, whereas `IR Excess' events have been deemed \textit{optically thin dust shells}, where newly formed dust is present but does not cover the bulk of the optical pseudo-photosphere of the nova.

\begin{figure}
    \centering
    \includegraphics[width=1.0\linewidth]{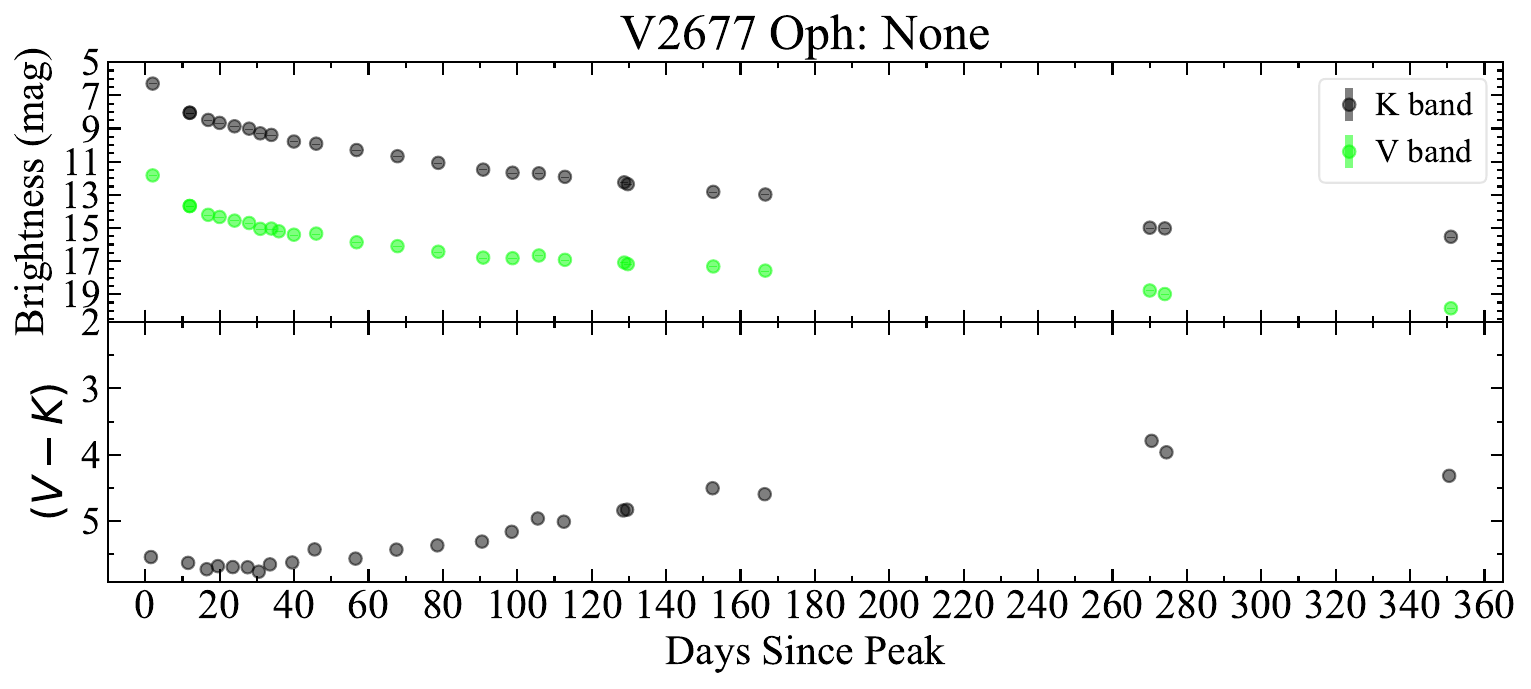}
    \caption{Example $V$ and $K$ light curves, and ($V-K$) colour curve, of a nova that has been classified as `None': V2677~Oph. Both the $V$ and $K$ light curves monotonically decline from maximum and ($V-K$) smoothly evolves towards bluer colours.}
    \label{fig:v2677oph}
\end{figure}

The `None' label refers to novae that do not display evidence of a dust formation event in their $V$- and $K$-band light curves, and for which there is enough data to assert that there was not a dust event with a reasonable degree of confidence. Figure~\ref{fig:v2677oph} shows an example; in typical cases categorized as `None', the $V$- and $K$-band light curves track one another smoothly; Moreover, the change of ($V-K$) colours as a function of time does not show an episode of reddening, and is instead observed to evolve towards the blue.

\begin{figure}
    \centering
    \includegraphics[width=1.0\linewidth]{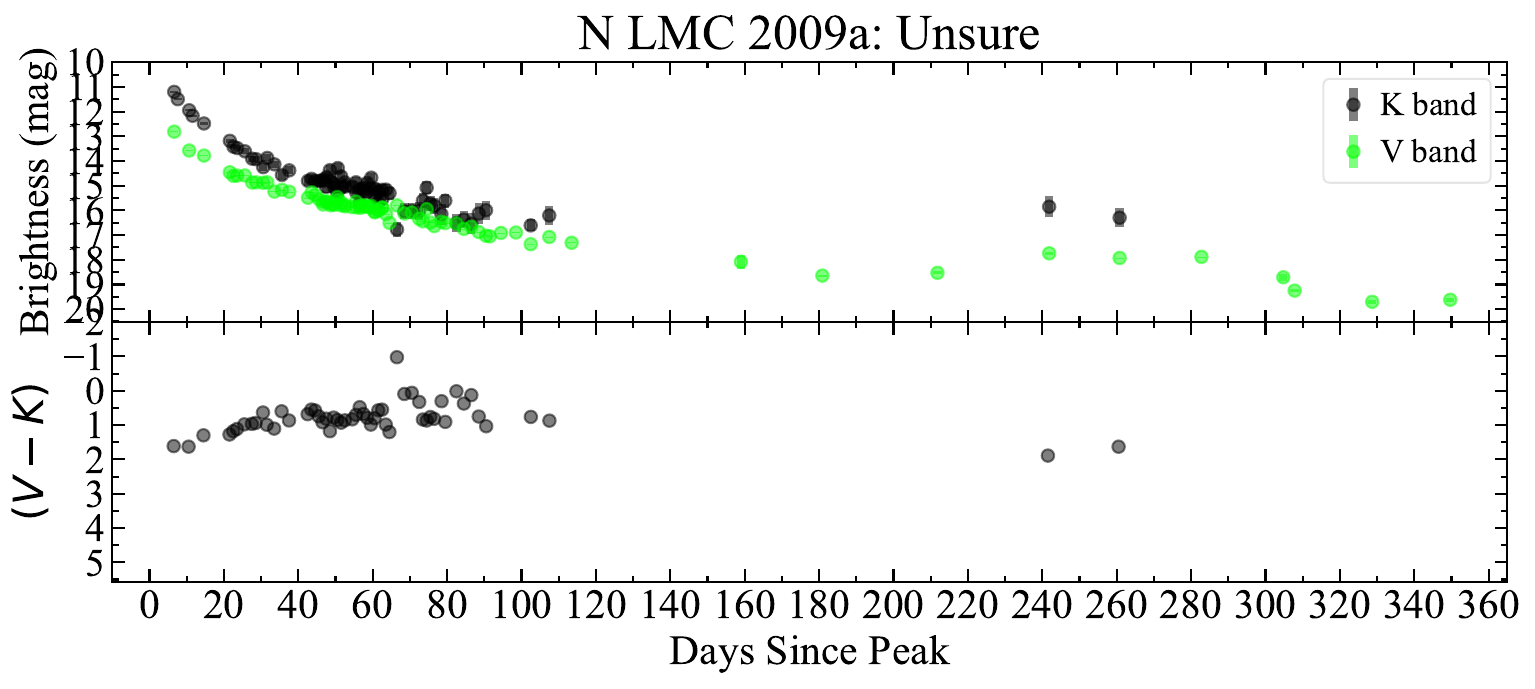}
    \caption{Example $V$ and $K$ light curves, and ($V-K$) colour curve, of a nova that has been classified as `Unsure': N LMC 2009a. It has a gap in photometric coverage during which a dust formation event could have occurred.}
    \label{fig:nlmc2009a}
\end{figure}

\begin{figure}
    \centering
    \includegraphics[width=1.0\linewidth]{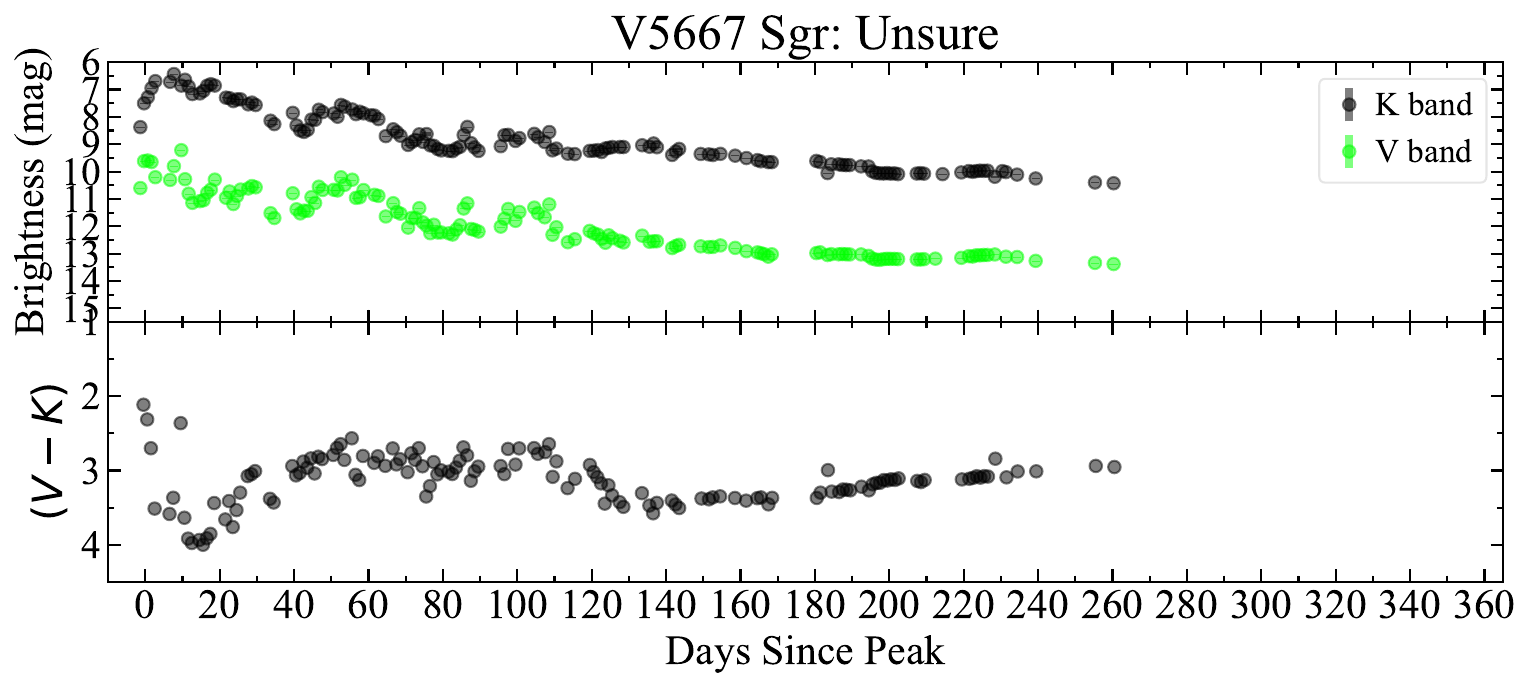}
    \caption{Example $V$ and $K$ light curves, and $V-K$ colour curve, of another nova that has been classified as `Unsure':   
    V5667~Sgr. It shows some change in $V-K$ colour but has no clear features associated with the colour change in the $V$ and $K$ light curves that would distinguish a dust event.}
    \label{fig:v5667sgr}
\end{figure}

The `Unsure' label is assigned to a nova that may have features of the `IR Excess' or `Dust Dip' categories, but these features are not pronounced enough to categorize the nova with any certainty. Novae were sometimes categorized as `Unsure' if there was a gap in photometric coverage, often due to solar conjunction (e.g. N LMC 2009a; Figure~\ref{fig:nlmc2009a}), during which a dust formation event could have occurred, or if there were relatively mild features in the ($V-K$) colour evolution which were not accompanied by clear dips/excesses in the light curves (e.g. V5667 Sgr; Figure~\ref{fig:v5667sgr}). In categorizing the novae, we erred to an `Unsure' label when we were uncertain about classification, in order to avoid introduction of bias. 

\begin{figure}
    \centering
    \includegraphics[width=1.0\linewidth]{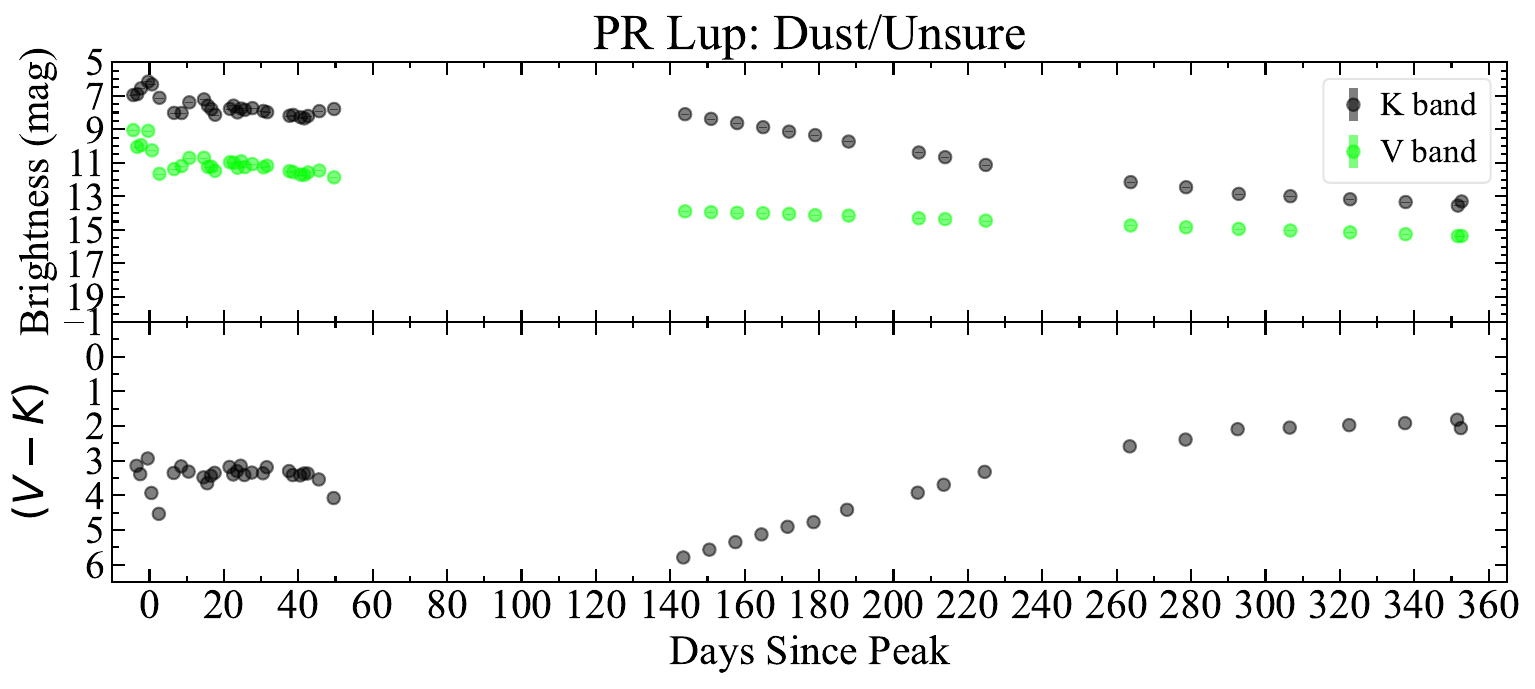}
    \caption{$V$ and $K$ light curves, and ($V-K$) colour curve, for PR~Lup. PR~Lup almost certainly formed dust, but it is unclear if it should be classified as `IR Excess' or `Dust Dip' due to a large gap in observations around solar conjunction.}
    \label{fig:prlup}
\end{figure}

\begin{figure}
    \centering
    \includegraphics[width=1.0\linewidth]{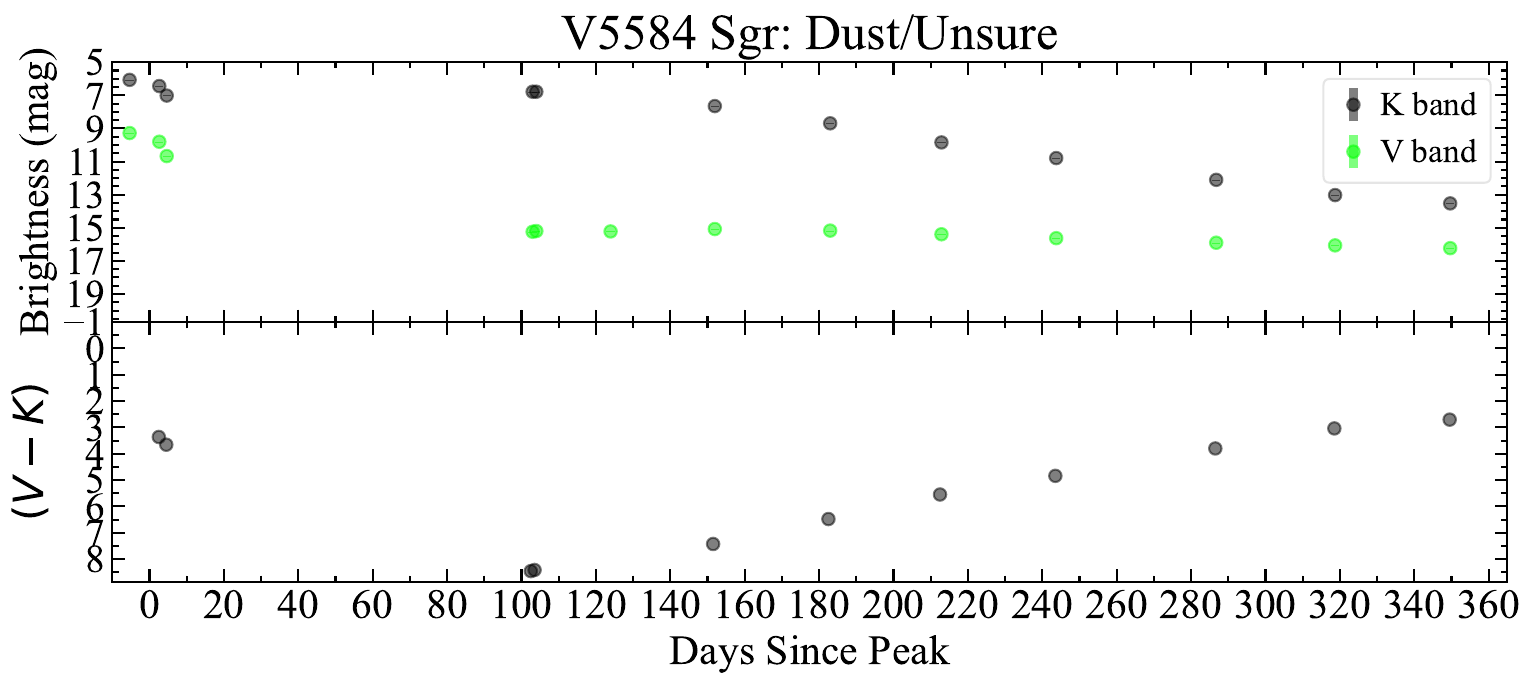}
    \caption{$V$ and $K$ light curves, and ($V-K$) colour curve, for V5584~Sgr. V5584~Sgr almost certainly formed dust, but it is unclear if it should be classified as `IR Excess' or `Dust Dip' due to a large gap in observations around solar conjunction.}
    \label{fig:v5584sgr}
\end{figure}

In the cases of most `Unsure' novae, we are unsure if they formed dust. V5584~Sgr and PR~Lup are exceptions in that they almost certainly formed dust, but due to large observation gaps caused by solar conjunction, we did not capture the dust formation episodes. We therefore classify them as `Dust/Unsure', but they are almost certainly either `Dust Dip' or `IR Excess', based on the large change in the ($V-K$) colour during the solar conjunction. The light curves and colour plots of PR~Lup and V5584~Sgr can be found in Figure~\ref{fig:prlup} and Figure~\ref{fig:v5584sgr} respectively.

\begin{figure*}
    \centering
    \includegraphics[width=1\linewidth, trim= 0 0.8cm 0 2.5cm,clip]{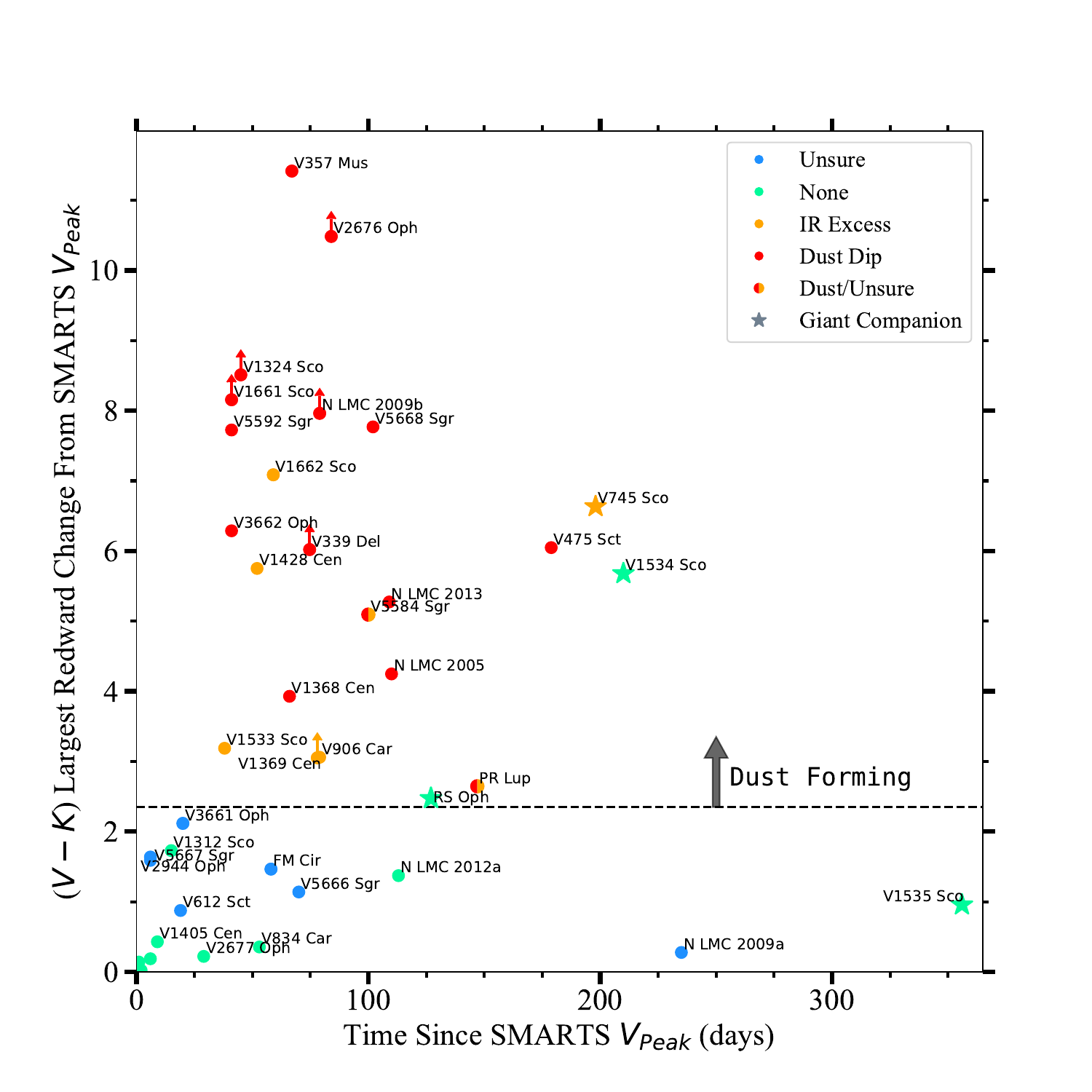}
    \caption{A plot summarizing our recommended ($V-K$) dust formation diagnostic. On the y-axis is plotted the largest redward change in the nova's ($V-K$) colour compared to its colour at SMARTS $V$-band light curve peak. On the x-axis is plotted the time for this colour change to occur. Novae with dust features in their light curves (red and orange dots) are clustered in the upper left corner of the plot, while novae without dust features (cyan dots) are mostly in the lower portion of the plot. The horizontal line at a ($V-K$) colour change of 2.35\,mag does a good job separating novae that form dust from novae that do not. Novae with giant companions complicate this separation, and are denoted as stars rather than dots. Novae in the bottom left corner of the plot are too crowded for labels, and include V574~Pup, V5591~Sgr, V5589~Sgr, and V5588~Sgr.}
    \label{fig:Color_Change_VK}
\end{figure*}

In cases where there is additional evidence from the literature (i.e., spectroscopy) that supports a nova forming dust or not, we take this information into account in our dust classification. Take for example, the recurrent nova RS~Oph; while dust has been detected in the binary, IR spectroscopy and imaging suggests that the silicate grains have been processed and therefore likely originate in the red giant companion's wind \citep{Evans_etal_2007, Barry_etal_2008}; we therefore classify it as `None'.  On the other hand, another recurrent nova with a red giant companion, V745~Sco, was suggested to have formed dust based on detection of broad CO lines in IR spectra, which are concurrent with a relatively subtle increase in the IR brightness between day 8 and 10 after visible peak \citep{Banerjee_etal_2023}. While we do not believe V745~Sco shows strong evidence of dust formation from the SMARTS data alone, Banerjee et al.\ point out that `all novae that have shown CO in emission have invariably proceeded to form dust', so we classify it as `IR Excess' here. Yet another nova with a likely giant companion, V1534~Sco, was monitored with IR spectroscopy and showed no CO emission or other signs of dust formation \citep{Joshi_etal_2015}; it is classified here as `None'.

\begin{figure}
    \centering
    \includegraphics[width=1\linewidth]{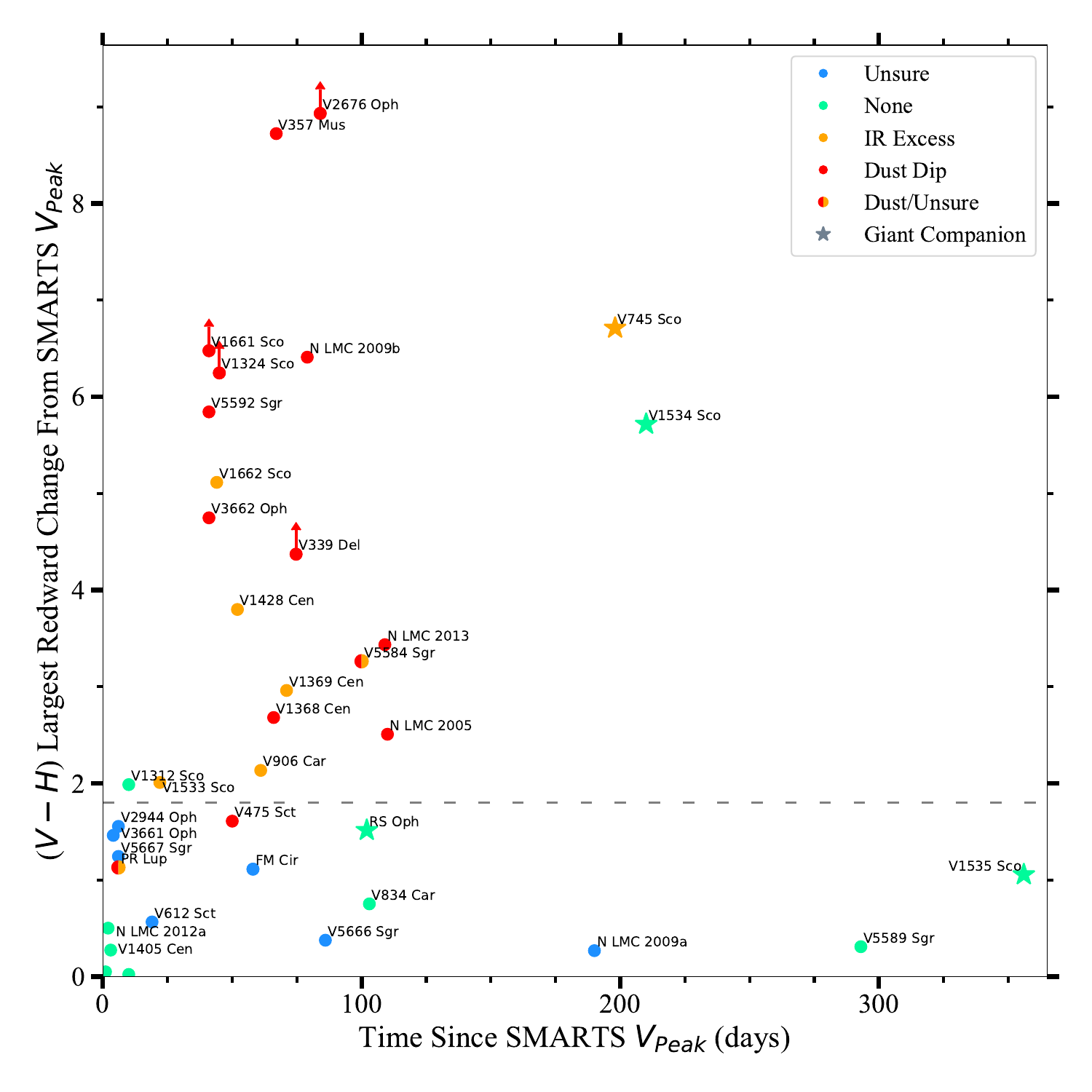}
    \caption{A plot similar to Figure~\ref{fig:Color_Change_VK}, except with the $H$-band in place of the $K$-band.  We include a dotted line at a ($V-K$) colour change of 1.8 that mostly separates the dust-forming from the non-dust-forming novae. Note that the light curves for both PR Lup and V475~Sct do not capture the `bottom' of the dust event in both $V$-band and $K$-band, and as such both novae appear lower on the plot than they would with better coverage. V5668~Sgr is not included due to $H$-band saturation near peak and during its dust dip. Novae in the bottom left corner of the plot are too crowded for labels, and include V2677~Oph, V5588~Sgr, V5591~Sgr, and V574~Pup.}
    \label{fig:Color_Change_VH}
\end{figure}

In summary, our sample contains 13 `Dust Dip' novae (32.5~per~cent), 6 `IR Excess' novae (15~per~cent), 2 `Dust/Unsure' novae (5~per~cent), 12 `None' novae (30~per~cent), and 7 `Unsure' novae (17.5~per~cent). Therefore, out of the 40 novae in our sample, 21 show signatures of dust formation in their light curves (including PR~Lup and V5668~Sgr). Plots like Figures~\ref{fig:v357mus}--\ref{fig:v5584sgr}, showing the $V$ and $K$ light curves and colour curves, are available for all 40 novae in Appendix \ref{sec_full} (Figures~\ref{fig:app_v475sct}--\ref{fig:app_v906car}).



\section{Diagnostics of Dust Formation}
\label{sec_diagnostic} 

\subsection{($V-K$) Colour Diagnostic}
\label{sec_vk}

Colour curves, comparing an optical band with a near-IR band as a function of time, are particularly useful for distinguishing dust formation events in novae. Dust should cause the optical band to dim and the near-IR band to plateau or brighten, so their difference should be even more affected by dust than the optical or near-IR light curves alone. From visual inspection of the near-IR $JHK$-band light curves for our novae, it is clear that the $K$-band is the most likely to show an excess in flux at the time of dust formation, and provides the strongest contrast with the $V$-band light curve (e.g. Figure~\ref{fig:v357mus}). We therefore begin our search for a simple diagnostic of dust formation in novae by analysing the ($V-K$) colour curves. 
How the ($V-K$) colour changes with time has potential to be a powerful indicator of dust formation, especially because it is insensitive to interstellar reddening. 

\begin{figure}
    \centering
    \includegraphics[width=1\linewidth]{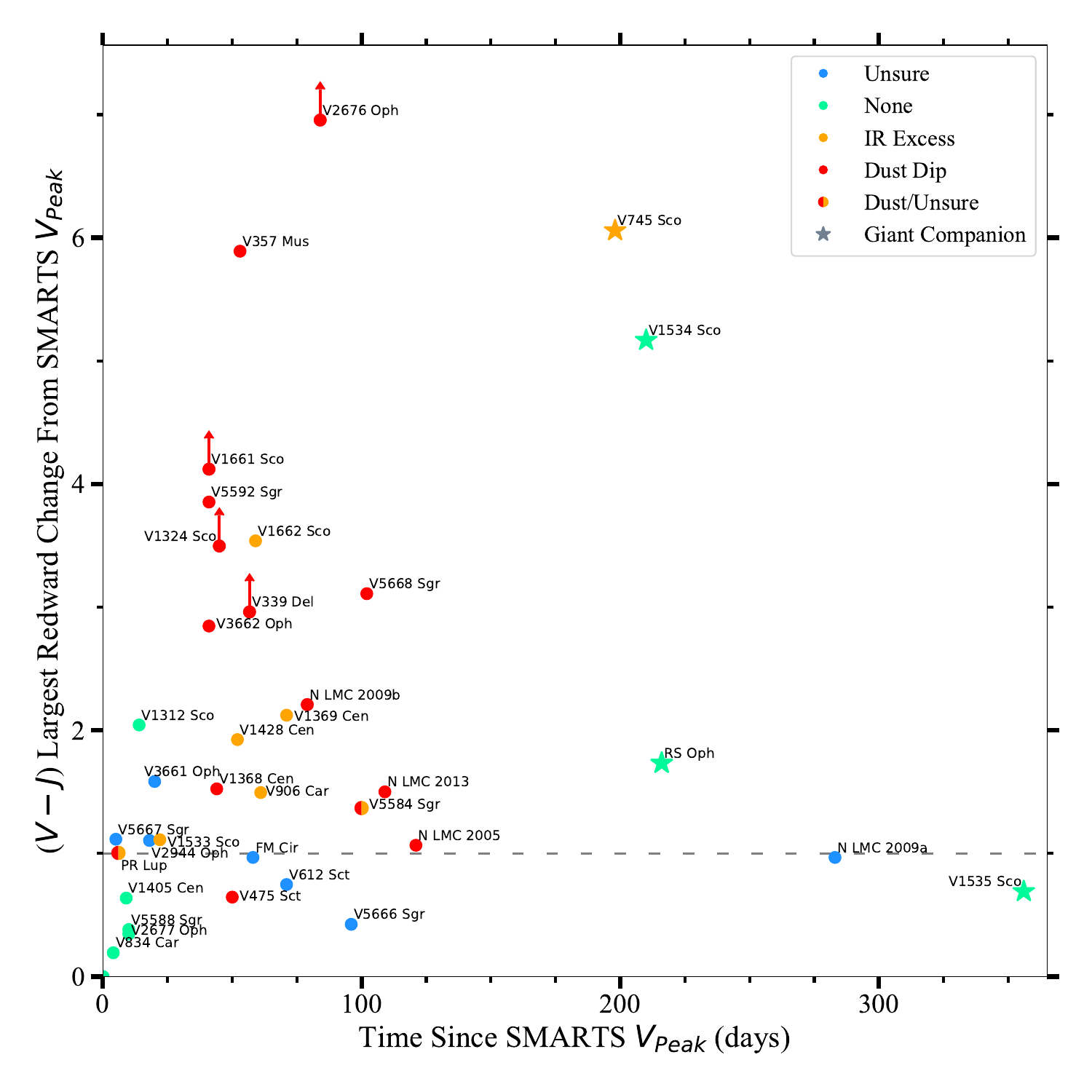}
    \caption{Same as Figure~\ref{fig:Color_Change_VK} but for $(V-J)$ colours. There is no longer a clear separation between the `None'-dusty novae and novae with dust features. We include a dotted line at a ($V-K$) colour change of 1.0 that mostly separates the dust-forming from the non-dust-forming novae. Note that the light curves for both PR~Lup and V475~Sct do not capture the `bottom' of the dust event in both $V$-band and $K$-band, and as such both novae appear lower on the plot than they would with better coverage. Novae in the bottom left corner of the plot are too crowded for labels, and include N LMC 2012a, V5589~Sgr, V5591~Sgr, and V574~Pup.}
    \label{fig:Color_Change_VJ}
\end{figure}

Using the ($V-K$) colour curves, we have separated the dust-forming novae and novae with no evidence of dust formation into distinct, non-overlapping groups, as shown in Figure~\ref{fig:Color_Change_VK}. To calculate the vertical axis of Figure~\ref{fig:Color_Change_VK}, we take the ($V-K$) colour around $V$-band light curve peak, as measured in the SMARTS data (see Figures~\ref{fig:app_v475sct}-\ref{fig:app_v906car}). Because of the limited cadence of the SMARTS data, sometimes these are offset in time from the true light curve peak by several days, with several novae having greater offsets; V574~Pup has the largest offset of 30 days. The average offset is 5.9 days, and the median is 3.1 days. We explore in Appendix \ref{sec:colorchange} how quickly and dramatically the ($V-K$) colour changes around peak, and conclude that the SMARTS photometry is of sufficient cadence to constrain the colour near peak.
We then measure the greatest redward excursion from this peak colour, as observed over the first year of eruption. It is this largest redward change in ($V-K$) colour that is plotted on the y-axis of Figure~\ref{fig:Color_Change_VK}. Novae where the $K$-band light curve saturates or the $V$-band light curve drops below detectability are plotted as lower limits on the ($V-K$) change.
On the horizontal axis of Figure~\ref{fig:Color_Change_VK} we plot the time it took for this change in colour to occur. V1428~Cen (Figure~\ref{fig:v1428cen}), for example, is at 5.75 mag on the vertical axis and 52 days on the horizontal axis, as the most extreme change in ($V-K$) colour is 5.75\,mag between day $\sim$ 6 (the time of SMARTS $V$-band maximum) and day 58.

Novae that form dust, be they `Dust Dip' novae (red dots) or `IR Excess' novae (orange dots), dwell in the upper portion of Figure~\ref{fig:Color_Change_VK}, implying a large redward colour diversion. This region of the plot is characterized by a change in ($V-K$) from peak of $> 2.35$ mag (marked as a horizontal dashed line). Most of the `None' novae (cyan dots) are significantly removed from this upper cluster of dust-forming novae and are found near the bottom of the plot. \textbf{Our best single diagnostic for dust formation in novae is an observation of ($V-K$) colour that is $>$ 2.35\,mag redder than at peak.}

V1534~Sco and RS~Oph, the two cyan `None' points found in the top (dust-forming) portion of the plot, are both systems that have a red giant companion star and as such redden significantly when approaching quiescence \citep{Brandi_etal_2009,Joshi_etal_2015, Munari_Banerjee_2018}. 
We also note evidence suggesting that V1535~Sco, found on the bottom-far right of Figure~\ref{fig:Color_Change_VK}, also has a red giant companion star \citep{Linford_etal_2017}. 

We note that sometimes novae are missed during light curve peak, due to solar conjunction or other gaps in coverage. 
A recent extreme example is the 2020 nova YZ\,Ret, 
which peaked at visual magnitude 3.7 but was noticed only a week after the eruption 
\citep{2022MNRAS.514.2239S}. 
Our recommended dust diagnostic can still be used in these cases if later $V$ and $K$ light curves are obtained, the interstellar reddening to the nova is measured, and the typical intrinsic $(V-K)_0$ colour of a nova at peak is known. For this reason, we estimate $(V-K)_{\rm 0,peak}$ in Appendix \ref{sec_intrinsic}, and find it, on average, to be $1.1$ with a standard deviation of 0.8 mag. The implication is that, if a nova eruption exhibits an intrinsic $(V-K)_0 > 5.1$ during its evolution, it is likely (at the 2$\sigma$ level) to have formed dust.

%



\subsection{Can dust formation be diagnosed from \textit{J}- and \textit{H}-bands?} \label{sec_vj}

We do not expect the excess in IR flux that accompanies dust formation to be as dramatic in the $J$- and $H$-bands compared to $K$-band (Figure \ref{fig:v357mus}; e.g., \citealt{Szkody_etal_1979,Bode_etal_1984,Arai_etal_2010,Aydi_etal_2019_I}). And yet, wide-field time-domain surveys like WINTER and PRIME are being carried out in the $J$- and $H$-bands respectively, and should produce high-quality near-IR light curves for a large fraction of Galactic novae. Can we use these $J$- and $H$-band data sets to diagnose dust formation in novae? 

Figures~\ref{fig:Color_Change_VH} and \ref{fig:Color_Change_VJ} are plots of the largest redward colour change from $V$-band peak vs $\Delta t$ (similar to our ($V-K$) diagnostic plot; Figure~\ref{fig:Color_Change_VK}) for ($V-H$) and ($V-J$), respectively. As in Figure~\ref{fig:Color_Change_VK}, the dust-forming novae tend to sit towards the tops of these plots, and the `None'  novae dwell towards the bottoms. However, the vertical separation between dust-formers and non-dust-formers is not as clear in these plots. A ($V-H$) reddening from peak of $>$1.8 captures nearly all dust-forming novae, but also includes the `None' nova V1312~Sco (Figure~\ref{fig:app_v1312sco}). A ($V-J$) reddening of $>$1.0 similarly includes nearly all dust-forming novae, but also includes V1312~Sco and three `Unsure' novae. Note that the light curves for both PR Lup and V475 Sct do not capture the 'bottom' of the dust event in $J$-band and $K$-band, and as such both novae appear lower on the plot than they would with better coverage. Therefore, while the $J$- and $H$-bands can be used in concert with optical photometry to assess whether a nova is likely to have formed dust, the false positive rate is likely to be higher than if one instead uses ($V-K$) as a diagnostic.

Condensation temperature estimates for dust commonly seen in nova ejecta
range from 1150\,K to 1720\,K, with a median of 1325 K \citep{Ebel_2000,Evans_Rawlings_2008}, which we can use as a rough estimate for the temperature at which grains will form. It is not unreasonable to expect multiple species of dust in the wind of a single nova which can generate a complex IR spectral energy distribution (SED), but the condensation temperature can give us an upper bound on the peak of the 
dust thermal re-emission SED. The blackbody curve for 1325 K peaks at 2.19 \textmu m, which is within the $K$-band ($1.99-2.30$ \textmu m) but redward of the $J$- and $H$-bands ($1.16-1.33$ \textmu m and $1.47-1.79$ \textmu m, respectively). Thermal radiation at that temperature would still cause an increase in the $J$- and $H$-bands, but not as large or easily measured as the change in the $K$-band. At cooler temperatures, dust re-emission may not be detectable in the $J$-band, while still having a 
noticeable effect in the $K$-band.

As in Section~\ref{sec_vk}, it may be useful to know the intrinsic $(V-J)_{0}$ and $(V-H)_{0}$ of nova eruptions at peak; this could enable the assessment of whether a nova forms dust even if no photometry is obtained around light curve peak. We estimate these colours around peak for novae in our sample with good measurements of interstellar reddening in Appendix \ref{sec_intrinsic}, finding $<(V-J)_{\rm 0,peak}> = 0.8 \pm 0.6$ and $<(V-H)_{\rm 0,peak}> = 0.7 \pm 0.6$. Combined with the color-change diagnostics of Figures~\ref{fig:Color_Change_VH} and \ref{fig:Color_Change_VJ}, we conclude that if a nova reaches an intrinsic $(V-J)_0 > 3.0$ or $(V-H)_0 > 3.7$ during its eruption, it likely ($>$95\%) formed dust.





\section{Discussion}
\label{sec_disc}

\subsection{Do slower novae form dust more often than faster novae?} \label{sec_speed}

\begin{figure}
    \centering
    \includegraphics[width=1\linewidth]{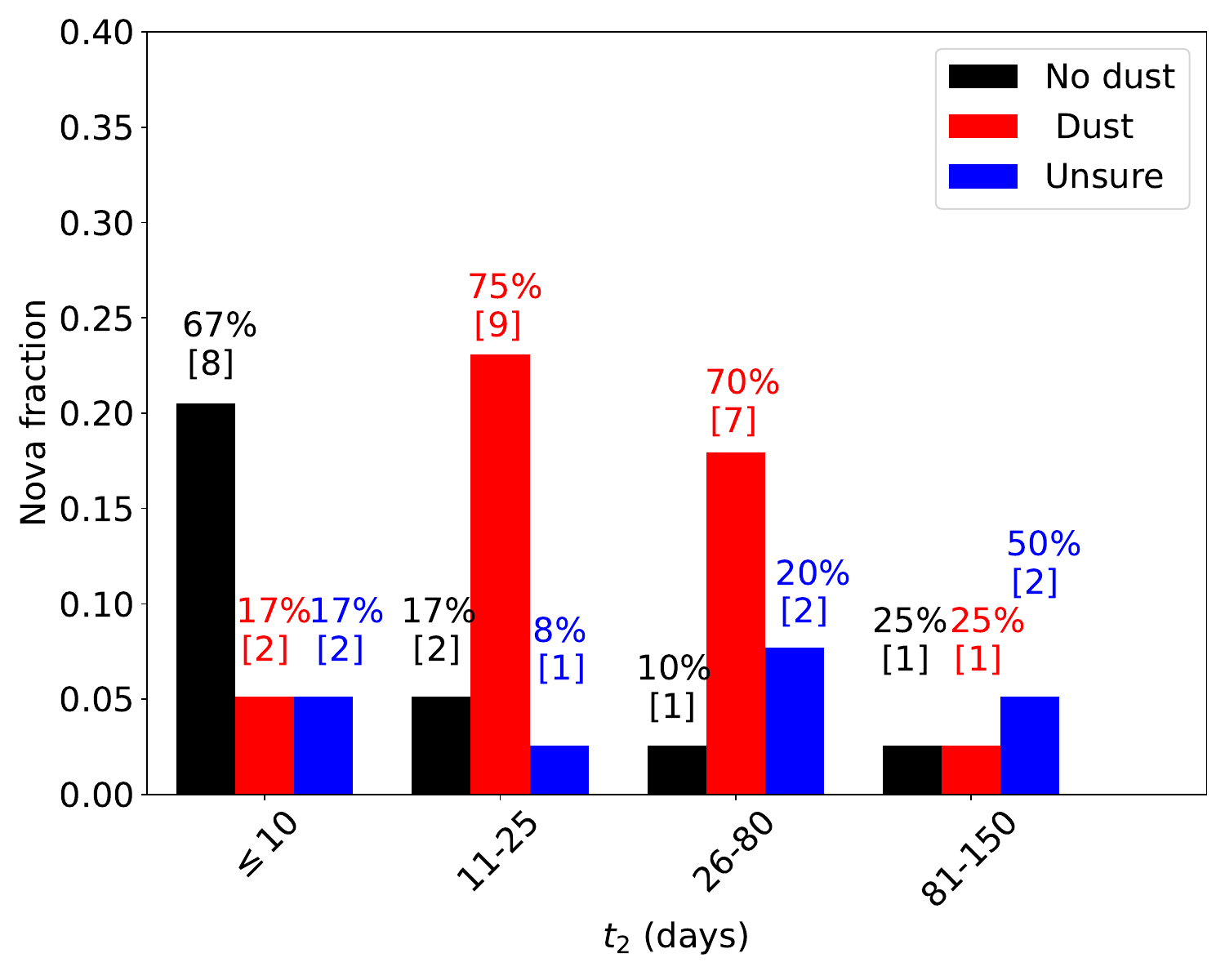}
    \caption{Bar chart comparing the fraction of novae that form dust as a function of speed class. Black bars represent novae that do not show dust formation, red bars represent novae with evidence for dust formation, and blue bars represent novae that are unsure. The errors on the percentages are the following, from left to right: $67^{+10}_{-15}$\%, $17^{+16}_{-6}$\%, $17^{+16}_{-6}$\%, $17^{+16}_{-6}$\%, $75^{+8}_{-16}$\%, $8^{+15}_{-3}$\%, $10^{+17}_{-3}$\%, $70^{+10}_{-17}$\%, $20^{+17}_{-7}$\%, $25^{+27}_{-10}$\%, $25^{+27}_{-10}$\%, and $50^{+20}_{-20}$\%
(see Section~\ref{sec_speed}).}
    \label{fig:speedclass}
\end{figure}

We assess how the likelihood that a nova forms dust depends on its speed class, $t_2$, in Figure~\ref{fig:speedclass}. It is a histogram that divides the novae in our sample into four speed class bins---\textit{very fast}: $t_2 \leq$ 10 days; \textit{fast}: $t_2$ = 11--25 days; \textit{moderately fast}: $t_2$ = 26--80 days; and \textit{slow}: $t_2$ = 81--150 days \citep{Payne-Gaposchkin_1957}. Our results show that for the very fast nova group, $67^{+10}_{-15}$~per~cent do not show dust, with only $17^{+16}_{-6}$~per~cent of them showing clear signatures of dust formation (error bars assume binomial uncertainty with Bayesian confidence intervals calculated as in \citealt{Cameron2011}). In contrast, for the fast and moderately fast novae, the majority
($75^{+8}_{-16}$~per~cent and $70^{+10}_{-17}$~per~cent, respectively) show signatures of dust formation. Slow novae, with $t_2$ between 81 and 150 days, tend to be rare, and there are not enough data to draw clear conclusions about this class. 

\begin{figure}
    \centering
    \includegraphics[width=1\linewidth,trim={0 50 0 100},clip]{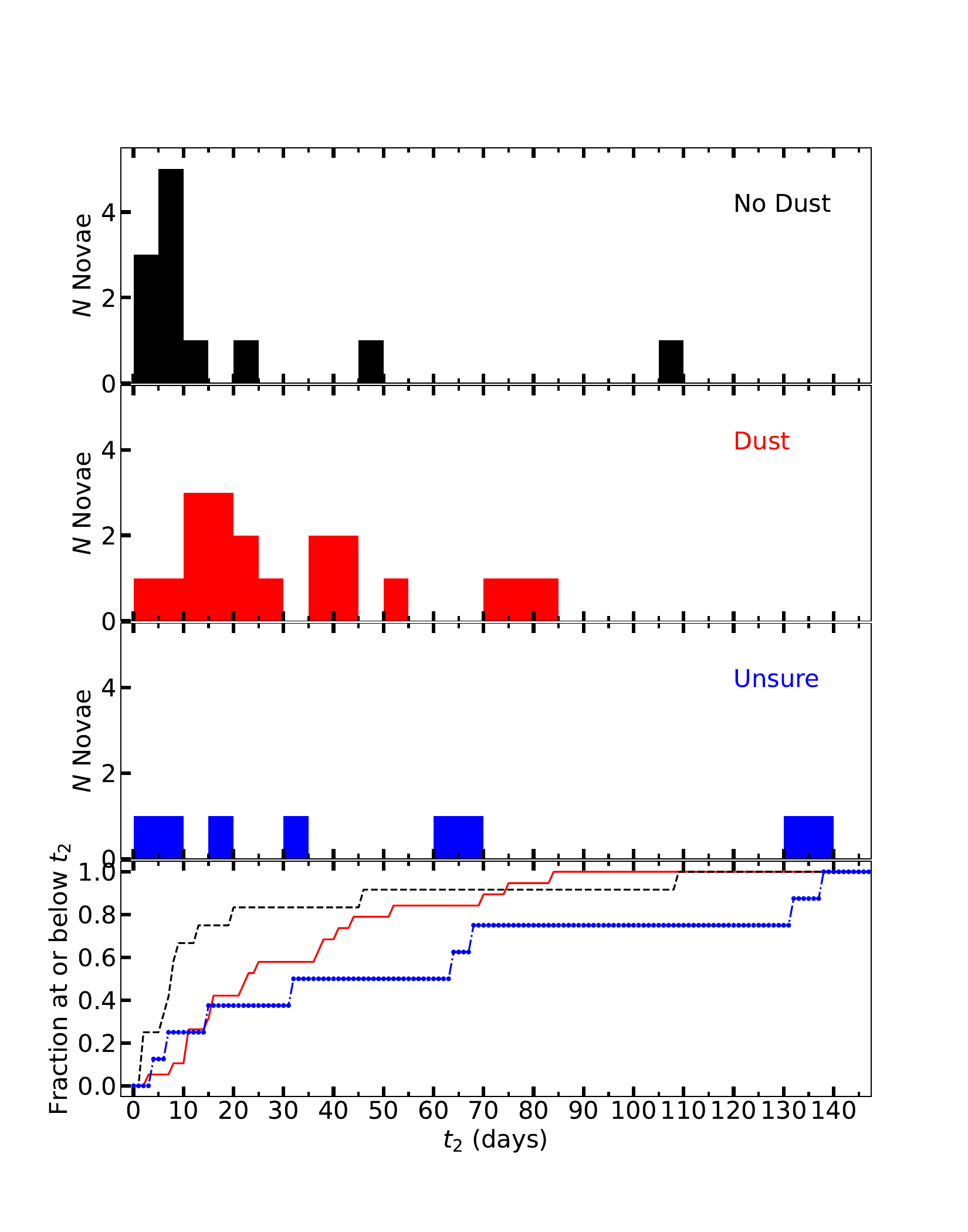}
    \caption{\textit{Top 3 panels}: histograms showing the number of novae as a function of $t_2$, divided into 5-day bins. The novae with no evidence of dust formation are presented in the \textit{top} panel using black bars.
    The dust-forming novae (including both dust dip and IR excess) are presented in the \textit{second-from-top} panel using red bars, and the `Unsure' novae are presented in the \textit{third-from-top} panel using blue bars. \textit{Bottom panel:} cumulative distributions of $t_2$ values, plotted for each of the three groups (non-dust forming nova are plotted as a black dashed line, dust forming novae as a red solid line, and unsure novae as a blue dotted line).}
    \label{fig:histogram}
\end{figure}

Figure~\ref{fig:histogram} shows the $t_2$ distributions of the different dust classes of novae in more detail. We see that there is a very strong transition from novae that do not form dust to those that do around the $t_2 = 10$ days mark. Our data are consistent with
85~per~cent of novae with $t_2 \leq 10$ days not hosting a dust formation episode (Table \ref{tab:sample}). While the general trend of the fastest novae being less likely to produce dust has been commented on for decades
\citep[e.g.][]{Gallagher_1977,Williams_etal_2013}, the data presented here are the clearest indication of this trend, incorporating high-quality IR data for a significant sample of novae, that we are aware of.

What might explain this trend of slower evolving novae being more likely to form dust than very fast novae? To first order, the trend seems reasonable given that the ejecta in very fast novae tend to have lower mass and more rapid expansion velocities, implying lower densities \citep{Shara_1981,Yaron_etal_2005, Aydi_etal_2020b}. This means that it is more difficult for dust to condense in 
the ejecta of such novae \citep[e.g.][]{Rawlings_1988}. The lower density of the ejecta also means that it is difficult to shield the dust from the irradiation by the central source, which could lead to rapid dust destruction. In contrast, slower evolving novae where the ejecta are expected to be more massive and more slowly expanding, could allow dust to condense and to be shielded from the irradiation by the central white dwarf. 

There is the additional factor that very fast novae will be more likely to occur on massive white dwarfs composed of O and Ne, 
rather than C and O. 
These elements are mixed from the white dwarf into the nova ejecta,
as indicated by spectroscopy in the optical-IR  \citep{Gehrz_etal_1998,2012ApJ...755...37H} and X-ray \citep[e.g.,][]{2020MNRAS.497.2569S}.
There is a possibility that dust forms more readily in ejecta polluted with C/O (i.e. carbon grains) rather than O/Ne (i.e. silicate grains). Finally, there is the possibility that very fast novae host less energetic or less frequent shocks in their ejecta than slower novae, which could imply diminished dust formation if radiative shocks are the sites of dust formation in novae \citep{Derdzinski_etal_2017}.
However, the fastest nova observed to date, the 2021 eruption of V1674~Her with $t_2=1.1$ days, displayed prominent shock signatures in both
$\gamma$-rays and radio \citep{2023MNRAS.521.5453S}, and it was not reported to form dust
\citep{2021ATel14741....1W,2024MNRAS.527.1405H}.

There is also the possibility that more 
very fast novae form dust than are currently recognized, because the signatures of dust formation are subtle and short-lived. The dust formation in the very fast nova V745~Sco was only recognized because of IR spectroscopy, wherein \cite{Banerjee_etal_2023} observed broad molecular CO emission lines associated with the nova ejecta. They also associate the presence of CO with a small bump in the K-band light curve that lasts between one and two weeks, but such inferences would have been near impossible from photometry alone. Relatively high cadence IR spectroscopy of very fast novae is needed to constrain the prevalence of rapid, subtle dust-forming episodes like that seen in V745~Sco. In addition, optical spectral line profiles also have the potential to reveal dust in nova ejecta, even long after the optical dust dip and/or IR excess have passed \citep{Shore_etal_2018}.


\subsection{The fraction of novae that form dust}\label{sec_fraction}


Among the 40 novae in our sample, we can confirm that 
a bit over half 
of them show evidence for dust formation, whether through a dip in the optical light curve
(13 novae) or an IR excess (8 novae, 
including PR~Lup and V5584~Sgr). For another
7 of the novae in our sample, we cannot confirm whether they formed dust or not due to them not having the typical smoothly-declining light curves we associate with non-dust-producing novae while still lacking clear features distinguishing dust formation. For the remaining 12 novae, we see no evidence for dust formation in the light curves. This means that at least $53 \pm 8$~per~cent of the novae in our sample form dust (again, taking binomial uncertainty with Bayesian confidence intervals). Taking into account the novae classified as `Unsure',  the fraction of novae that form dust is at most $70^{+6}_{-8} $~per~cent. 
Excluding LMC novae and focusing on Milky Way novae alone, we find the same result: at least 53$\pm$8~per~cent of the Galactic novae in our sample 
formed dust. These statistics imply that a significant number of novae---likely the majority---do form dust.

\citet{Strope_etal_2010} analysed the light curves of 93 novae and found that only 18~per~cent of them show dust dips, which is less than half of what our results show. One reason for this discrepancy could be due to the method used to classify a dust event in a light curve. \citet{Strope_etal_2010} relied on $V$-band light curves to classify dust dips, making it challenging to identify dust formation episodes unless they are pronounced in optical bands. Here, we instead use multi-band light curves (in both optical and near-IR) and make use of colours to help us better identify dust dips, enabling us to identify dust formation episodes that are not as pronounced or obvious. $V$-band light curves alone are not ideal for deriving an estimate of the total percentage of dust-forming novae in a sample, since our results show that around 15~per~cent of novae show IR excess only, without a significant dust dip in the optical. However, even if we limit our sample to the ``Dust Dip" novae which show clear $V$-band dips more similar to the selection criteria in \citet{Strope_etal_2010}, this is still about $35\pm8$ per cent, higher than their value of 18 per cent at about $2\sigma$. 

50--70~per~cent of novae in our sample form dust, and now we consider whether this fraction holds for classical novae generally. In principle, a selection bias in the Stony Brook/SMARTS-observed sample of novae could affect our estimate of the overall fraction of novae that form dust.
In the previous subsection (Section~\ref{sec_speed}) we show that the fastest novae are less likely to form dust,
suggesting that a bias in our sample's distribution of nova speed-classes could affect our derived dust fraction.
The Stonybrook/SMARTS Atlas, from which we draw our sample, deliberately targets He/N,
recurrent, and LMC novae \citep{Walter_etal_2012}.
Since these types typically exhibit faster optical light curves,
our sample may over-represent the fastest novae.
To test if we are biased in speed class, we compare the distribution of speed classes in our sample with 
that of Craig et al.\ 2025 (in preparation), who take a sample of novae that erupted during 2008--2021 
and consistently measure $t_2$ values (primarily from AAVSO data; \citealt{AAVSODATA}); 
the Craig et al.\ sample is a reasonable representation of the population of novae discovered during 
the last two decades. Our sample consists of 33/31/26/10~per~cent in the very fast, fast, moderately fast, 
and slow speed classes respectively, whereas the 50 novae in the Craig et al.\ sample are distributed 
as 28/32/26/14~per~cent amongst these speed classes. These two distributions are essentially identical, within sampling errors, implying that the sample of novae considered here represents the currently discovered Galactic population well.
Hence, speed class distribution is unlikely to be  significantly biasing our estimate of the fraction of novae that form dust.

\begin{figure*}
    \centering    \includegraphics[width=0.54\textwidth]{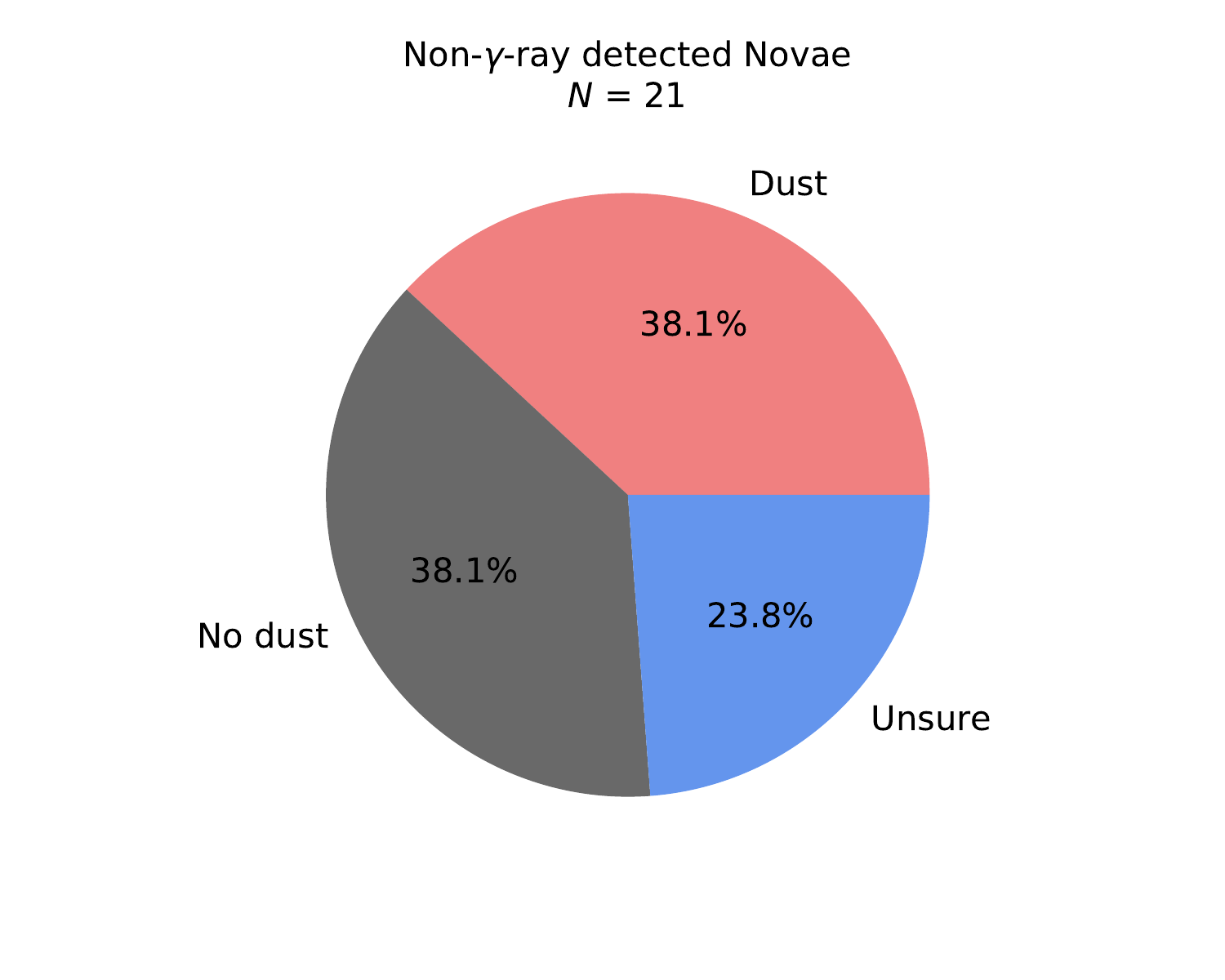}
    \hspace{-1.8cm}
    \includegraphics[width=0.54\textwidth]{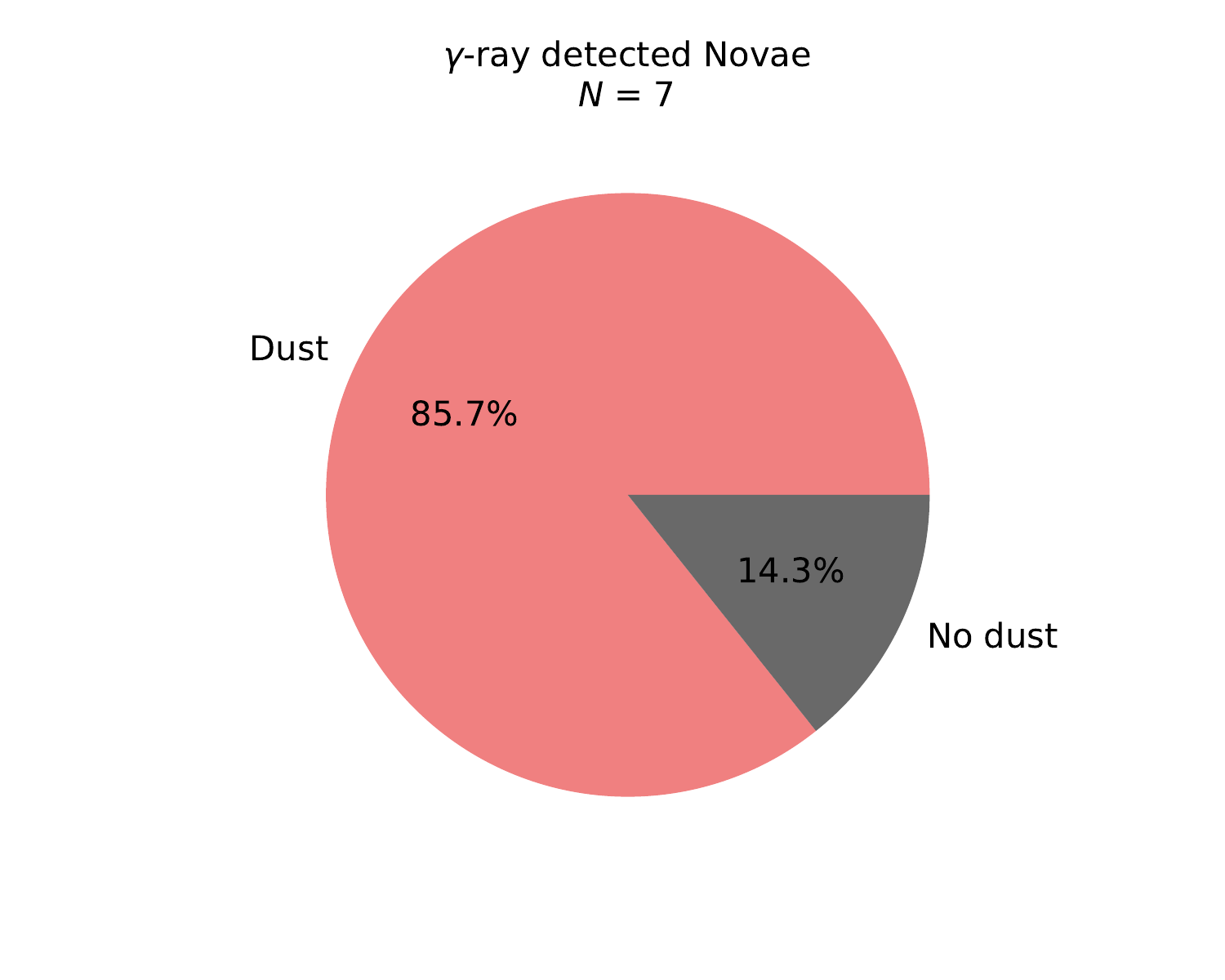}
    \vspace{-1cm}
    \caption{\textit{Left:} a pie chart representing the 21 $\gamma$-ray non-detected novae in our sample, divided into non-dust forming (grey), dust forming (light red), and unsure (light blue).
    \textit{Right}: same as \textit{left} but for the 7 
    $\gamma$-ray detected novae in our sample. There are no `unsure' novae in the $\gamma$-ray detected sample.}
    \label{fig:pie_charts}
\end{figure*}

Not only could the sample we use for this study be biased in terms of $t_2$, but the overall population of Galactic novae our broad community discovers could also be biased in terms of $t_2$. In recent years there have been several works arguing that faint, fast novae comprise a significant fraction of all novae, and are routinely missed by surveys \citep[e.g.][]{Kasliwal_etal_2011,Shara_etal_2016}---and according to Section \ref{sec_speed}, we might expect such novae to be less likely to form dust. However, a high-cadence survey of novae in the Milky Way using the All-Sky Automated Survey for Supernovae
\citep[ASAS-SN;][]{Shappee_etal_2014,Kochanek_etal_2017} finds no such significant population of faint/fast Galactic novae, implying that our samples of novae detected in recent decades are a reasonable representation of the larger, underlying population \citep{Kawash_etal_2021a,Kawash_etal_2022}. 
On the other hand, IR surveys have demonstrated that there is a significant population of heavily-extinguished novae at the lowest Galactic latitudes \citep{De_etal_2021, Zuckerman_etal_2023}, and these novae are systematically missed by optical surveys \citep{Kawash_etal_2021b}. These novae are associated with the Milky Way's thin disk, and therefore may be more likely to occur on relatively massive white dwarfs, yielding relatively short $t_2$, small ejecta masses, and high expansion velocities \citep{DellaValle_etal_1992, DellaValle_Livio_1998}. It is possible, therefore, that a population of fast novae that is relatively unlikely to form dust is currently mostly hidden by the Galactic dusty foreground.

Currently, $\lesssim$\hphantom{i}half of Galactic novae are discovered by a patchwork of surveys and efforts \citep{Shafter_2017,De_etal_2021, Kawash_etal_2022, Zuckerman_etal_2023}, and only a subset are observed in the IR to enable assessment of their dust formation properties. 
%
The next generation of all-sky surveys,
including the Legacy Survey of Space and Time (LSST; \citealt{Ivezic_etal_2019}), WINTER, and PRIME,
will feature red/near-IR bands capable of penetrating the Galactic dusty foregrounds
that currently limit the discovery of Galactic novae \citep{Kawash_etal_2021b}.
These red/IR bands will also reveal whether individual novae have formed dust,
improving estimates of dust formation rates among novae.


\subsection{Are $\gamma$-ray detected novae more likely to form dust?}
Galactic novae are routinely detected in GeV $\gamma$-rays by \emph{Fermi}-LAT \citep{Abdo_etal_2010, Ackermann_etal_2014, Cheung_etal_2016, Franckowiak_etal_2018}, but the $\gamma$-ray properties of novae are diverse, with at least a two orders of magnitude spread in their $>$100 MeV luminosities \citep{Chomiuk_etal_2021}. This diversity implies there is a range of shock luminosities in novae, which could in turn translate to a diversity in dust formation properties, if dust is formed in radiative shocks \citep{Derdzinski_etal_2017}.

In Figure~\ref{fig:pie_charts}, we compare the dust formation statistics of novae that were detected by \emph{Fermi}-LAT (left panel) with a group of novae that were searched in the \textit{Fermi}-LAT data but did not lead to significant detections (right panel). Some novae in our sample erupted before the launch of \textit{Fermi}, while others were discovered after the launch of \textit{Fermi} but do not have $\gamma$-ray analyses published; we do not include these novae in the pie-charts (see Table \ref{tab:sample} for references to $\gamma$-ray analyses, or lack thereof). The pie-charts are divided into `slices' of our nova dust classes (dust forming, non dust forming, and unsure). 

Our results show that among the $\gamma$-ray detected novae,
$86^{+5}_{-21}$~per~cent of them form dust, while 1 out 7 (RS~Oph) is believed to not form dust in its eruptions (although dust is observed in the symbiotic binary host; \citealt{Evans_etal_2007}). This stands in remarkable contrast with novae that are not detected in $\gamma$-rays, where only
$38^{+11}_{-9}$~per~cent form dust, while the same fraction do not show features of dust formation. The rest of the non-$\gamma$-ray detected novae
($24^{+11}_{-7}$~per~cent) are classified as unsure. 

Does this mean that $\gamma$-ray luminous novae are more likely to form dust? 
\citet{Derdzinski_etal_2017} linked strong $\gamma$-ray emitting shocks with dust formation, 
suggesting that the regions behind radiative shock fronts provide ideal conditions for dust formation.
The rapid cooling of the shocked gas and the dense environment allow the dust to efficiently condense while shielding it from the irradiation
by the central white dwarf. Our statistics show that there is an obvious correlation between dust formation and $\gamma$-ray detection, which is consistent with the theory of \citet{Derdzinski_etal_2017}. However, it is worth noting that $\gamma$-ray non-detections do not necessarily imply the absence of shocks. Novae detected in $\gamma$-rays by \textit{Fermi}-LAT tend to be optically bright (brighter than 8 mag in the $V$-band; Craig et al.\ 2025, in prep.), implying that these novae are nearby, and suggesting that
distance could be more important than $\gamma$-ray luminosity in determining if a nova is detected by \emph{Fermi}-LAT. This means that energetic shocks could still be present in novae not detected in $\gamma$-rays, if they are distant. The connection between dust formation and shocks in novae could be further tested by comparing $\gamma$-ray luminosity (or upper limits therein) with dust formation properties; this will be a subject of investigation in Craig et al.\ (2025, in preparation). Increasing the sample of $\gamma$-ray detected novae with dedicated optical and IR follow up will also improve our statistics and help us better understand this connection. 


\subsection{What do the properties of dust formation episodes correlate with?}\label{sec_correlate}

A correlation between the time-scale on which dust forms and nova speed class ($t_2$) has been remarked upon for decades \citep{Gallagher_1977, Bode_Evans_1982}, finding that the onset of dust formation starts earlier for faster novae. More recently, \citet{Shafter_etal_2011_spitzer} confirmed a clear correlation between the time for dust to condense (which they take to be the start of the dust dip) and $t_2$ for a sample of novae in the Milky Way and M31.  \citet{Shafter_etal_2011_spitzer} and \citet{Williams_2013} argue that this correlation is a bit unexpected, as the dust condensation time $t_{\rm cond}$ depends on nova luminosity $L$ and ejecta velocity $v_{\rm ej}$ as $t_{\rm cond} \propto L^{1/2}\,v_{\rm ej}^{-1}$ \citep{Evans_Rawlings_2008}. Substituting in scalings of $L$ and $v_{\rm ej}$ with $t_2$ (\citealt{Warner_2008}; including the controversial maximum magnitude--rate of decline relation; \citealt{Kasliwal_etal_2011}; \citealt{2023RNAAS...7..191S}), they find that $t_{\rm cond}$ should be largely independent of $t_2$.  \citet{Williams_2013} adapts this model to account for the fact that a nova's spectral energy distribution hardens over the course of its eruption, and find they can explain the correlation.

\begin{figure}
    \centering
    \includegraphics[width=1\linewidth]{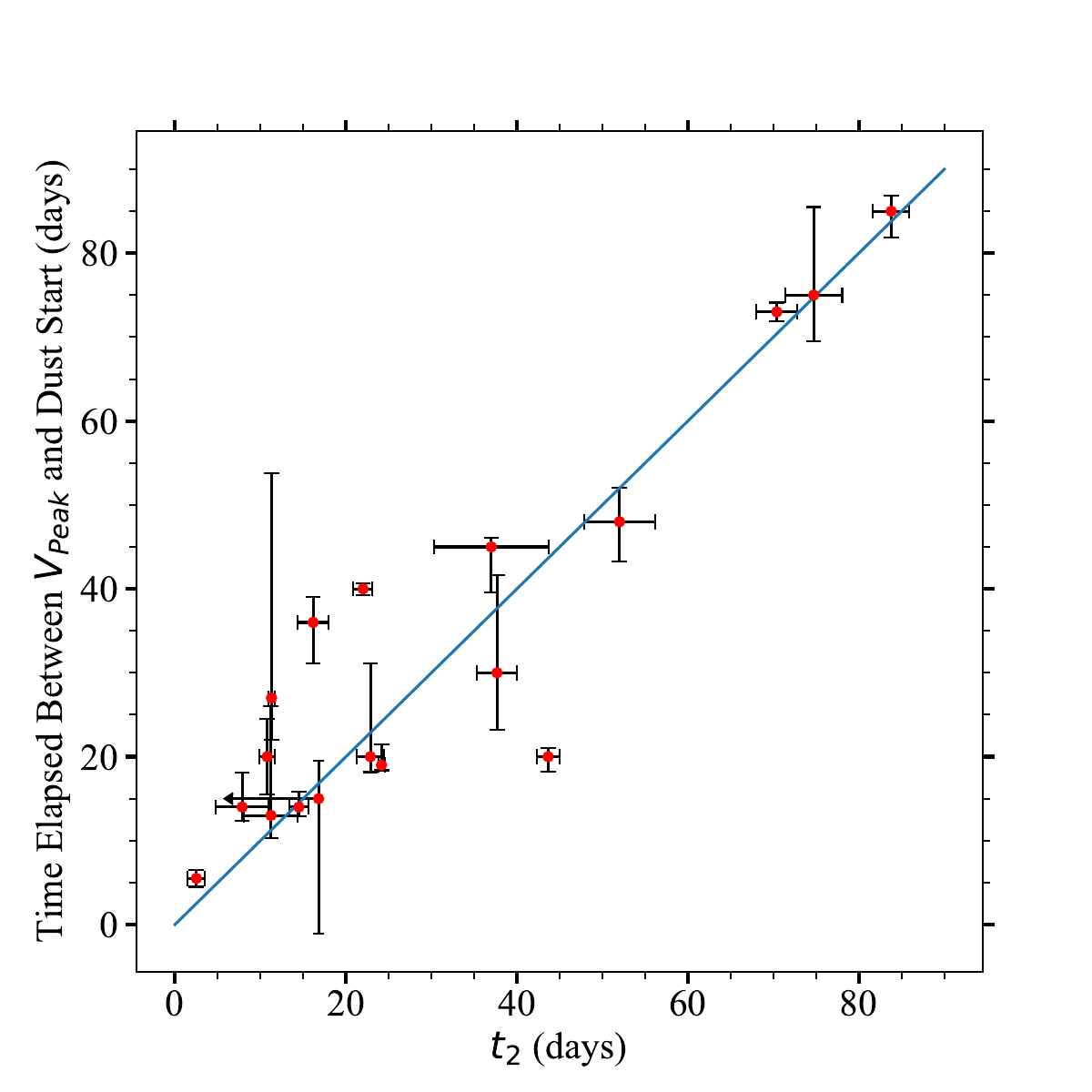}
    \caption{The time elapsed between $V_{\mathrm{peak}}$ and the onset of dust formation plotted against $t_2$ for the dust forming novae in our sample (V5584~Sgr and PR~Lup are excluded due to gap in photometric coverage during dust event, N LMC 2009b is excluded due to lack of $t_{2}$ data). The blue line is a unity line for which $x=y$.}
    \label{fig:t2_vs_Duststart}
\end{figure}

We revisit this correlation with our data set in Figure~\ref{fig:t2_vs_Duststart}, where we plot the time elapsed between $V_{\mathrm{peak}}$ and the beginning of the optical dip/IR-excess (t$_{\rm dust,start}$ in Table \ref{tab:sample}) against $t_2$. Consistent with other studies, we find a striking correlation. Overplotted on the data is a unity line, showing where the time of onset of dust formation and $t_2$ are identical, and we find that the data adhere to it quite closely. Therefore, the correlation between the time of dust onset and $t_2$ may simply be driven by the fact that the nova declines below 2 mag from peak \emph{because} it forms dust. Take, for example, V357~Mus (Figure~\ref{fig:v357mus}); its $V$-band light curve gently and gradually declines from maximum  for $\sim$3 weeks, before it plummets due to the dust formation episode starting around day 20 (blue bar in Figure~\ref{fig:v357mus}). The $t_2$ value for this nova is 23 days, after the onset of dust formation, implying that the dust formation episode played a role in setting $t_2$. This effect can lead to some unintuitive results, as in the dusty nova V1280~Sco, which is one of the slowest novae ever observed in terms of its expansion velocities and time for the optical spectrum to go nebular \citep{Naito_etal_2012}, but which has a $V$-band $t_2$ of just 21.3 days \citep{Hounsell_etal_2010}, placing it in the fast nova speed class. This fast $t_2$ is driven by the fact that V1280~Sco formed dust, and experienced a dramatic optical dust dip, just a few weeks into its massive eruption \citep{Das_etal_2008, Naito_etal_2012,Sakon_etal_2016}. While $t_2$ is sometimes used as a proxy for expansion velocity \citep[e.g.,][]{Warner_2008, Shafter_etal_2011}, the fact that it is dust formation that determines $t_2$ in $\sim$1/3 of novae undermines this use case. We instead encourage investigators to make more direct measurements of expansion velocity using spectroscopy. Alternatively, if investigators have the goal of using light curve shape to parameterize the properties of the ejecta (i.e., mass, velocity, density), it may be beneficial to use a redder band than $V$, where dust absorption will have less of an impact.

\begin{figure}
    \centering
    \includegraphics[width=\columnwidth]{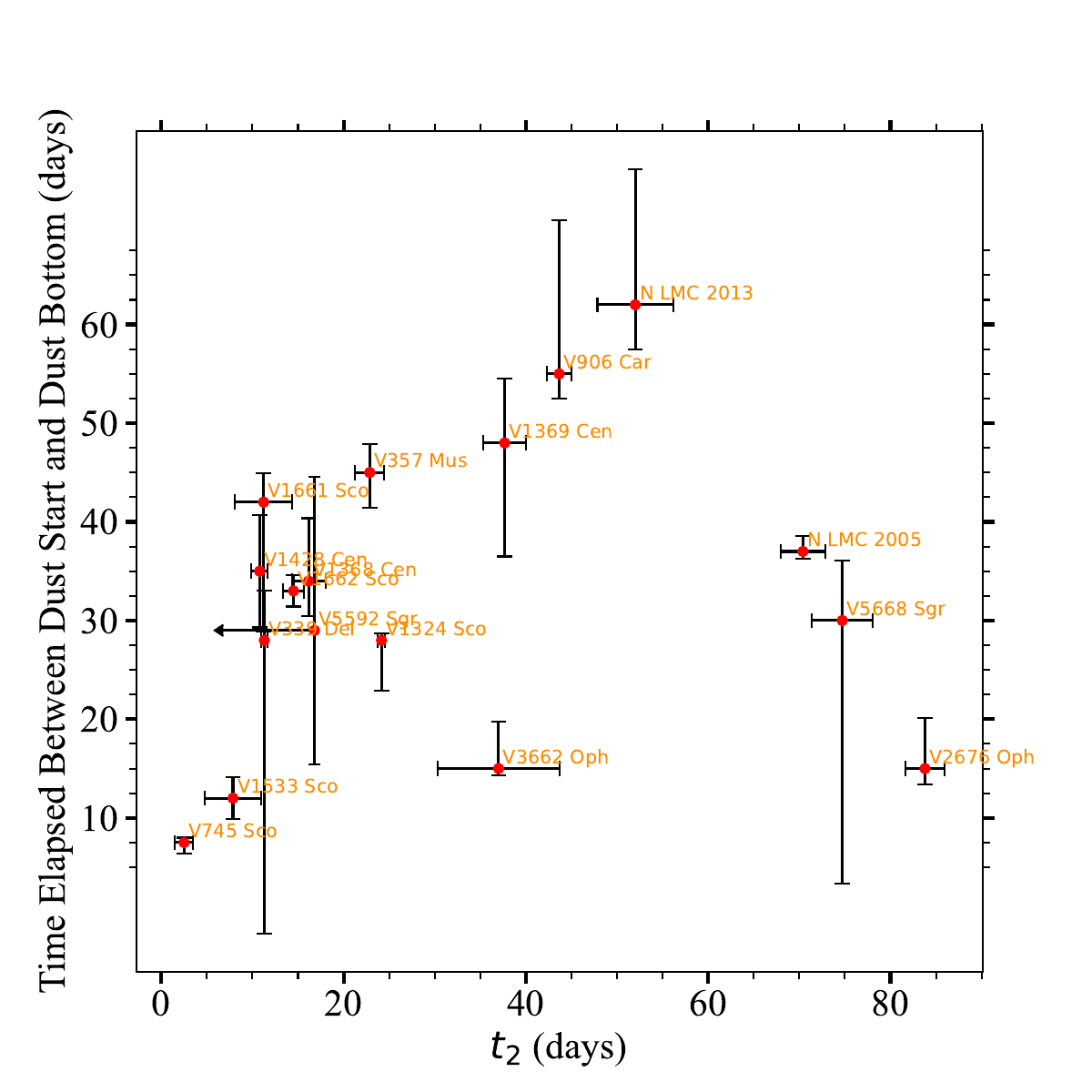}
    \caption{The time elapsed between the onset of dust formation episode and the bottom of the dust event plotted against $t_2$ for the dust-forming novae in our sample (V475~Sct, V5584~Sgr and PR~Lup are excluded due to gap in photometric coverage during dust event, N LMC 2009b is excluded due to lack of $t_{2}$ data).  The correlation coefficient between the two parameters is $r=-0.03$}
    \label{fig:t2_vs_StarttoBottom}
\end{figure}

Does the duration of the dust episode also correlate with $t_{2}$? If the recovery from dust dip is driven by the destruction of dust in the harsh radiation field of the nuclear-burning white dwarf, we might expect slower novae to have longer duration dust events, because their relatively slow/massive ejecta will take longer to unveil the harsh radiation field of the white dwarf. Figure~\ref{fig:t2_vs_StarttoBottom} plots $\Delta t_{\rm dust}$ (the time elapsed between the onset of dust formation and the bottom of the dust event) against $t_2$ for the dust-forming novae in our sample. Our results show that there is no obvious correlation between these two parameters, with a correlation coefficient of just $r = -0.03$. Apparently, additional factors beyond nova speed class are at work in determining the duration of dust events, or perhaps, as suggested by \citet{Shore_etal_2018}, the recovery from the dust dip does not require significant destruction of grains, but could just be an effect of the thinning out of the ejecta. We also searched for a correlation between the time of the onset of the dust dip and the duration of the dust dip, and similarly found no correlation. 

\begin{figure}
    \centering
    \includegraphics[width=1\linewidth]{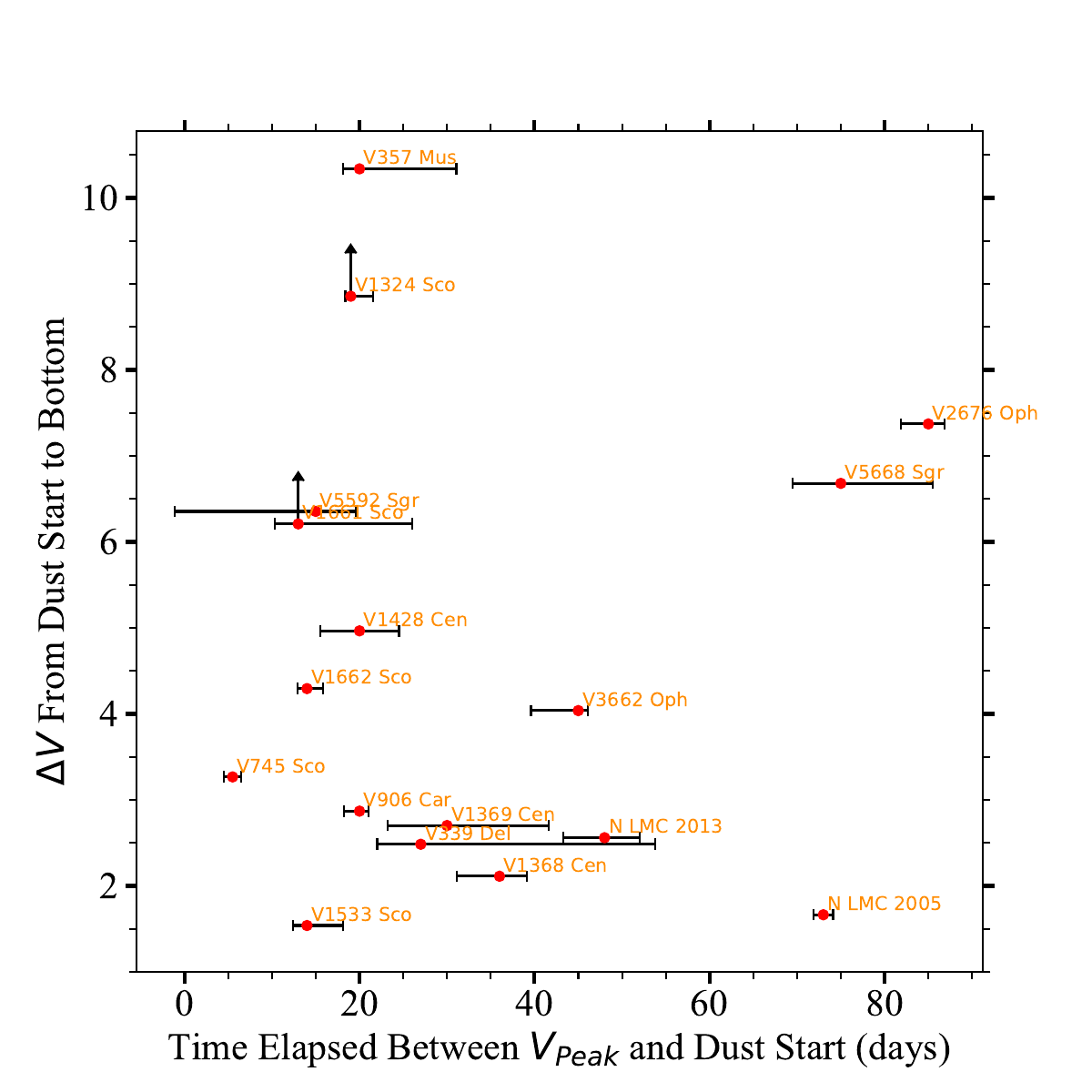}
    \caption{The depth of the $V$-band dust dip (in magnitudes) plotted against the time lag between optical light curve peak and the onset of dust formation for the dust-forming novae in our sample (V475~Sct, V5584~Sgr and PR~Lup are excluded due to gap in photometric coverage during dust event, N LMC 2009b is excluded due to lack of $V_{\mathrm{Peak}}$ data.). The correlation coefficient between the two parameters is $r=0.02$.}
    \label{fig:depth_vs_StarttoBottom}
\end{figure}

\citet{Strope_etal_2010} claim that the deeper a nova's dust dip (as observed in their $V$-band light curves), the later this dust dip will occur. We test this assertion in Figure~\ref{fig:depth_vs_StarttoBottom}, where on the $x$-axis we plot t$_{\rm dust,start}$, and on the $y$-axis, we plot the change in the $V$-band brightness over the course of the dust event, $\Delta V_{\rm dust}$.
We see no correlation between the parameters, with a correlation coefficient of $r = 0.01$. Take, for example, the deepest dust dip in our sample, V357~Mus with $\Delta V_{\rm dust} = 10.3$, which begins forming dust after just 20 days. We find a similar lack of correlation when we plot the depth of the dust dip against the duration of the dust dip (e.g., the ordinate of Figure~\ref{fig:t2_vs_StarttoBottom}), or when we characterize the `depth' of the dust dip as the redward change in the ($V-K$) colour index over the course of the dust formation episode. We conclude that there is no clear correlation between the depth of a dust dip and its time-scale. A lack of correlation would be expected if the dust only covers a portion of the nova ejecta, and the depth of the dust dip (and indeed, whether it is an optically thick dust dip or an optically thin IR excess) is primarily determined by inclination and viewing effects, as suggested by \citet[][e.g., their figure 1]{Derdzinski_etal_2017}.

\section{Conclusions}
\label{sec_conc}
We performed a statistical analysis of dust formation in a
sample of 40 well-observed novae with extensive optical and
IR photometric coverage \citep{Walter_etal_2012}. Our results show
that between 50 and 70~per~cent of novae form
dust (Section~\ref{sec_fraction}), a fraction substantially higher than previous
estimates in the literature \citep[e.g.,][]{Strope_etal_2010}. While this fraction holds true for currently discovered Galactic novae, it could be biased if current surveys are systematically missing some sub-populations of Galactic novae.
The prevalence of dust formation in novae makes them ideal, nearby laboratories to explore the physics of dust condensation in astrophysical transients. 

We propose a simple diagnostic of dust formation in novae: the largest redward change in ($V-K$) relative to the colour at $V$-band light curve peak (Figure~\ref{fig:Color_Change_VK}). There is a clear separation between novae that form dust and those that do not, with dust-forming novae showing $\Delta (V-K) > 2.35$. We note that novae with red giant companions present a more complicated picture, and may show 
$\Delta (V-K) > 2.35$ even if they do not form dust.
($V-J$) and ($V-H$) colour indices are also useful for identifying dust formation episodes, but are likely to have higher false positive rates than ($V-K$) (Section \ref{sec_vj}). Time-domain surveys in IR bands, like WINTER and PRIME, have the potential to find every nova eruption in the Milky Way, and also determine which of them form dust.

By comparing our dust formation statistics with nova speed class (quantified by $t_2$, the time for the $V$-band light curve to decline by two magnitudes from peak), we find that very fast novae ($t_2 \leq 10$ days) are much less likely to form dust than their slower counterparts (Figures~\ref{fig:speedclass} and \ref{fig:histogram}). While this trend has been  noticed for decades \citep{Gallagher_1977}, our data are the clearest demonstration of the effect to date, with just 17~per~cent of very fast novae forming dust, compared with $\sim$70~per~cent of slower novae with $t_2 = 11-80$ days. We do, however, view the subtle, short-lived dust formation episode in the very fast nova V745~Sco as a cautionary tale \citep{Banerjee_etal_2023}, and encourage the use of IR spectroscopy to identify molecular species like CO (which are tracers of dust), to assess the prevalence of mild dust formation episodes that could be missed by photometry alone (these may be more common in very fast novae).

We also assess whether novae detected in GeV $\gamma$-rays by \emph{Fermi}-LAT are more likely to form dust, as might be expected if dust forms in radiative shock fronts \citep{Derdzinski_etal_2017}. We find a signal with 86~per~cent of $\gamma$-ray detected novae forming dust, vs. just 38~per~cent of novae that are not detected by \emph{Fermi} (Figure~\ref{fig:pie_charts}). Future studies should dig deeper into this trend by comparing $\gamma$-ray luminosity (or upper limits) with dust formation properties.


We confirm the correlation between $t_2$ and the time of the onset of dust formation studied by \citet{Shafter_etal_2011_spitzer} and \citet[][see Figure~\ref{fig:t2_vs_Duststart}]{Williams_etal_2013}. However, we find that it can simply be explained if $t_2 \approx t_{\rm cond}$ (the condensation time-scale of dust formation), i.e. the nova is observed to drop 2 magnitudes below light curve maximum \emph{because} it formed dust (Section \ref{sec_correlate}). Future work should search for a more physically robust connection between the timescale of dust formation and nova properties like luminosity, ejecta velocity, and eject mass, by comparing e.g., $t_{\rm cond}$ with $v_{\rm ej}$ directly.

How dust forms around explosive transients remains a perplexing puzzle in high-energy astrophysics. This work provides a first step towards a quantitative characterization of dust formation in a population of novae. A second future step would be to collect detailed multi-wavelength observations (e.g., spectroscopy, X-ray observations, etc.) in order to reveal what nova properties dust formation and survival depend upon. In addition, high-resolution imaging has the potential to reveal the locations of dust condensation within the nova ejecta, further testing if dust forms at nova shocks. This work also provides strong motivation for IR transient surveys and follow-up observations, all with the aim of uncovering the mystery of dust production in explosive transients. 

\section*{Acknowledgements}
This paper was inspired by conversations with Brian Metzger, Andrea Derdzinski, and Jeno Sokololski. A.C., E.A., P.C., L.C., and A.S.\ are grateful for support from NASA grants 80NSSC23K0497, 80NSSC23K1247, and 80NSSC25K7334 and NSF grants AST-2107070 and AST-2205628. E.A.\ acknowledges support by NASA through the NASA Hubble Fellowship grant HST-HF2-51501.001-A awarded by the Space Telescope Science Institute, which is operated by the Association of Universities for Research in Astronomy, Inc., for NASA, under contract NAS5-26555. This research has used data from the SMARTS 1.3\,m, 1.0\,m, and 0.9\,m telescopes, which are operated as part of the SMARTS Consortium.

\section*{Data availability}
The data are available as online material and can be found here:\\\indent\url{https://www.dropbox.com/scl/fo/vkkoodp7dtz65jugmrafp/ANUc6HAe8Cm_3-HGykZtg_A?rlkey=klle8ko4a8xy9ixzjehda9tcq&st=d8rhtnpy&dl=0}

\bibliographystyle{mnras_vanHack}
\bibliography{biblio}

\appendix

\renewcommand\thetable{\thesection.\arabic{table}}    
\renewcommand\thefigure{\thesection.\arabic{figure}}   
\setcounter{figure}{0}

\section{Full Sample of Nova Light Curves}\label{sec_full}

For convenience, we include individual light curves and colour curves for all 40 novae in our sample (Table \ref{tab:sample}). The top panel of each plot includes the $K$-band light curve in black and the $V$-band light curve in green. The bottom panel shows the ($V-K$) colour as a function of time. These values are observed, not intrinsic, and therefore not corrected for extinction. The captions include the names of the novae and their assigned dust classification. For `Dust Dip' and `IR Excess' novae, the dotted blue vertical line represents an estimate of the start time of the dust event, with the blue shaded region extending from the last $(V-K)$ data point before it to the first $(V-K)$ data point after it.
The vertical dotted red line represents an estimate of the `bottom' of the dust event, when the nova is most red as measured by $(V-K)$. The red shaded region extends from the last $(V-K)$ data point before the dotted red line to the first $(V-K)$ data point after it. These shaded regions are shown with the intent of providing a more objective representation of our confidence in the precision of our estimates, and are accounted for as error bars on the time-scales of dust formation in Figures~\ref{fig:t2_vs_Duststart}--\ref{fig:depth_vs_StarttoBottom}.

\newpage
\begin{figure}
    \centering
    \includegraphics[width=1\linewidth]{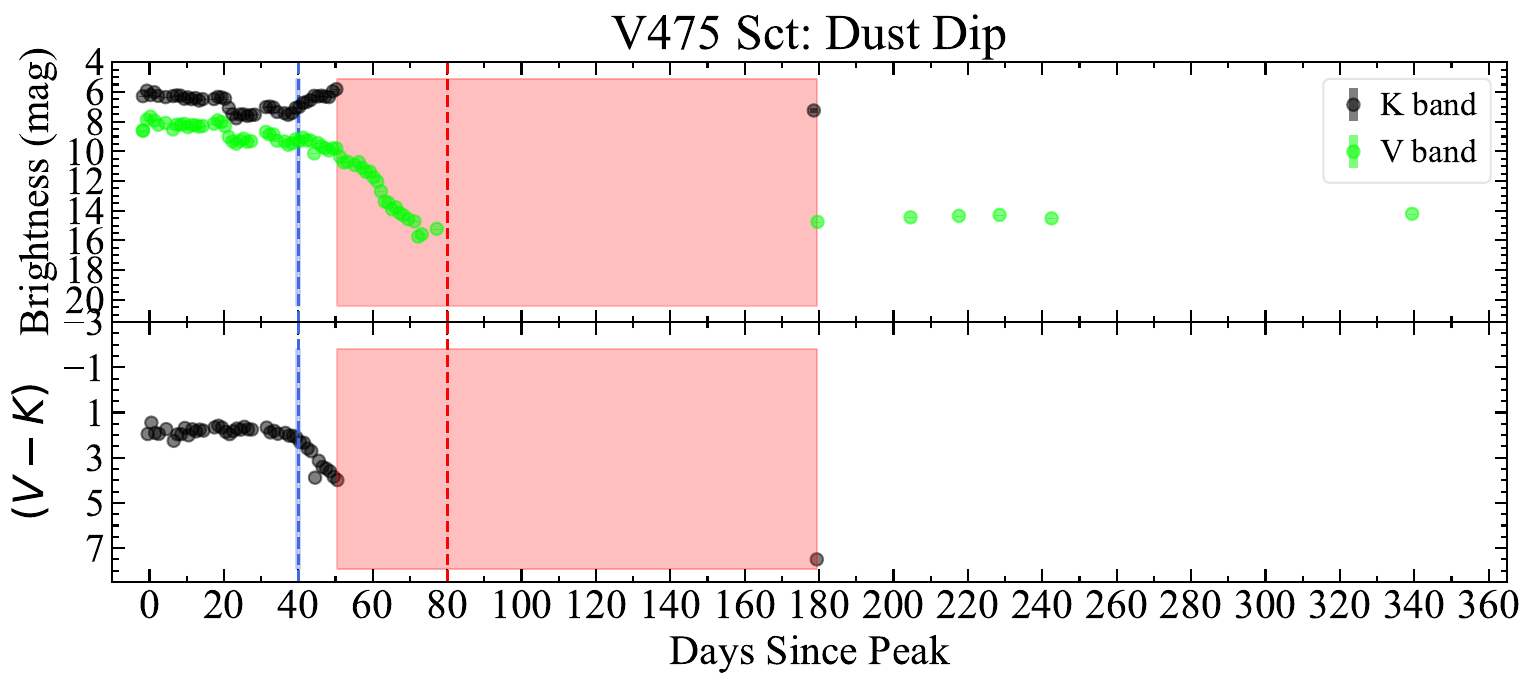}
    \caption{The $V$ and $K$ light curves (\textit{Top}) and $(V-K)$ colour curve (\textit{Bottom}) of nova V475~Sct, which we classify as `Dust Dip'. The blue vertical bar denotes the earliest feature attributable to the dust event (in this case, a rise in the $K$-band light curve), while the red vertical bar marks the `bottom' of the dust dip (the point immediately after the ($V-K$) colour appears to have finished the large redward shift caused by dust formation). The width of the bars is set by the cadence of the photometry at these times. We hypothesize that the `bottom' of the dust dip in V475~Sct occurs near the beginning of the solar conjunction based on optical light and colour curves from \citet{Chocol_etal_2006}.}
    \label{fig:app_v475sct}
\end{figure}

\begin{figure}
    \centering
    \includegraphics[width=1\linewidth]{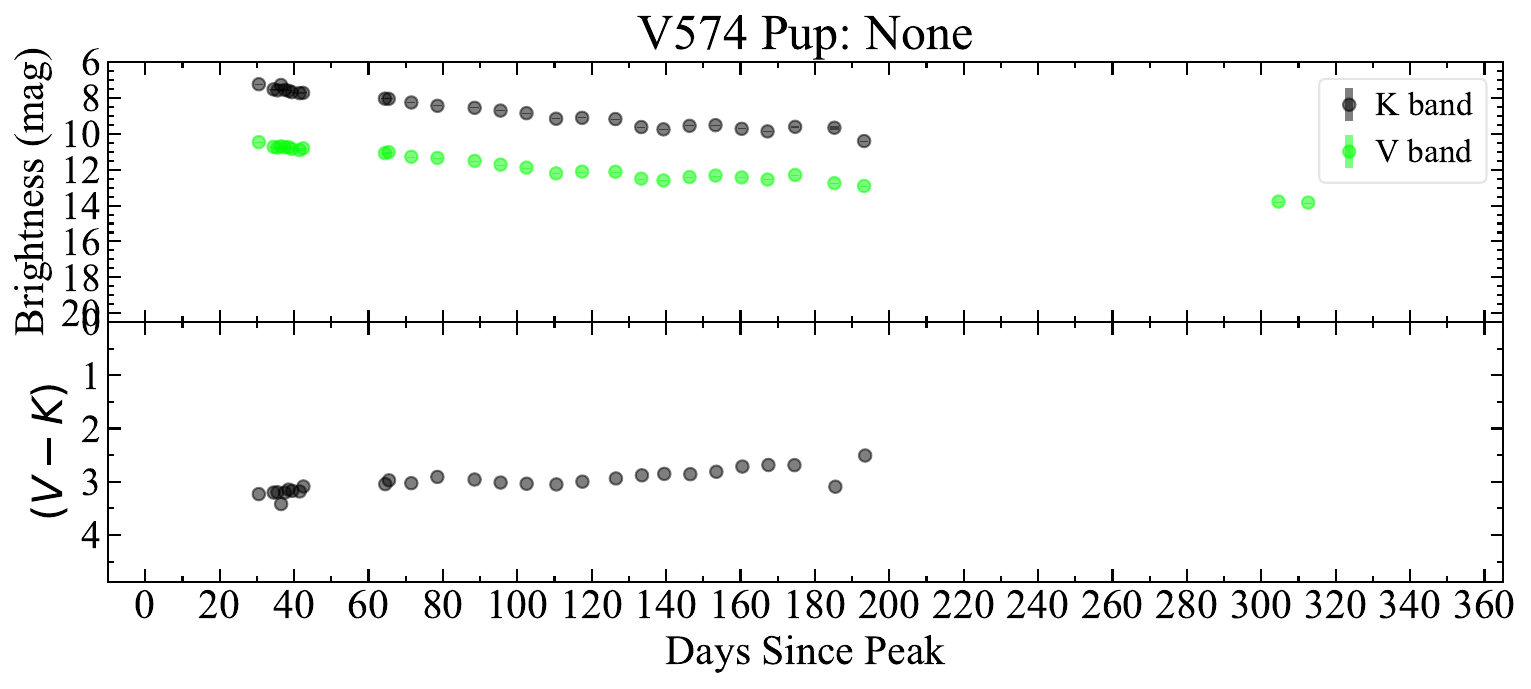}
    \caption{Same as Figure~\ref{fig:app_v475sct} but for nova V574~Pup, which we classify as `None'.}
    \label{fig:app_v574pup}
\end{figure}

\begin{figure}
    \centering
    \includegraphics[width=1\linewidth]{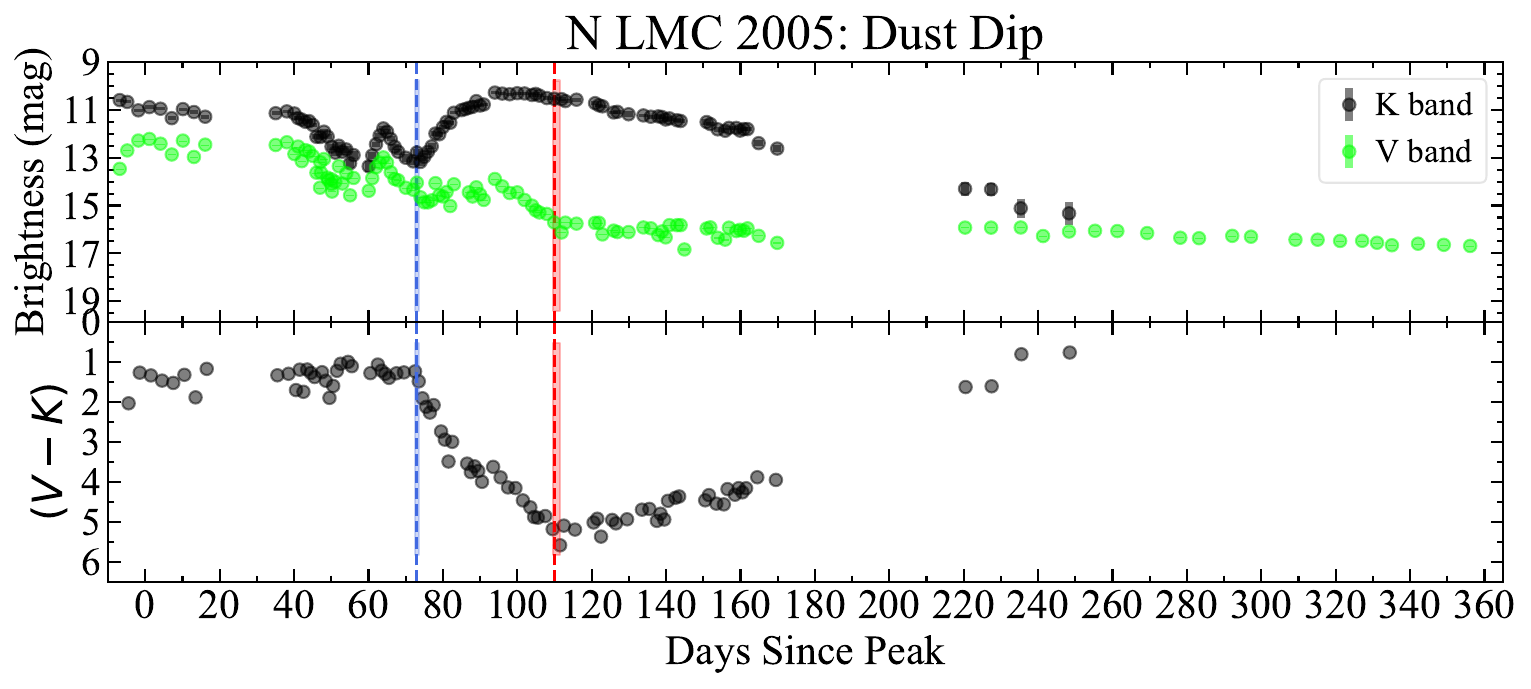}
    \caption{Same as Figure~\ref{fig:app_v475sct} but for nova N LMC 2005, which we classify as `Dust Dip'.}
    \label{fig:app_nlmc2005}
\end{figure}

\begin{figure}
    \centering
    \includegraphics[width=1\linewidth]{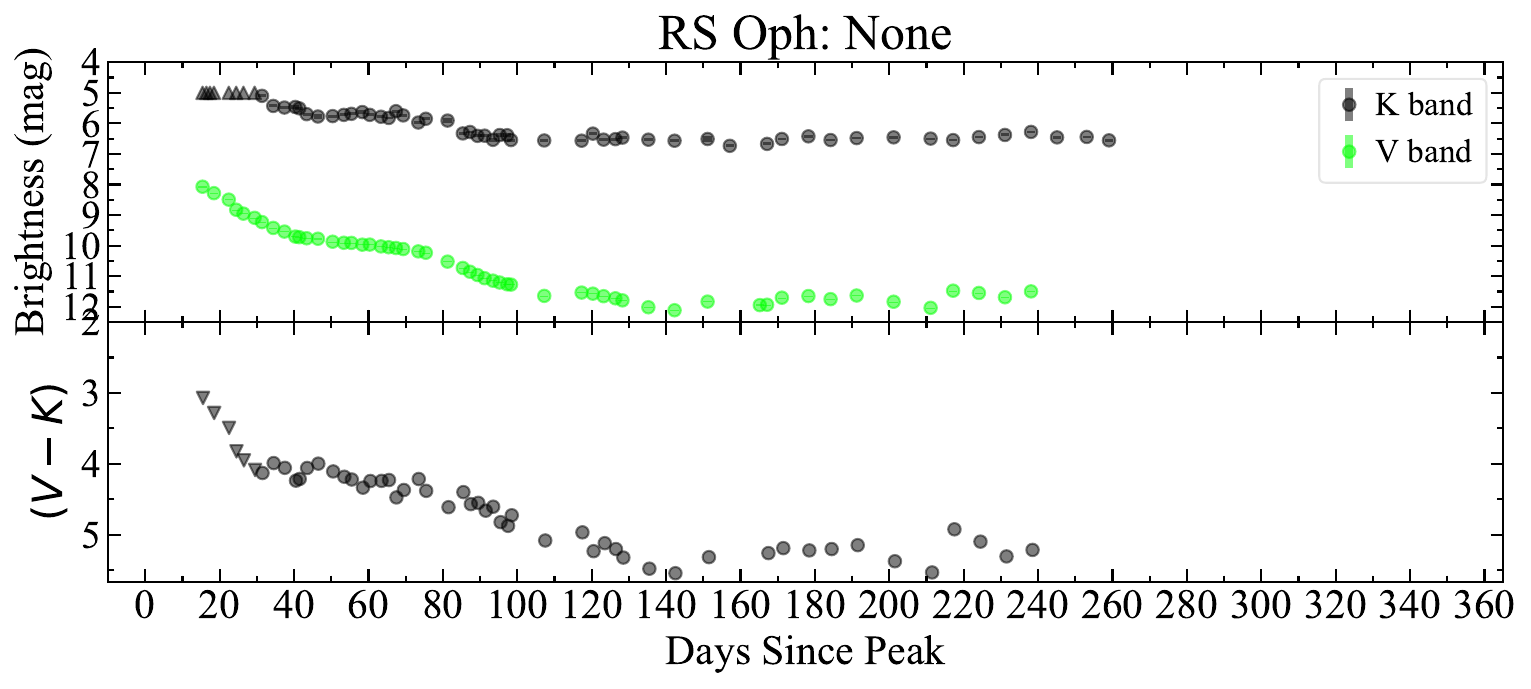}
    \caption{Same as Figure~\ref{fig:app_v475sct} but for nova RS~Oph (2006), which we classify as `None'.}
    \label{fig:app_rsoph}
\end{figure}

\begin{figure}
    \centering
    \includegraphics[width=1\linewidth]{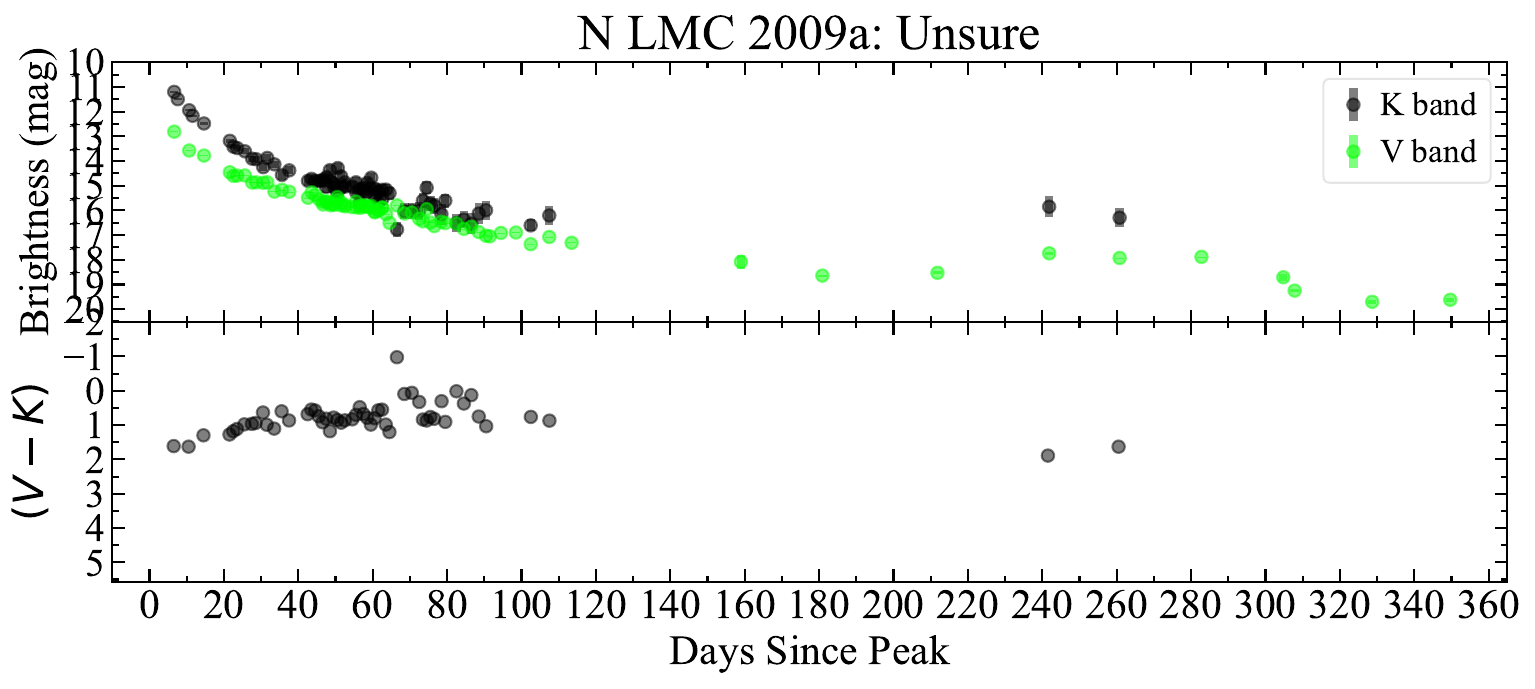}
    \caption{Same as Figure~\ref{fig:app_v475sct} but for nova N LMC 2009a, which we classify as `Unsure'.}
    \label{fig:app_nlmc2009a}
\end{figure}

\begin{figure}
    \centering
    \includegraphics[width=1\linewidth]{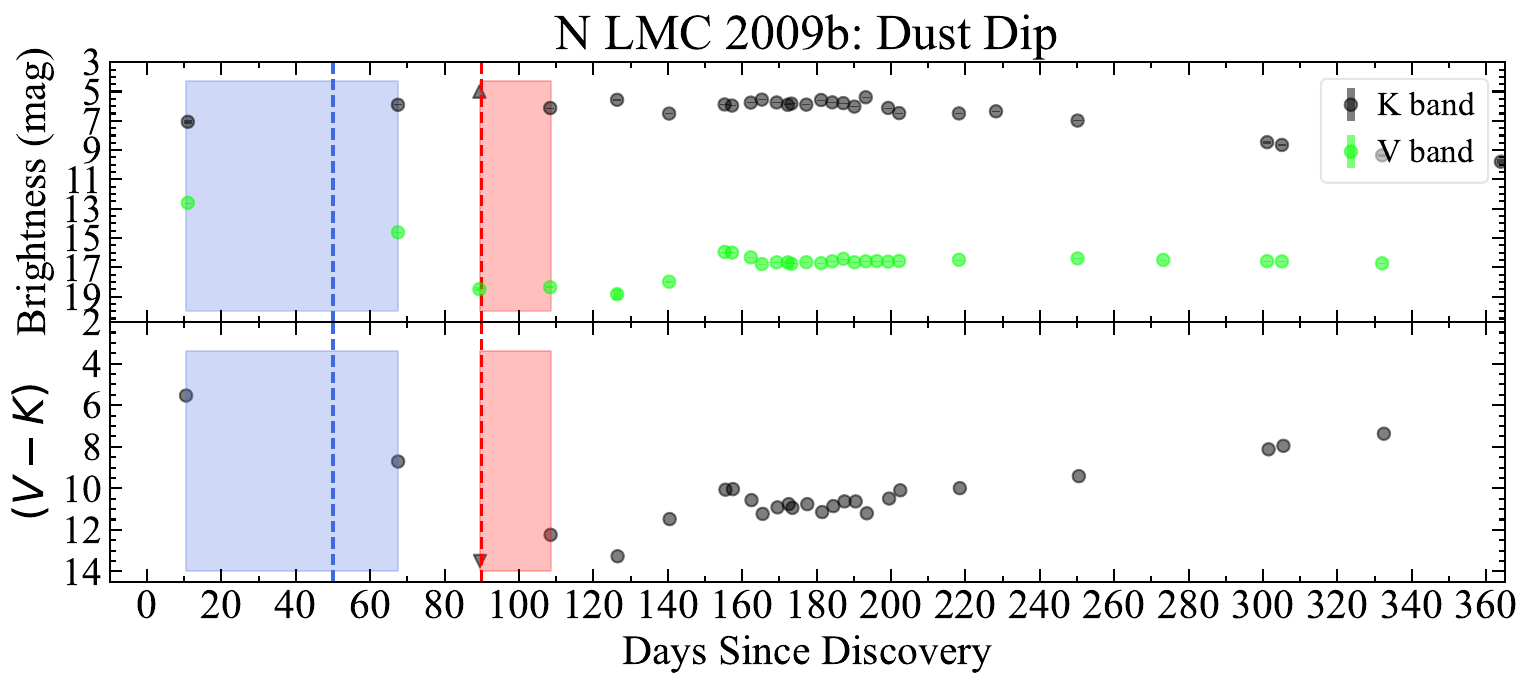}
    \caption{Same as Figure~\ref{fig:app_v475sct} but for nova N LMC 2009b, which we classify as `Dust Dip'.}
    \label{fig:app_nlmc2009b}
\end{figure}

\begin{figure}
    \centering
    \includegraphics[width=1\linewidth]{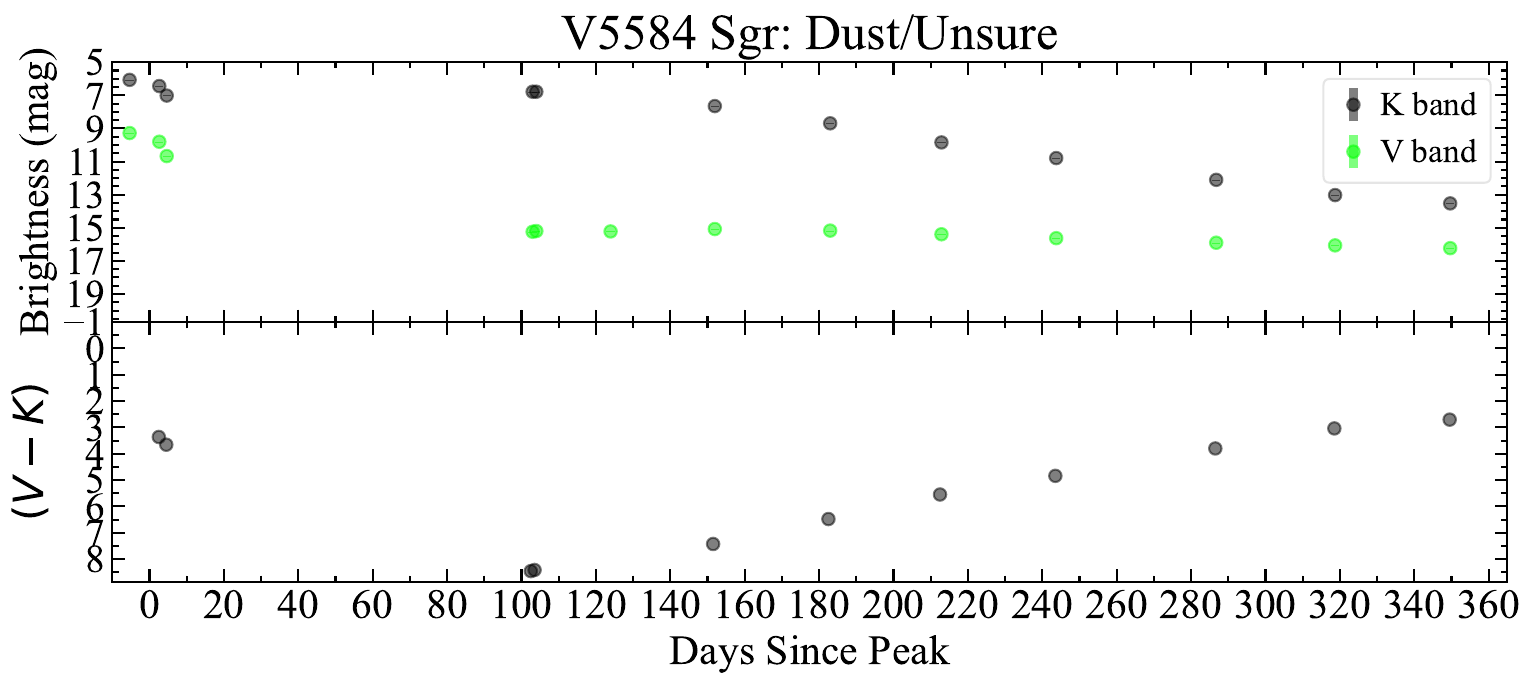}
    \caption{Same as Figure~\ref{fig:app_v475sct} but for nova V5584~Sgr, which we classify as `Dust Dip'.}
    \label{fig:app_V5584 Sgr}
\end{figure}

\begin{figure}
    \centering
    \includegraphics[width=1\linewidth]{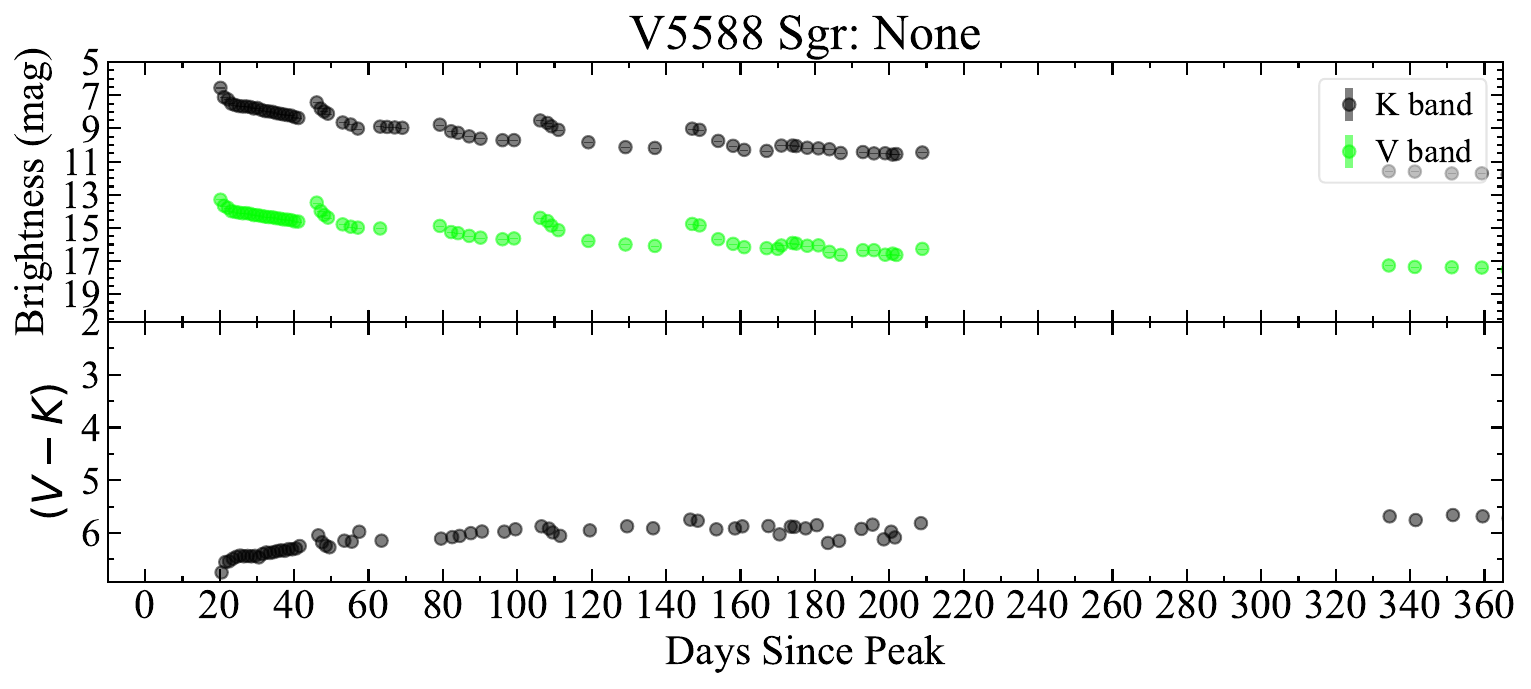}
    \caption{Same as Figure~\ref{fig:app_v475sct} but for nova V5588~Sgr, which we classify as `None'.}
    \label{fig:app_v5588sgr}
\end{figure}

\begin{figure}
    \centering
    \includegraphics[width=1\linewidth]{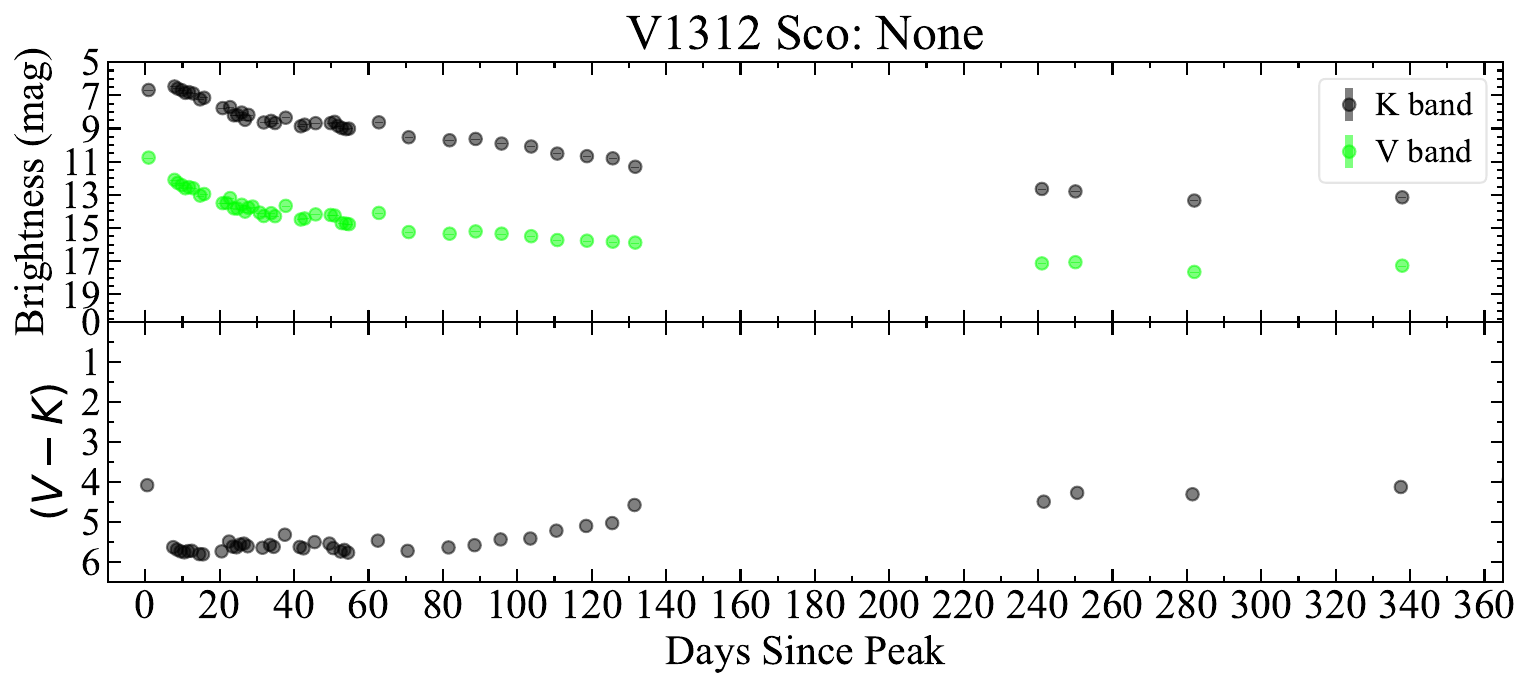}
    \caption{Same as Figure~\ref{fig:app_v475sct} but for nova V1312~Sco, which we classify as `None'.}
    \label{fig:app_v1312sco}
\end{figure}

\begin{figure}
    \centering
    \includegraphics[width=1\linewidth]{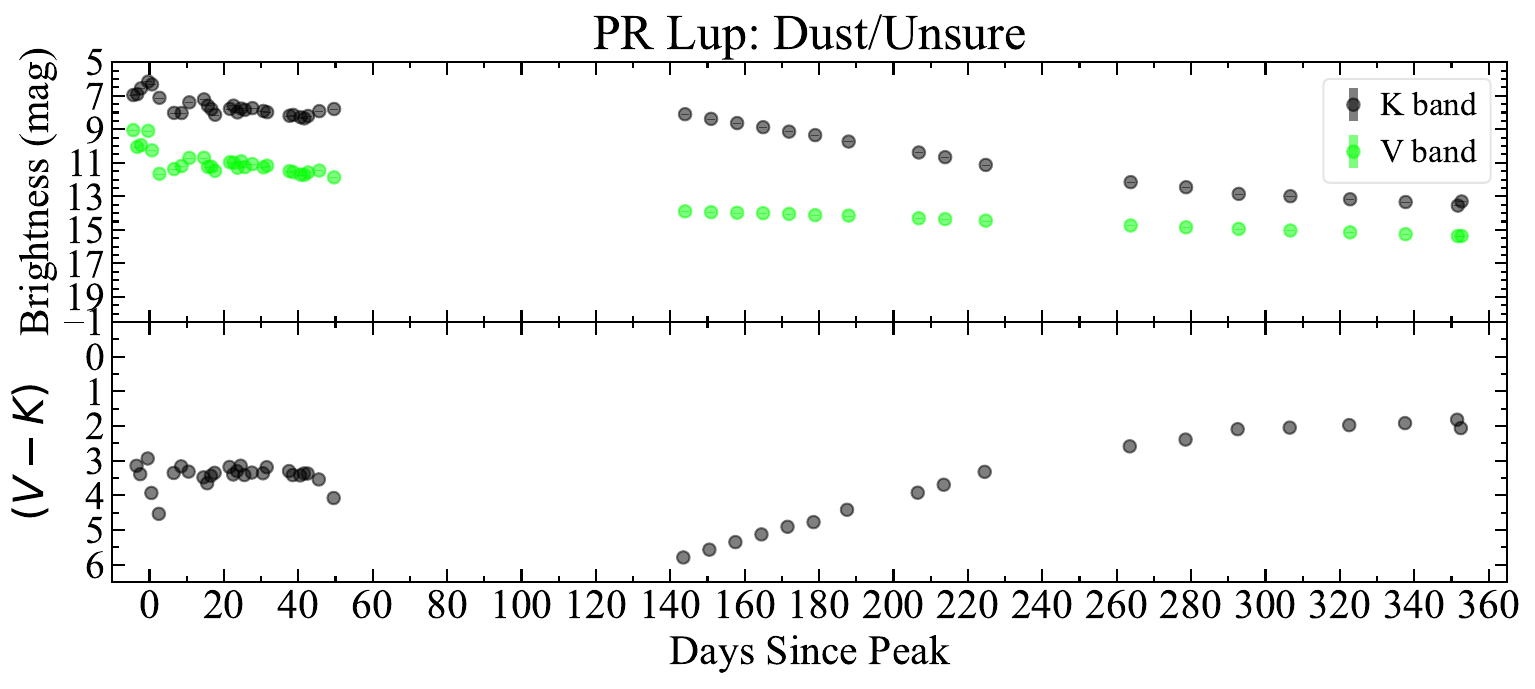}
    \caption{Same as Figure~\ref{fig:app_v475sct} but for nova PR~Lup, which we classify as `Dust/Unsure'.}
    \label{fig:app_prlup}
\end{figure}

\begin{figure}
    \centering
    \includegraphics[width=1\linewidth]{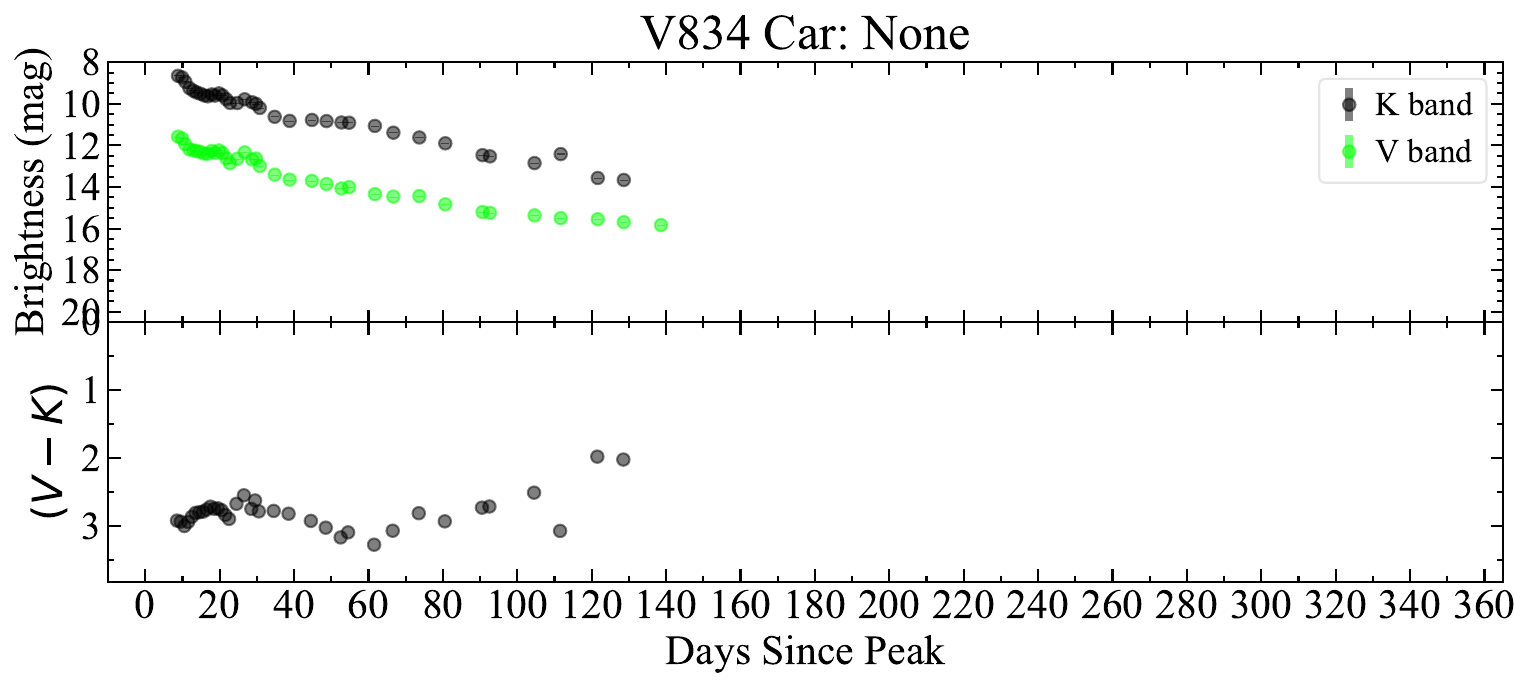}
    \caption{Same as Figure~\ref{fig:app_v475sct} but for nova V834~Car, which we classify as `None'.}
    \label{fig:app_v834car}
\end{figure}

\begin{figure}
    \centering
    \includegraphics[width=1\linewidth]{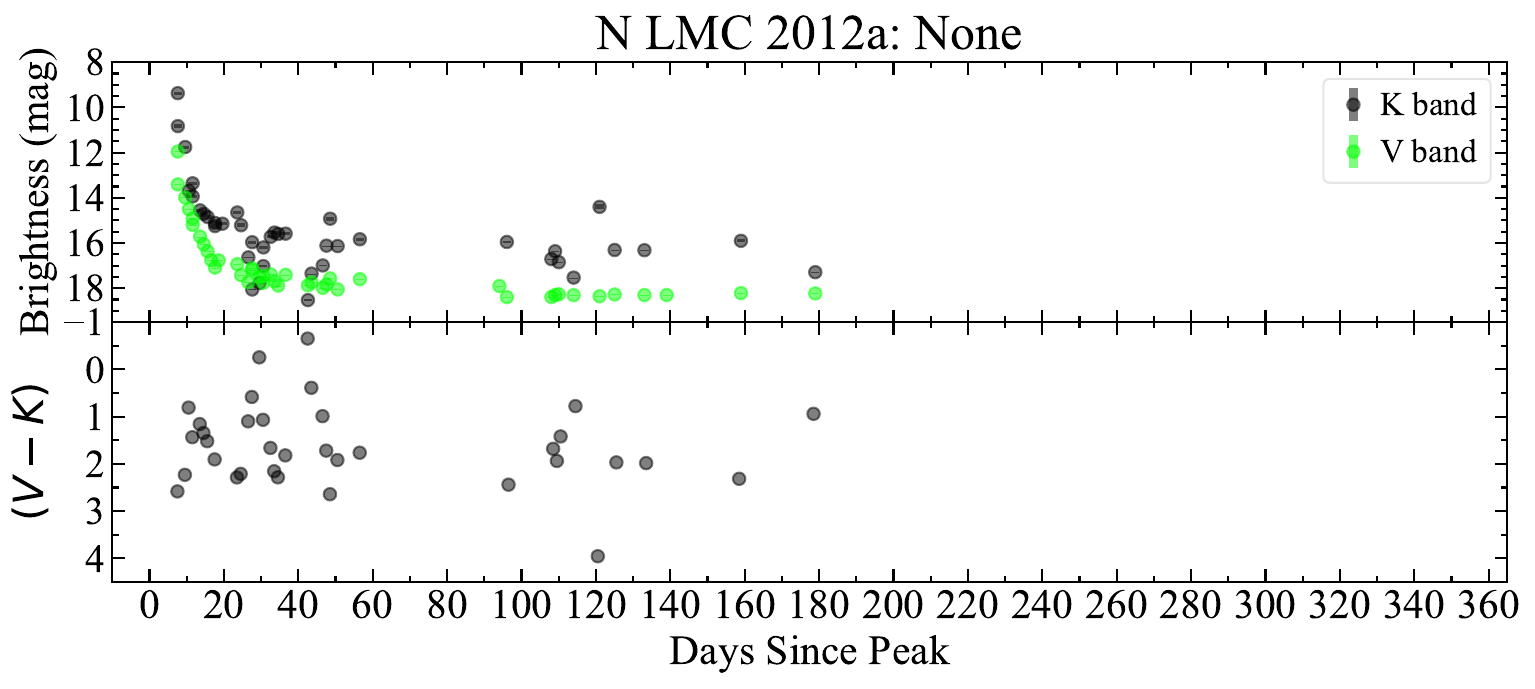}
    \caption{Same as Figure~\ref{fig:app_v475sct} but for nova N LMC 2012a, which we classify as `None'.}
    \label{fig:app_nlmc2012a}
\end{figure}

\begin{figure}
    \centering
    \includegraphics[width=1\linewidth]{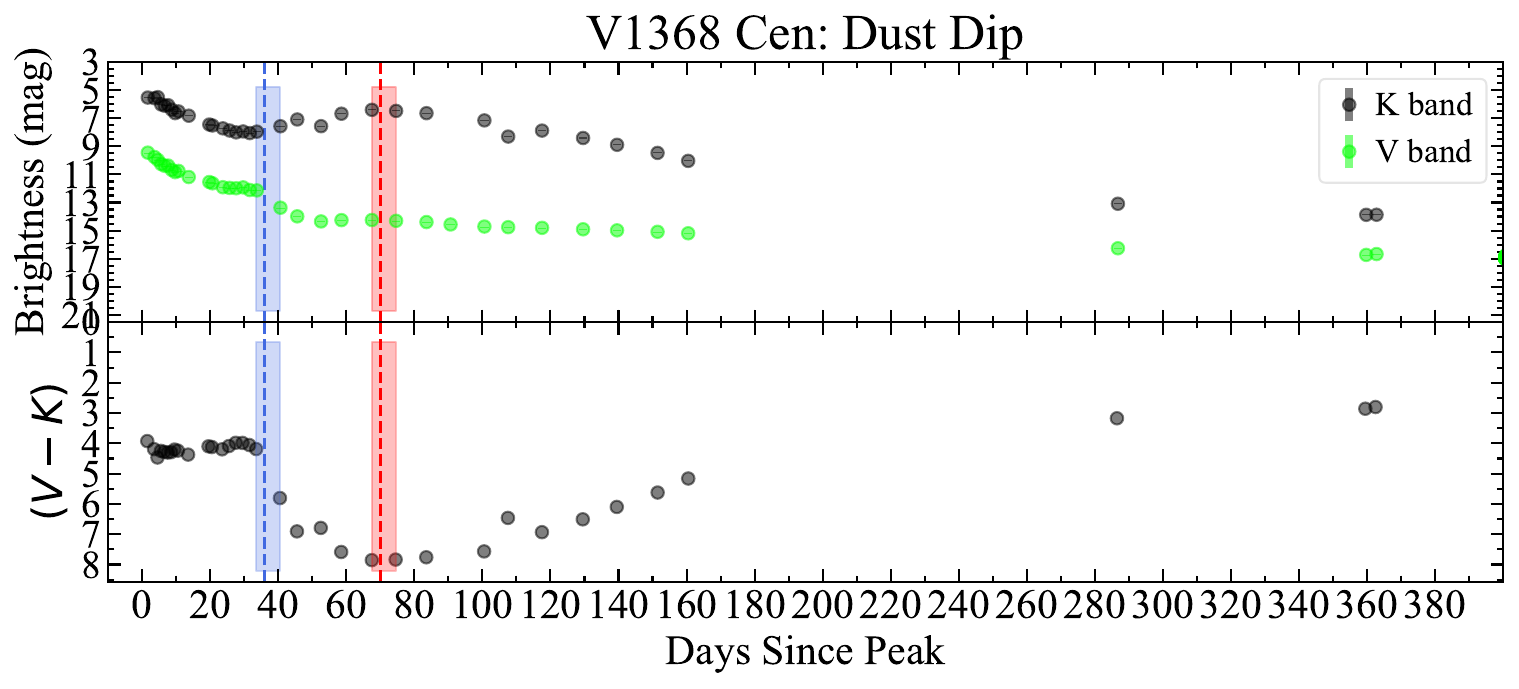}
    \caption{Same as Figure~\ref{fig:app_v475sct} but for nova V1368~Cen, which we classify as `Dust Dip'.}
    \label{fig:app_v1368cen}
\end{figure}

\begin{figure}
    \centering
    \includegraphics[width=1\linewidth]{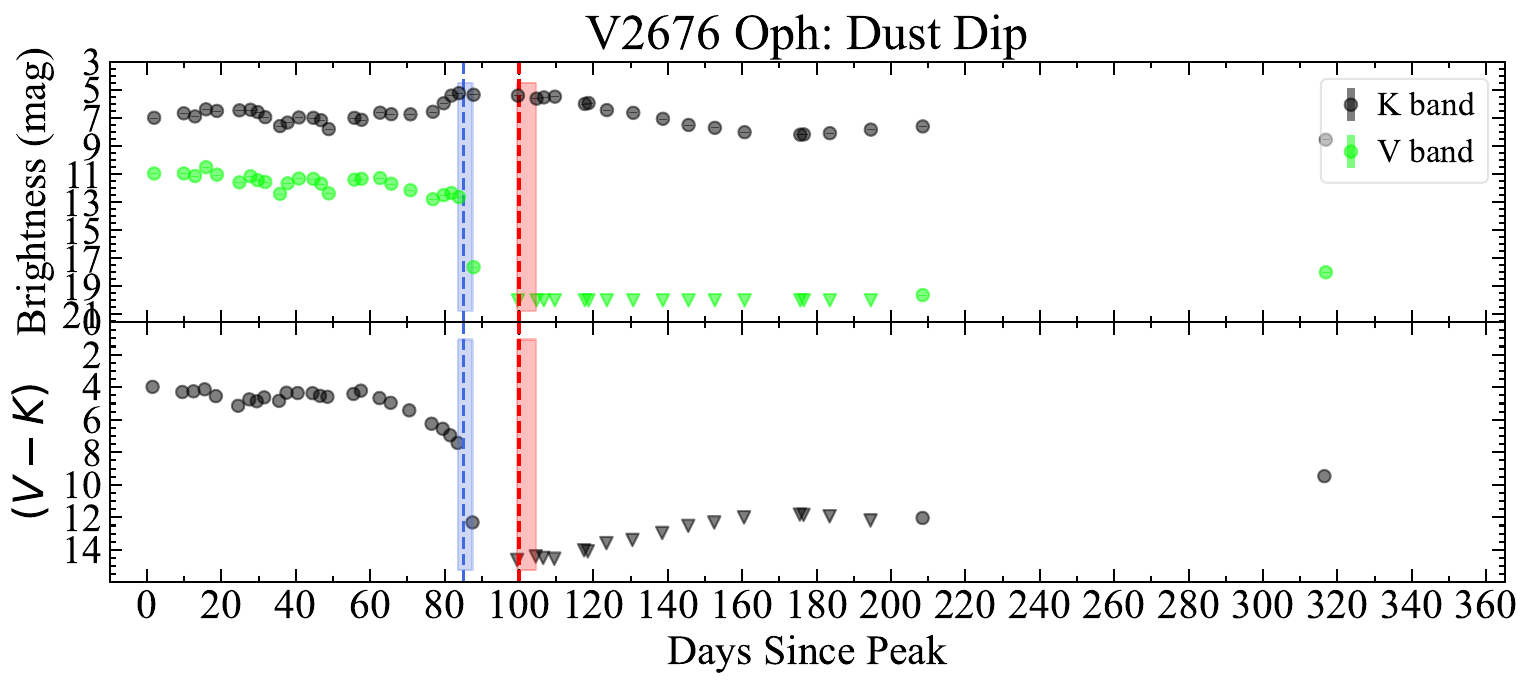}
    \caption{Same as Figure~\ref{fig:app_v475sct} but for nova V2676~Oph, which we classify as `Dust Dip'.}
    \label{fig:app_v2676oph}
\end{figure}

\begin{figure}
    \centering
    \includegraphics[width=1\linewidth]{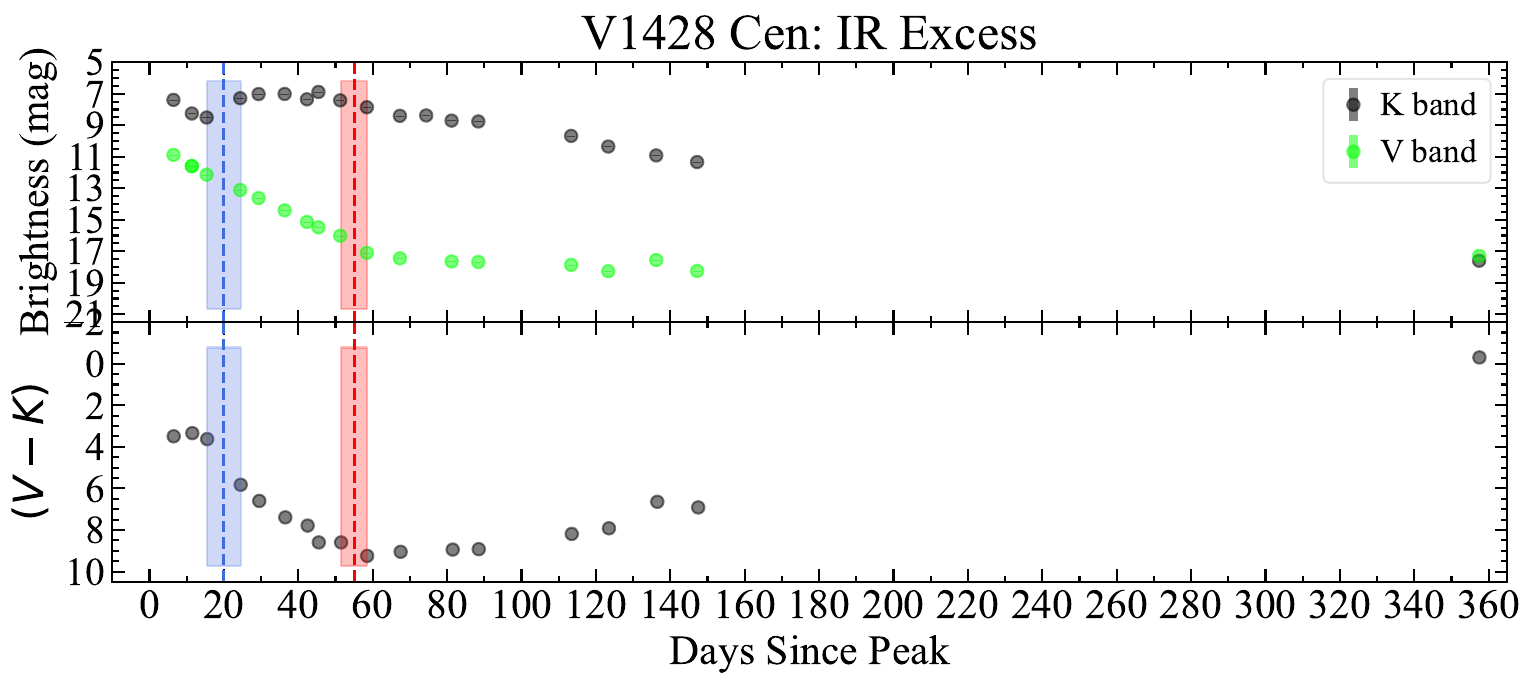}
    \caption{Same as Figure~\ref{fig:app_v475sct} but for nova V1428~Cen, which we classify as `IR Excess'.}
    \label{fig:app_v1428cen}
\end{figure}

\begin{figure}
    \centering
    \includegraphics[width=1\linewidth]{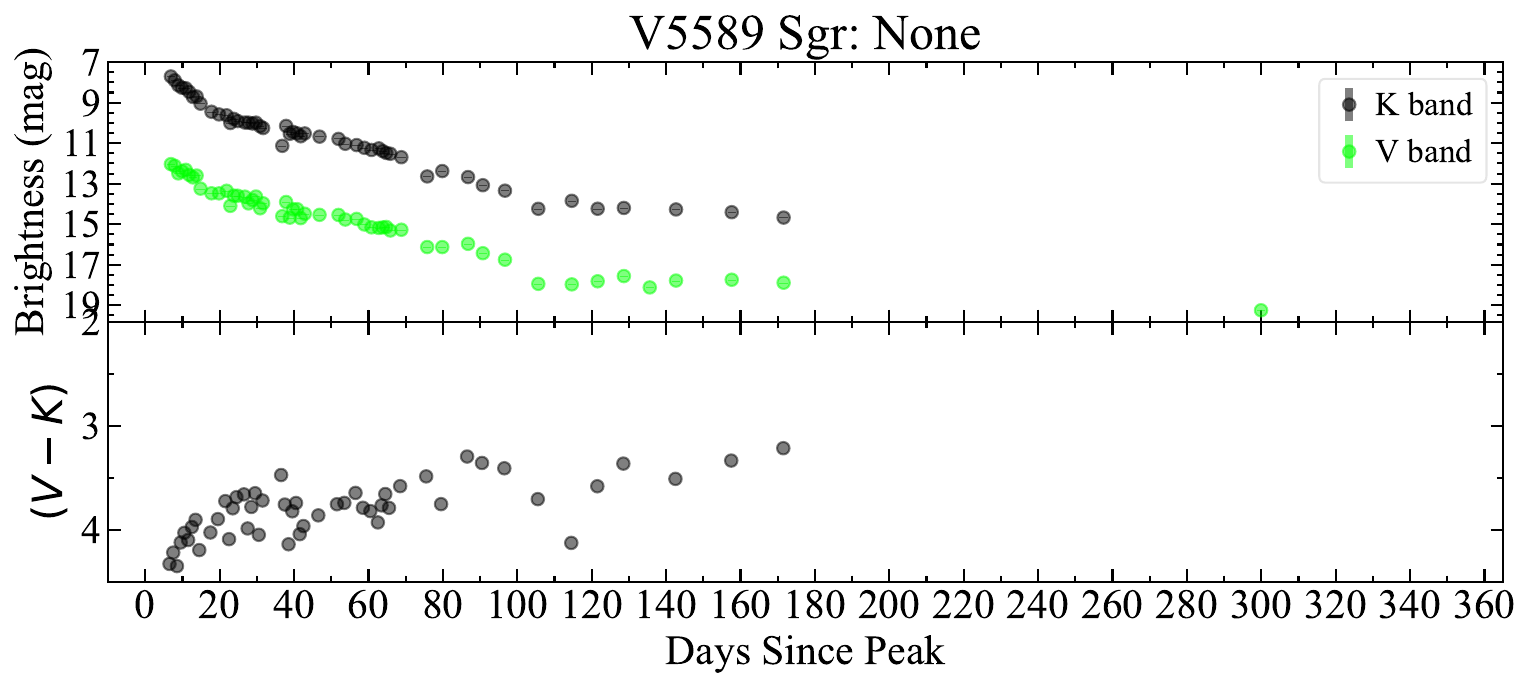}
    \caption{Same as Figure~\ref{fig:app_v475sct} but for nova V5589~Sgr, which we classify as `None'.}
    \label{fig:app_v5589sgr}
\end{figure}

\begin{figure}
    \centering
    \includegraphics[width=1\linewidth]{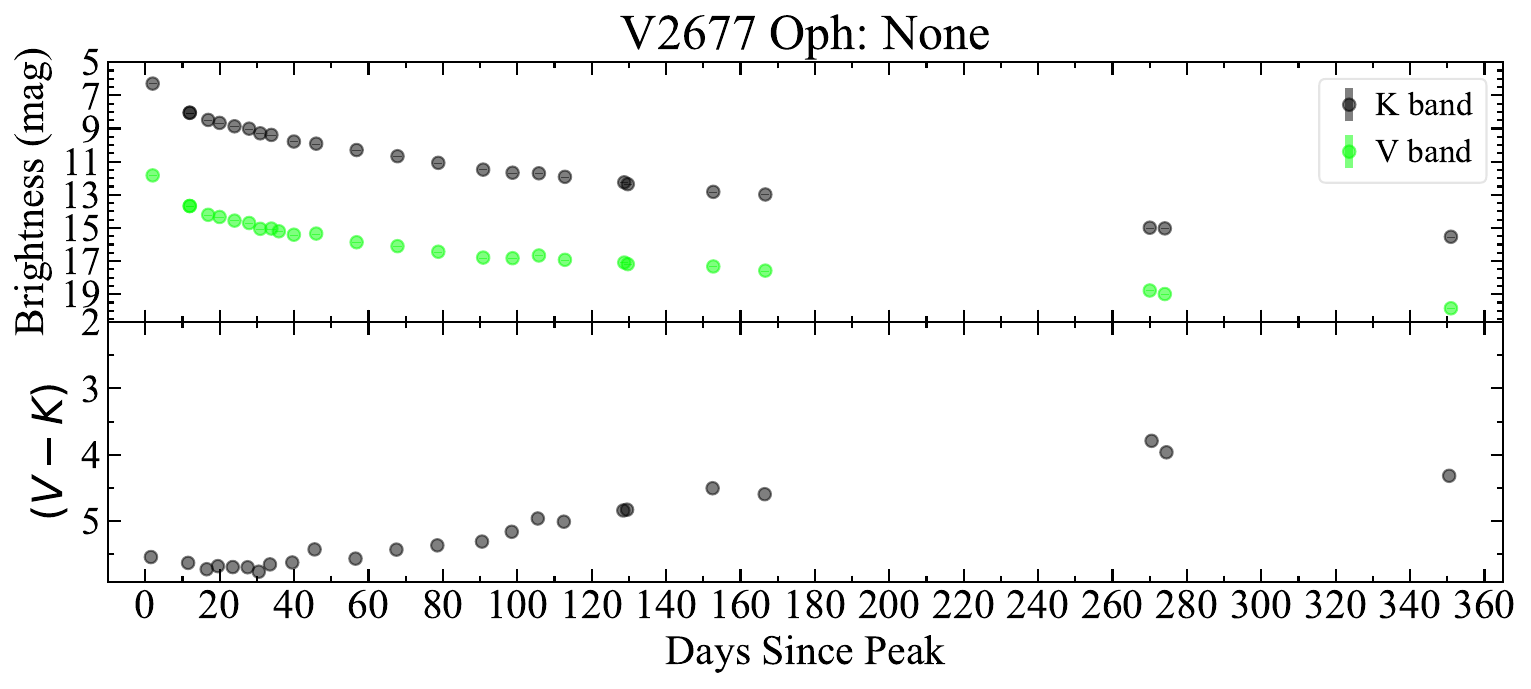}
    \caption{Same as Figure~\ref{fig:app_v475sct} but for nova V2677~Oph, which we classify as `None'.}
    \label{fig:app_v2677oph}
\end{figure}

\begin{figure}
    \centering
    \includegraphics[width=1\linewidth]{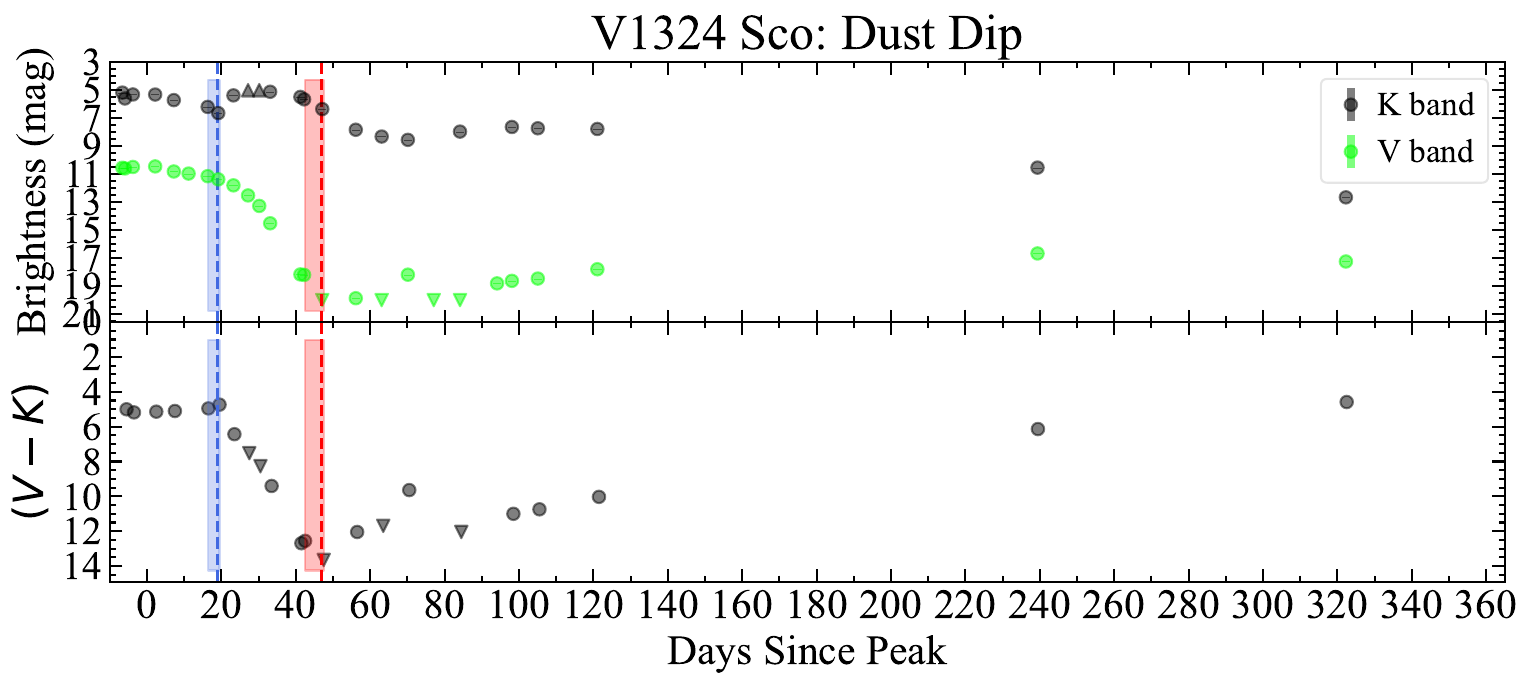}
    \caption{Same as Figure~\ref{fig:app_v475sct} but for nova V1324~Sco, which we classify as `Dust Dip'.}
    \label{fig:app_v1324sco}
\end{figure}

\begin{figure}
    \centering
    \includegraphics[width=1\linewidth]{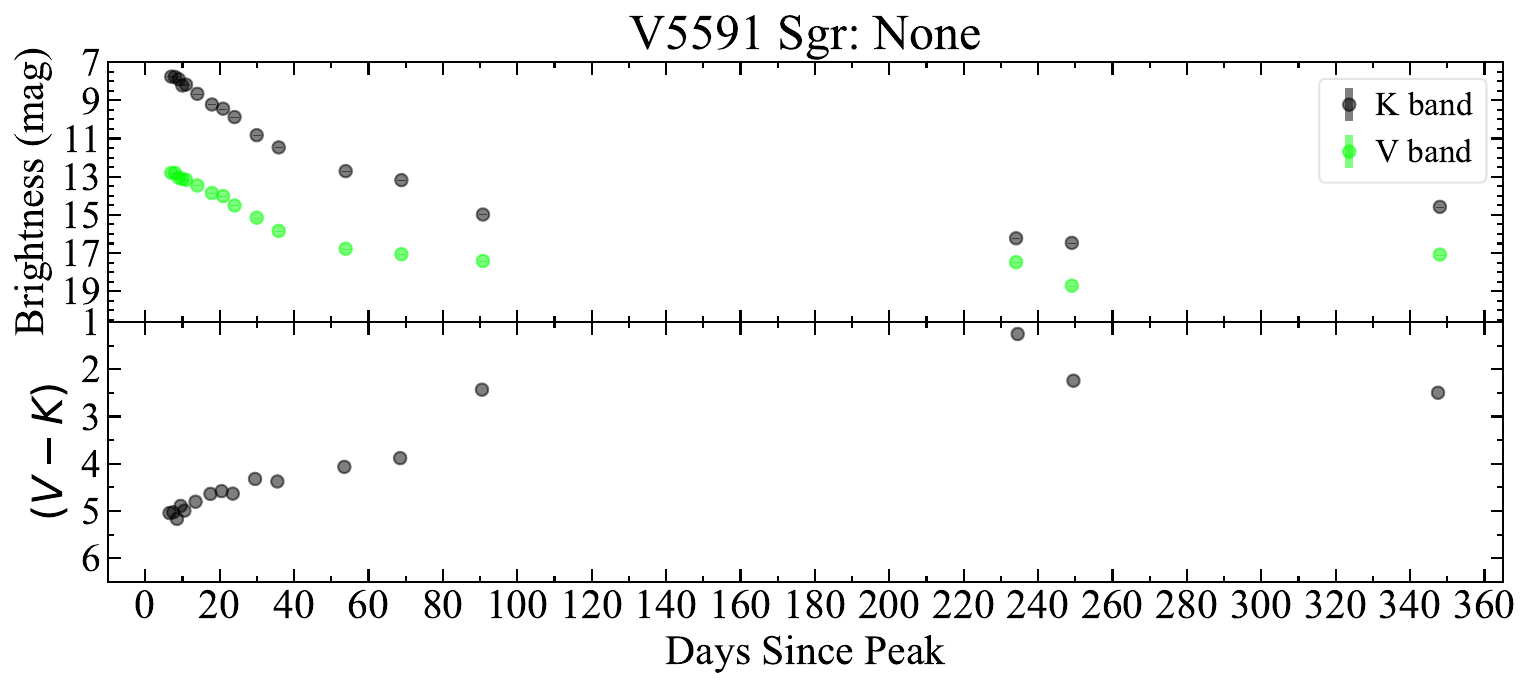}
    \caption{Same as Figure~\ref{fig:app_v475sct} but for nova V5591~Sgr, which we classify as `None'.}
    \label{fig:app_v5591sgr}
\end{figure}

\begin{figure}
    \centering
    \includegraphics[width=1\linewidth]{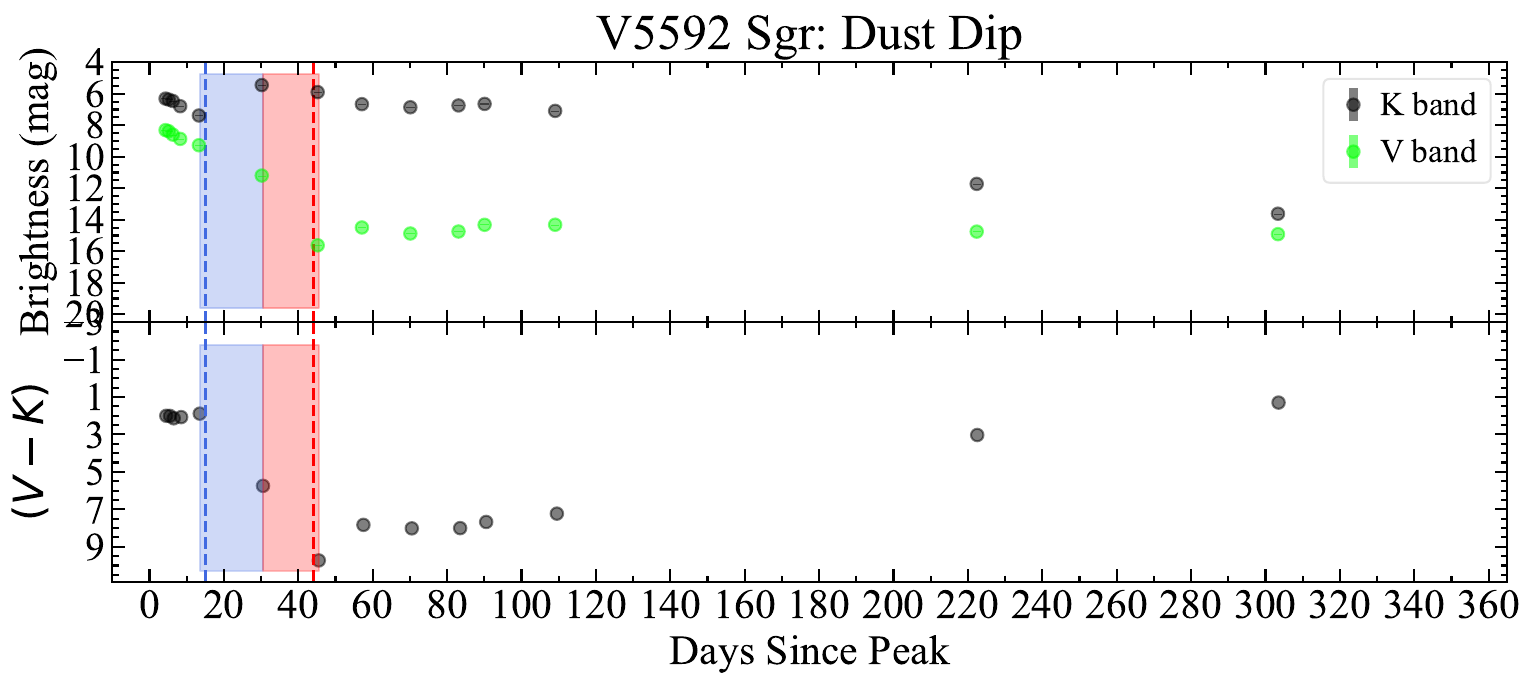}
    \caption{Same as Figure~\ref{fig:app_v475sct} but for nova V5592~Sgr, which we classify as `Dust Dip'.}
    \label{fig:app_v5592sgr}
\end{figure}

\begin{figure}
    \centering
    \includegraphics[width=1\linewidth]{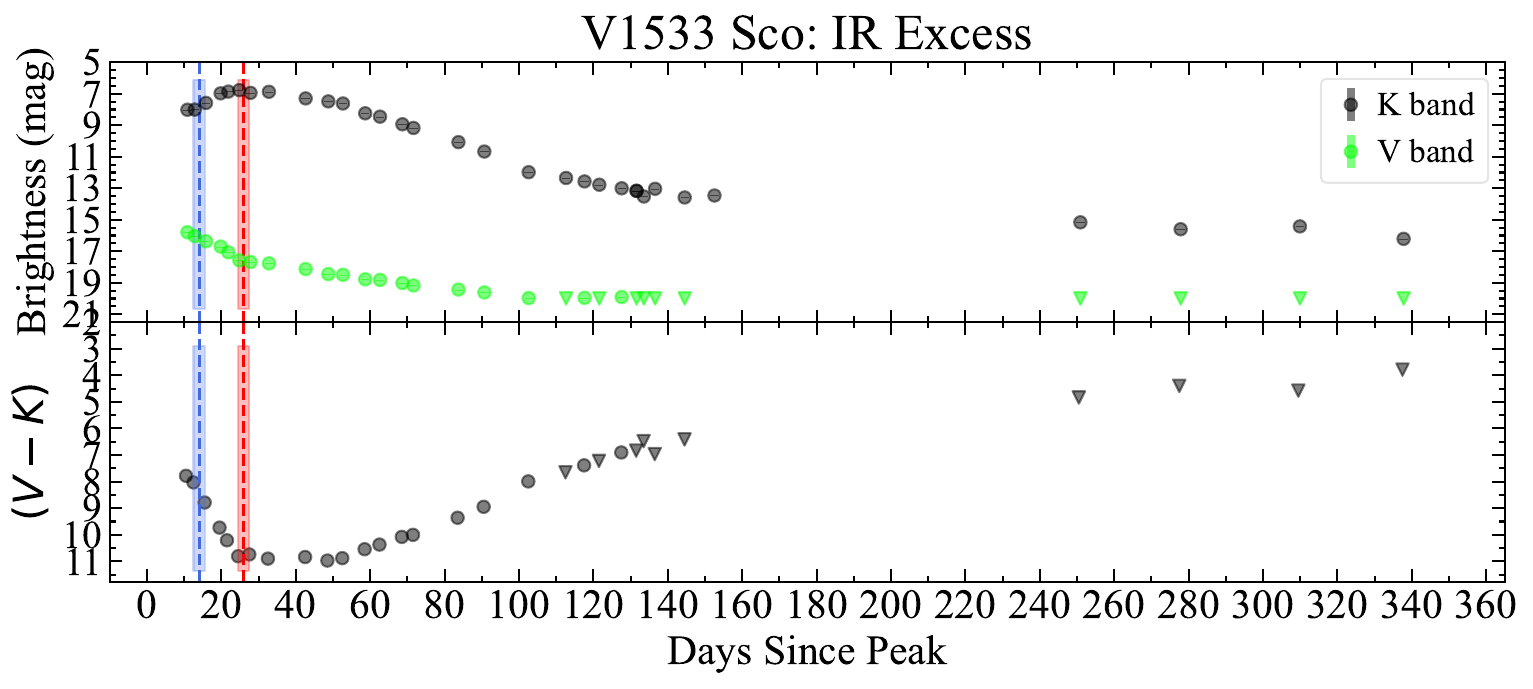}
    \caption{Same as Figure~\ref{fig:app_v475sct} but for nova V1533~Sco, which we classify as `IR Excess'.}
    \label{fig:app_v1533sco}
\end{figure}

\begin{figure}
    \centering
    \includegraphics[width=1\linewidth]{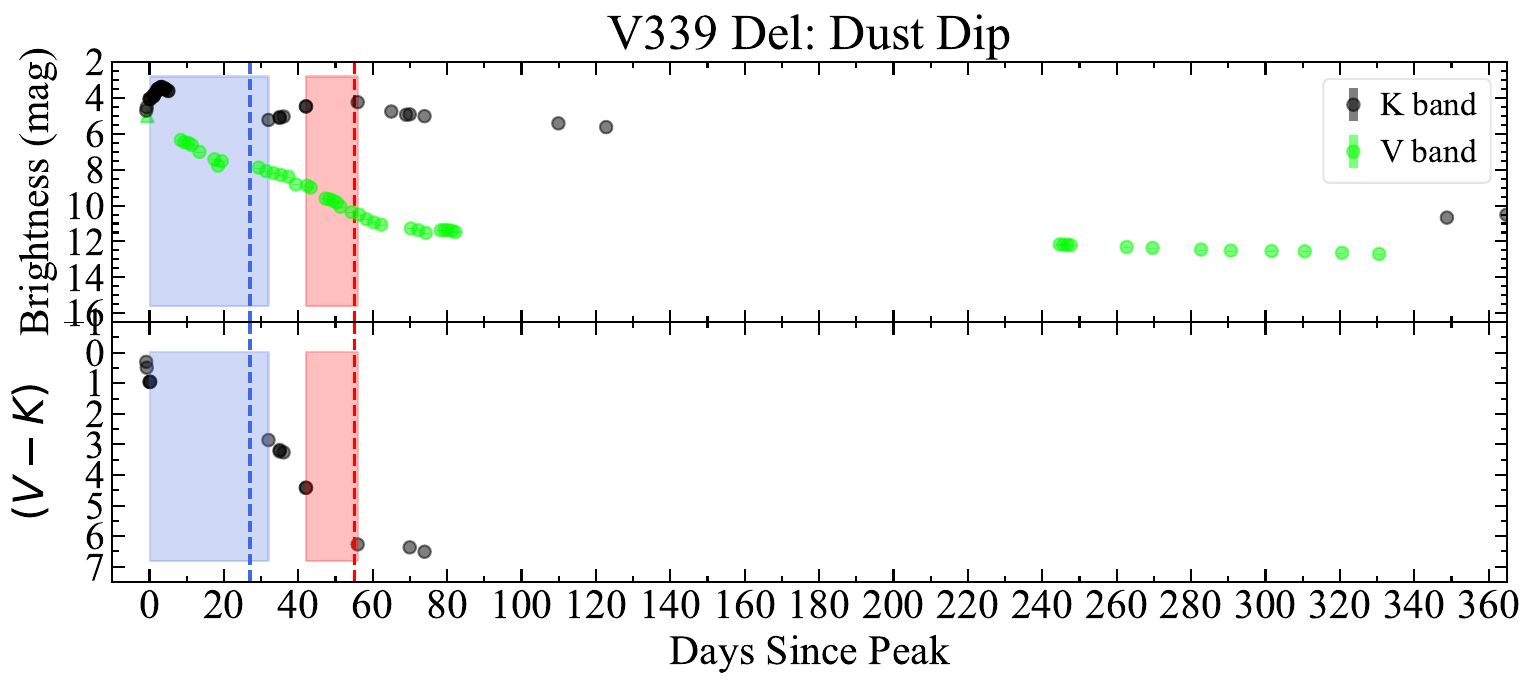}
    \caption{Same as Figure~\ref{fig:app_v475sct} but for nova V339~Del, which we classify as `Dust Dip'. \citet{Gehrz_etal_2015} shows a $V$-band light curve for V339 Del that has measurements out to day $\sim$ 180, in which the $V$-band curve starts to flatten between days 60-80, past which point it shows no appreciable dimming. As the $K$-band light curve does not continue to brighten past day $\sim$ 55, the nova likely continues to slowly get bluer over time, so we mark the dust event `bottom' at day $\sim$ 55.}
    \label{fig:app_v339del}
\end{figure}

\begin{figure}
    \centering
    \includegraphics[width=1\linewidth]{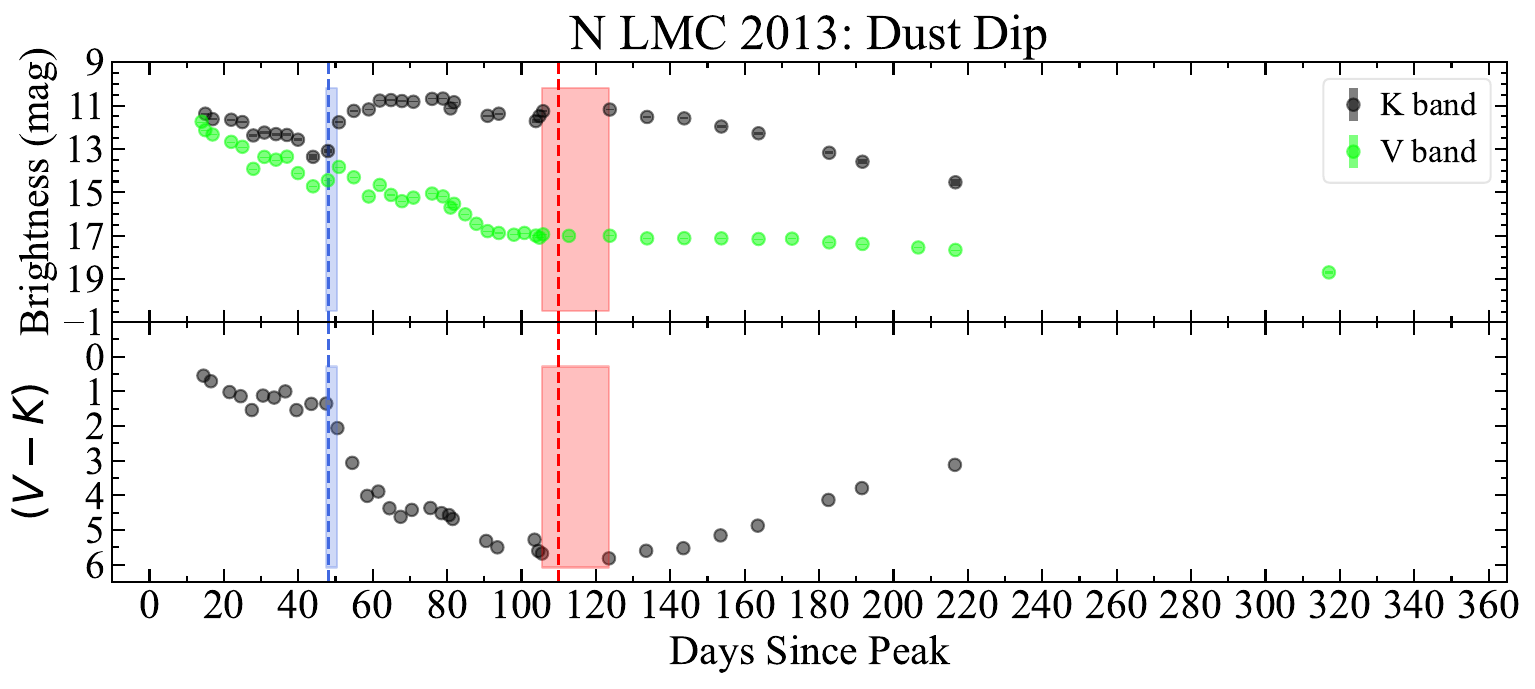}
    \caption{Same as Figure~\ref{fig:app_v475sct} but for nova N LMC 2013, which we classify as `Dust Dip'.}
    \label{fig:app_nlmc2013}
\end{figure}

\begin{figure}
    \centering
    \includegraphics[width=1\linewidth]{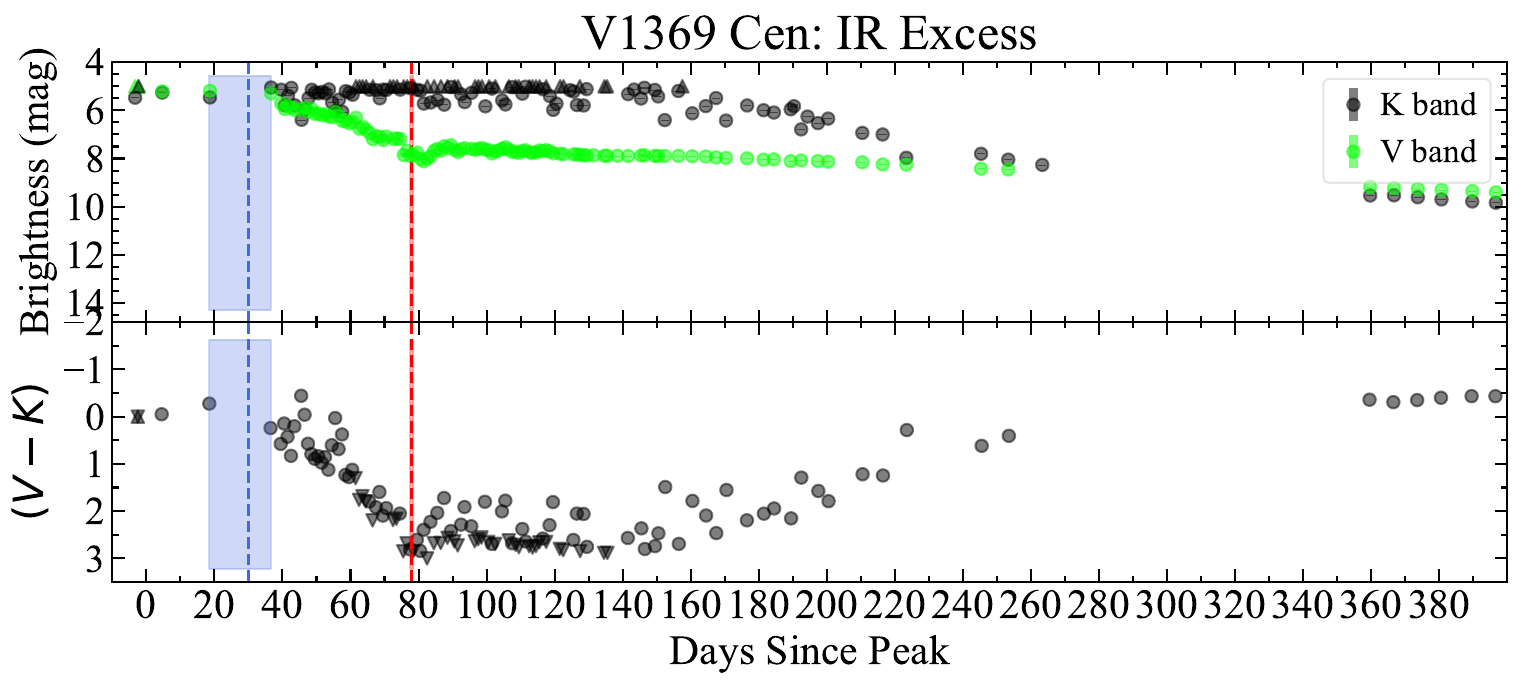}
    \caption{Same as Figure~\ref{fig:app_v475sct} but for nova V1369~Cen, which we classify as `IR Excess'. Because the first measurement in both the $K$-band and $V$-band is saturated, we have used the first $V$-band measurement after peak as our `$V$ peak in SMARTS Data' for figures \ref{fig:Color_Change_VK}, \ref{fig:Color_Change_VH}, and \ref{fig:Color_Change_VJ}. The two measurements are both close to peak ($<10$ days), so this decision should not significantly affect results.}
    \label{fig:app_v1369cen}
\end{figure}

\begin{figure}
    \centering
    \includegraphics[width=1\linewidth]{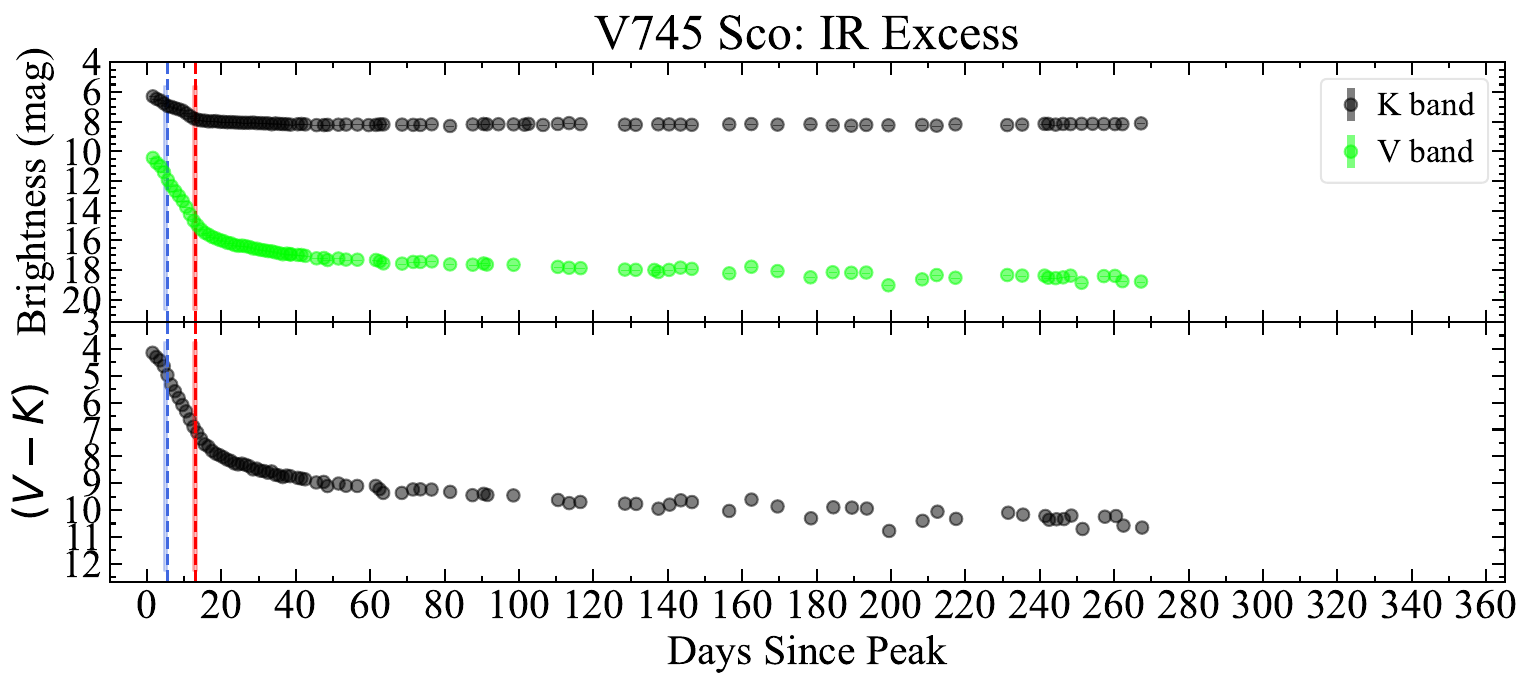}
    \caption{Same as Figure~\ref{fig:app_v475sct} but for nova V745~Sco (2014), which we classify as `IR Excess'.}
    \label{fig:app_v745sco}
\end{figure}

\begin{figure}
    \centering
    \includegraphics[width=1\linewidth]{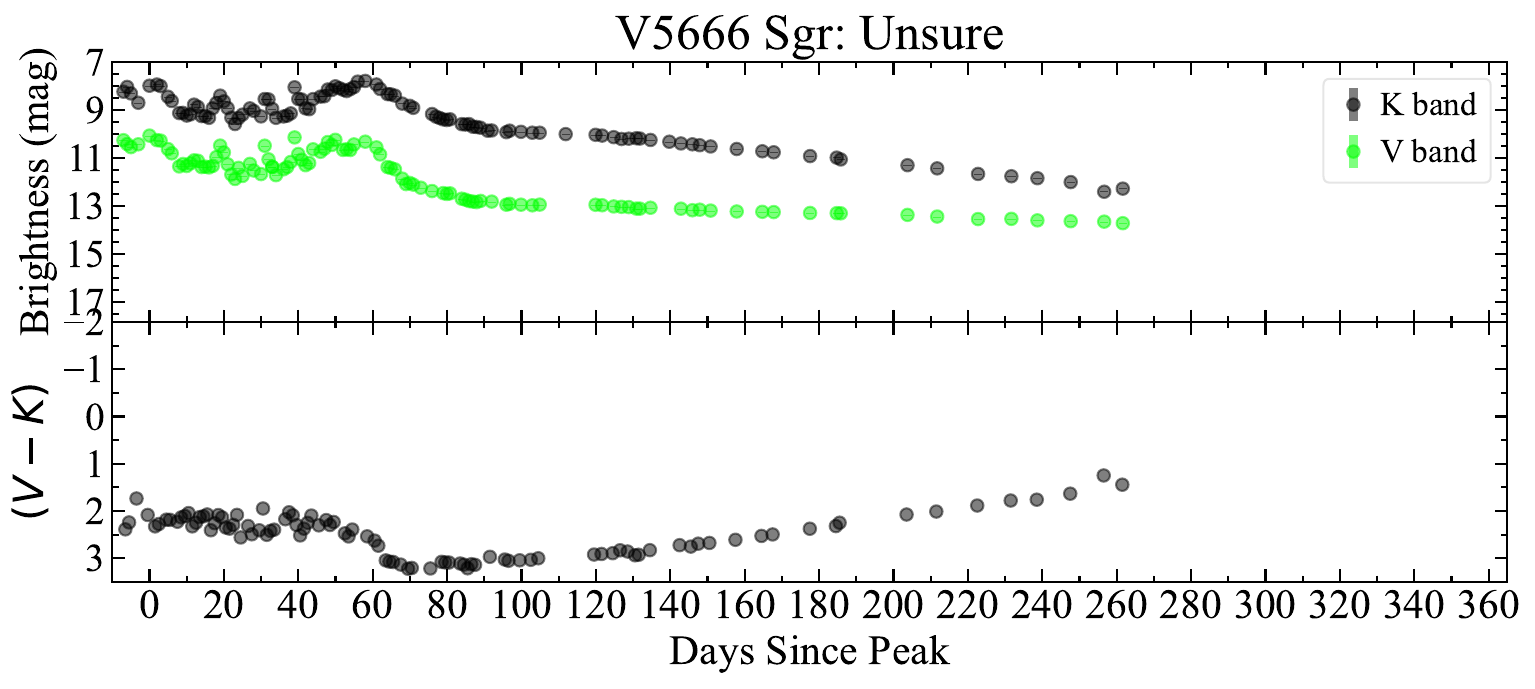}
    \caption{Same as Figure~\ref{fig:app_v475sct} but for nova V5666~Sgr, which we classify as `Unsure'.}
    \label{fig:app_v5666sgr}
\end{figure}

\begin{figure}
    \centering
    \includegraphics[width=1\linewidth]{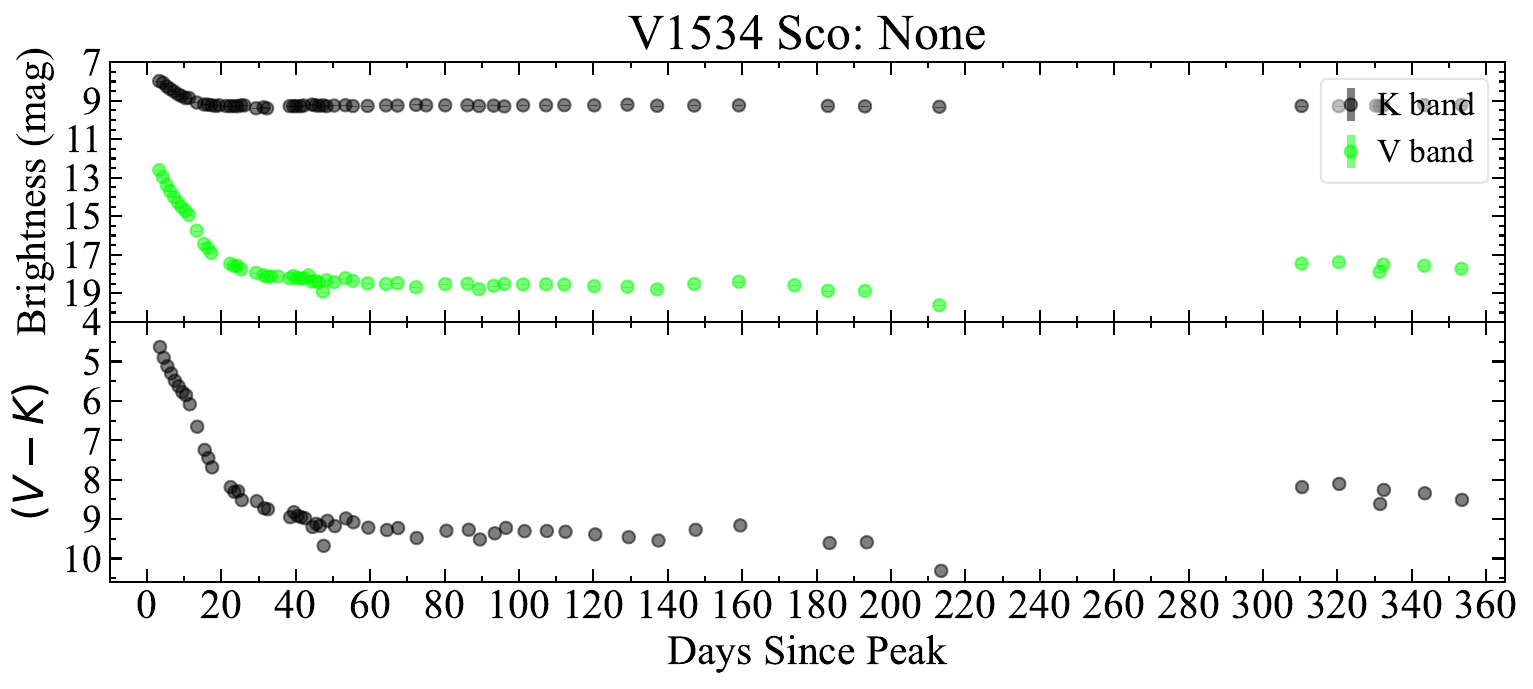}
    \caption{Same as Figure~\ref{fig:app_v475sct} but for nova V1534~Sco, which we classify as `None'.}
    \label{fig:app_v1534sco}
\end{figure}

\begin{figure}
    \centering
    \includegraphics[width=1\linewidth]{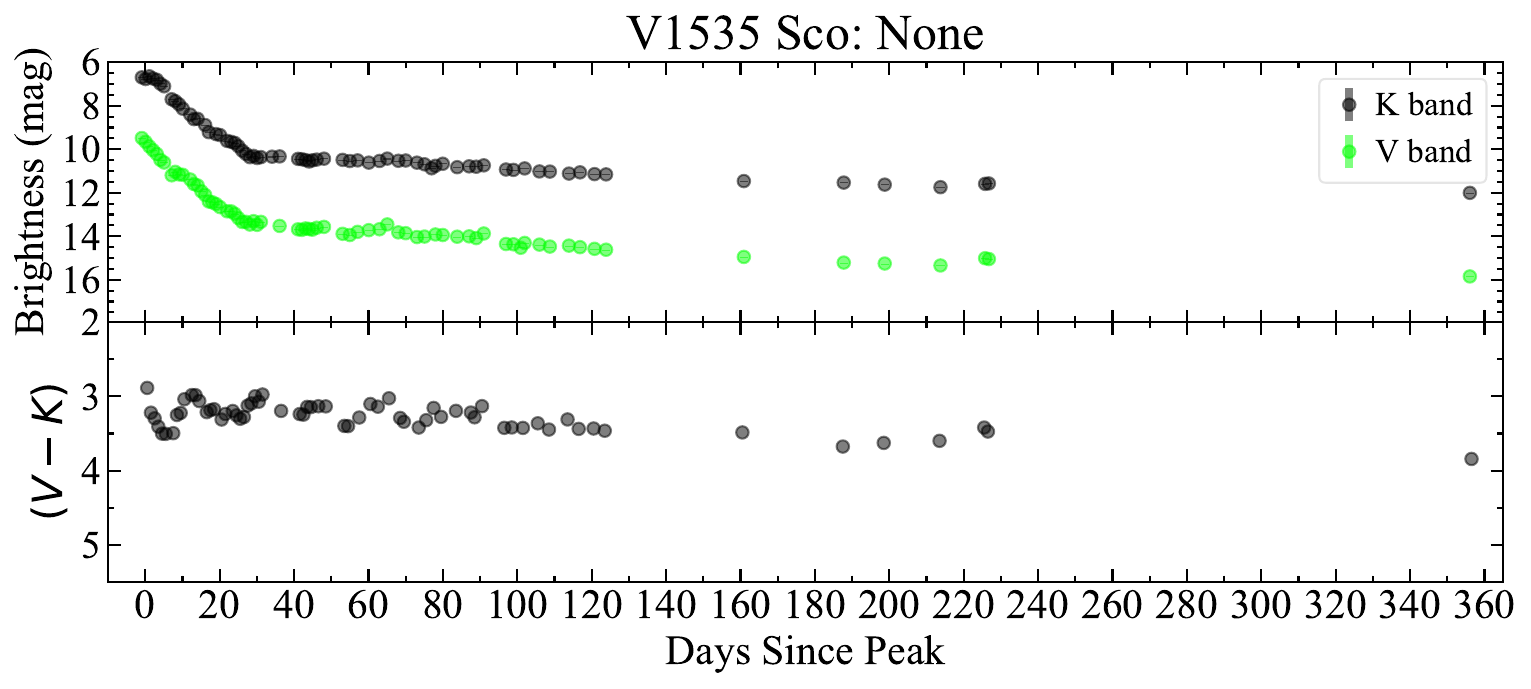}
    \caption{Same as Figure~\ref{fig:app_v475sct} but for nova V1535~Sco, which we classify as `None'.}
    \label{fig:app_v1535sco}
\end{figure}

\begin{figure}
    \centering
    \includegraphics[width=1\linewidth]{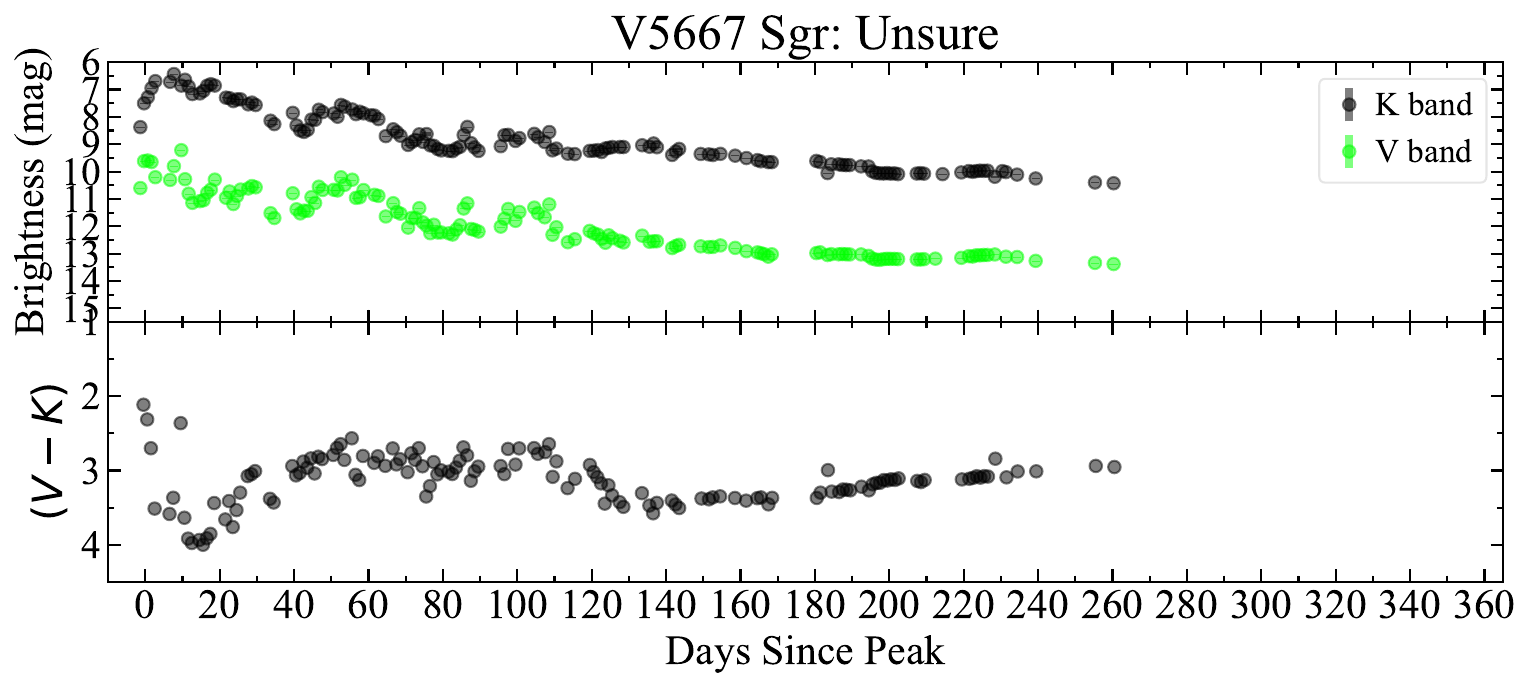}
    \caption{Same as Figure~\ref{fig:app_v475sct} but for nova V5667~Sgr, which we classify as `Unsure'.}
    \label{fig:app_v5667sgr}
\end{figure}

\begin{figure}
    \centering
    \includegraphics[width=1\linewidth]{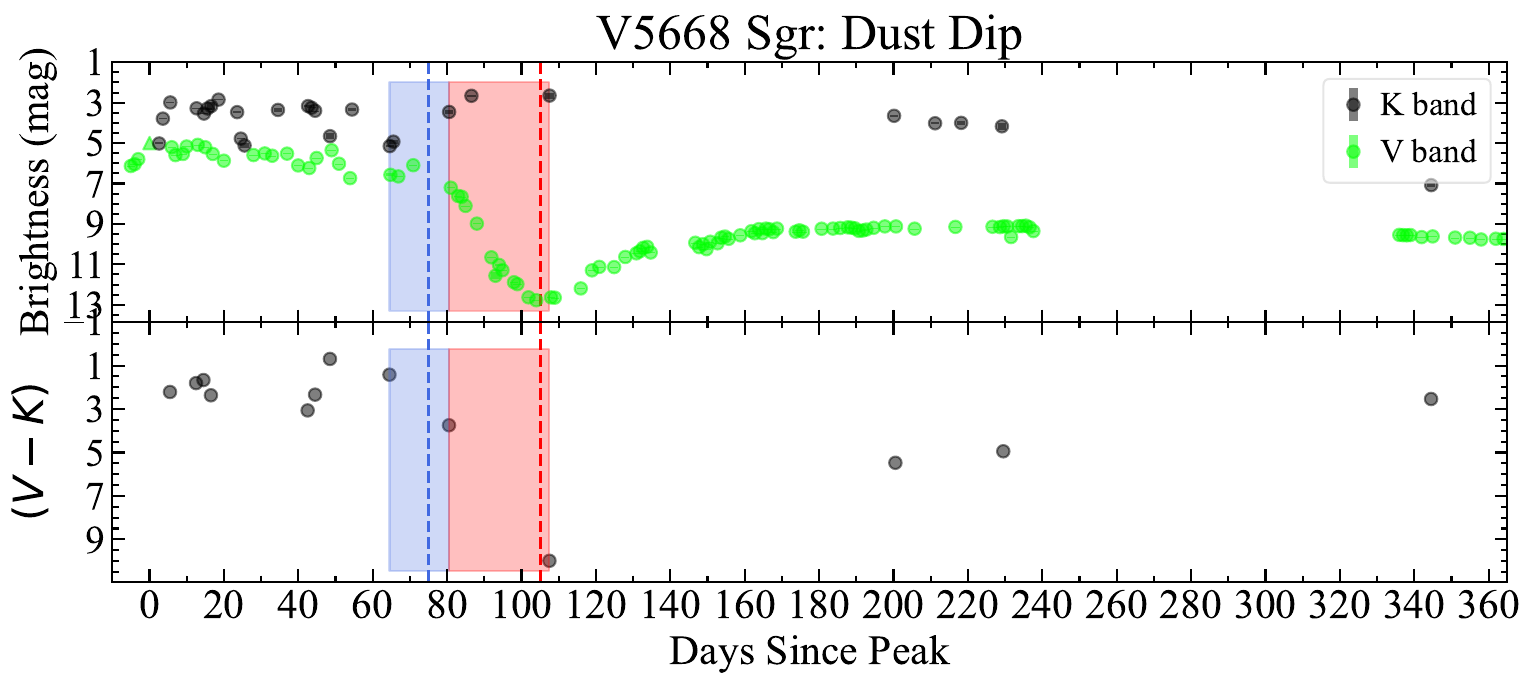}
    \caption{Same as Figure~\ref{fig:app_v475sct} but for nova V5668~Sgr, which we classify as `Dust Dip'.}
    \label{fig:app_v5568sgr}
\end{figure}

\begin{figure}
    \centering
    \includegraphics[width=1\linewidth]{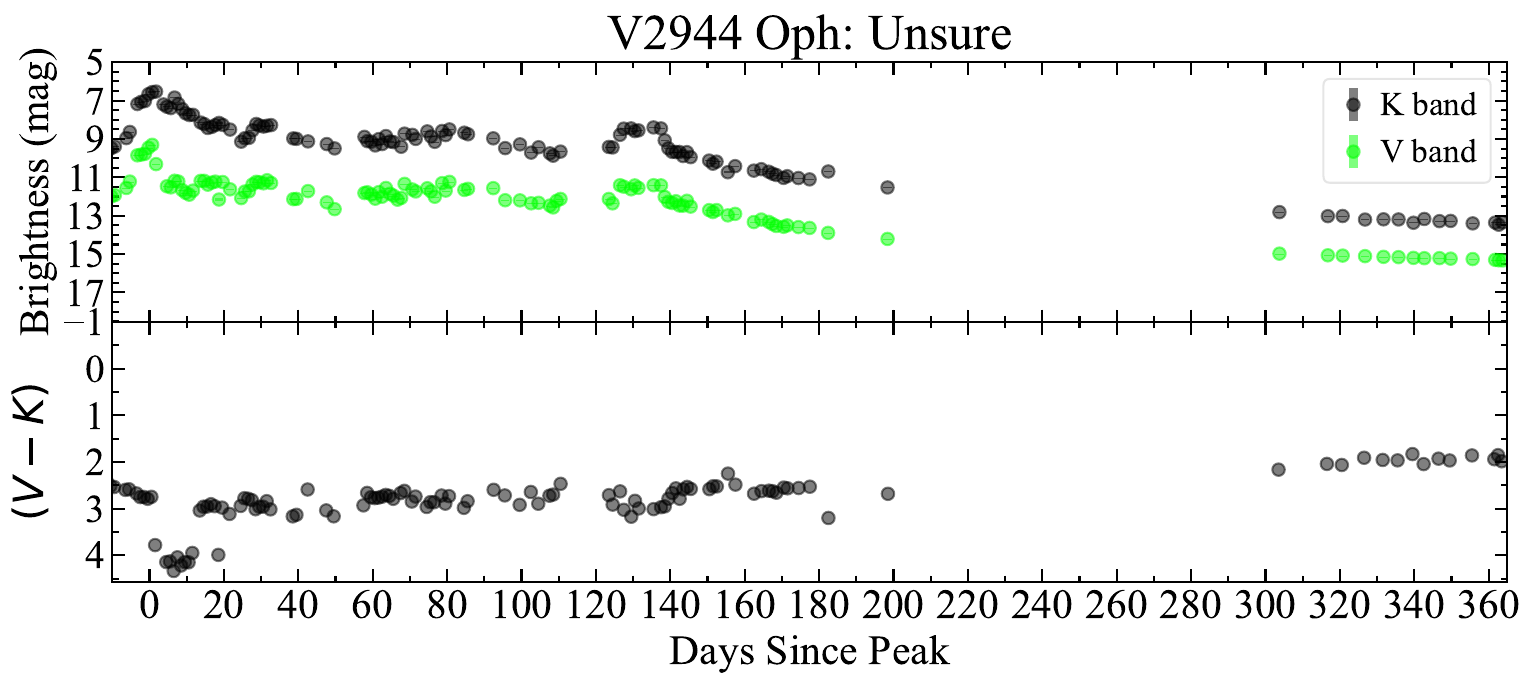}
    \caption{Same as Figure~\ref{fig:app_v475sct} but for nova V2944~Oph, which we classify as `Unsure'.}
    \label{fig:app_v2944oph}
\end{figure}

\begin{figure}
    \centering
    \includegraphics[width=1\linewidth]{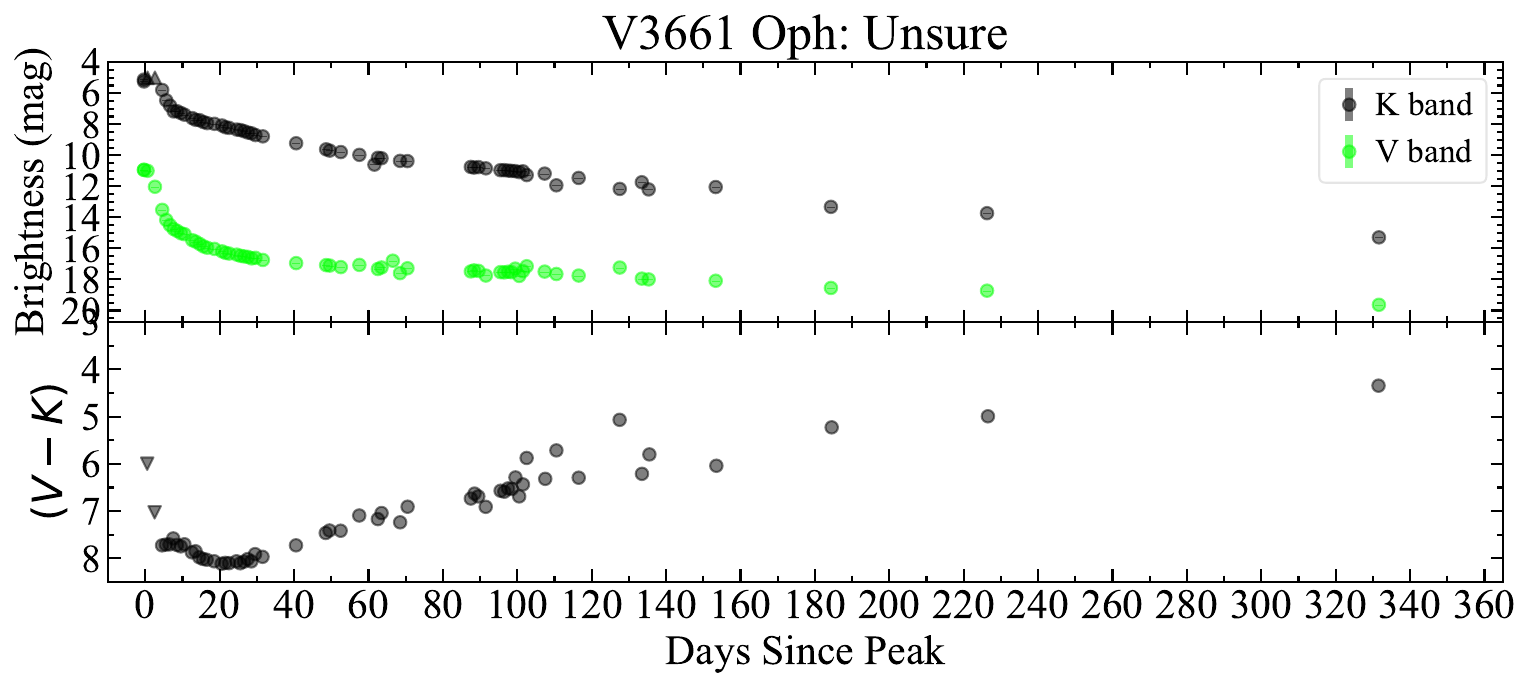}
    \caption{Same as Figure~\ref{fig:app_v475sct} but for nova V3661~Oph, which we classify as `Unsure'.}
    \label{fig:app_v3661oph}
\end{figure}

\begin{figure}
    \centering
    \includegraphics[width=1\linewidth]{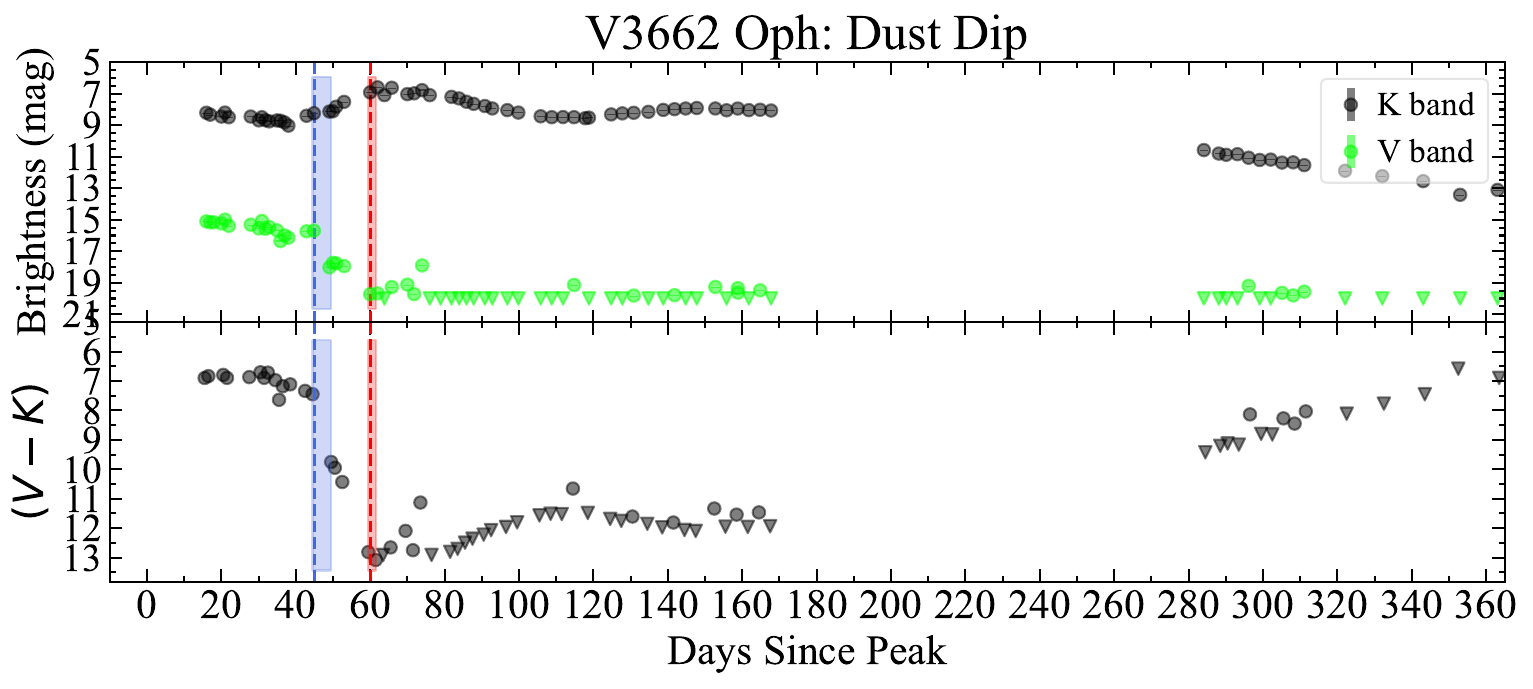}
    \caption{Same as Figure~\ref{fig:app_v475sct} but for nova V3662~Oph, which we classify as `Dust Dip'.}
    \label{fig:app_v3662oph}
\end{figure}

\begin{figure}
    \centering
    \includegraphics[width=1\linewidth]{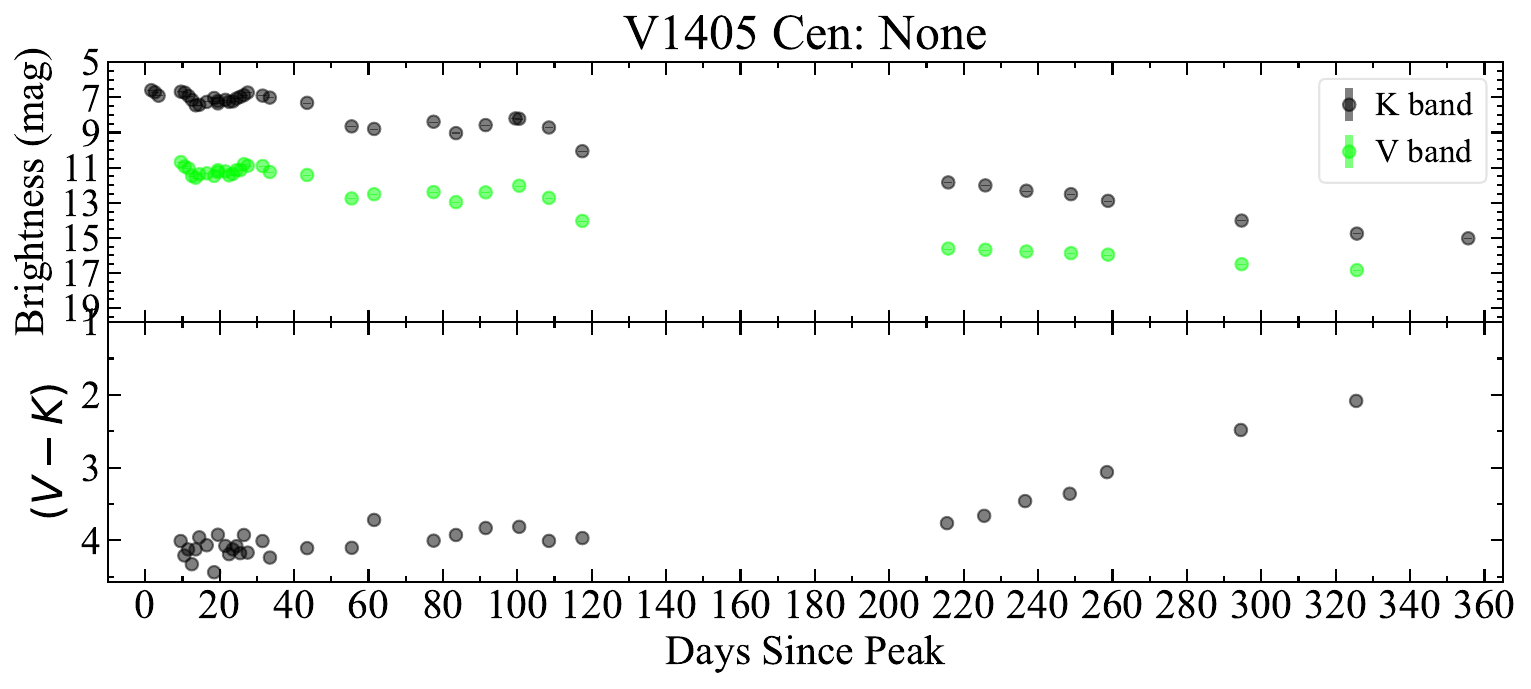}
    \caption{Same as Figure~\ref{fig:app_v475sct} but for nova V1405~Cen, which we classify as `None'.}
    \label{fig:app_v1405cen}
\end{figure}

\begin{figure}
    \centering
    \includegraphics[width=1\linewidth]{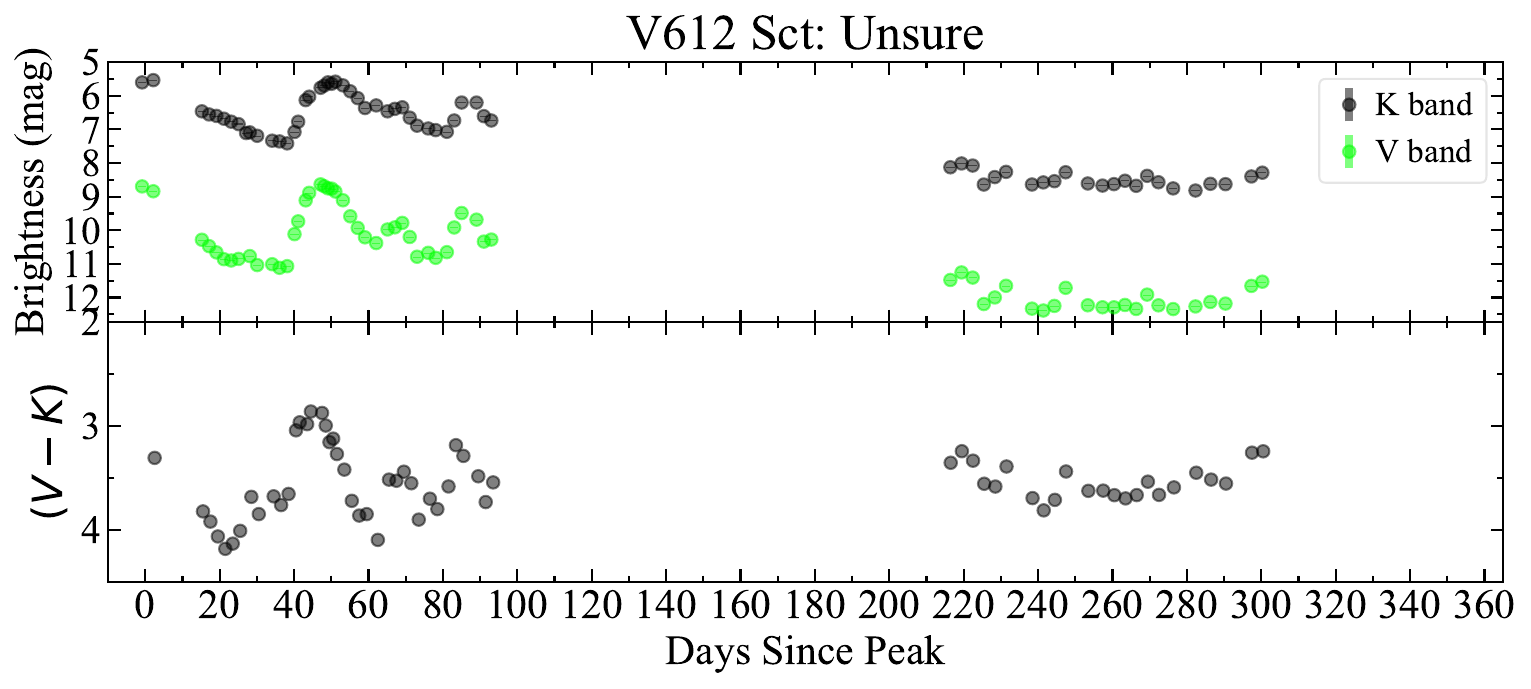}
    \caption{Same as Figure~\ref{fig:app_v475sct} but for nova V612~Sct, which we classify as `Unsure'.}
    \label{fig:app_v612sct}
\end{figure}

\begin{figure}
    \centering
    \includegraphics[width=1\linewidth]{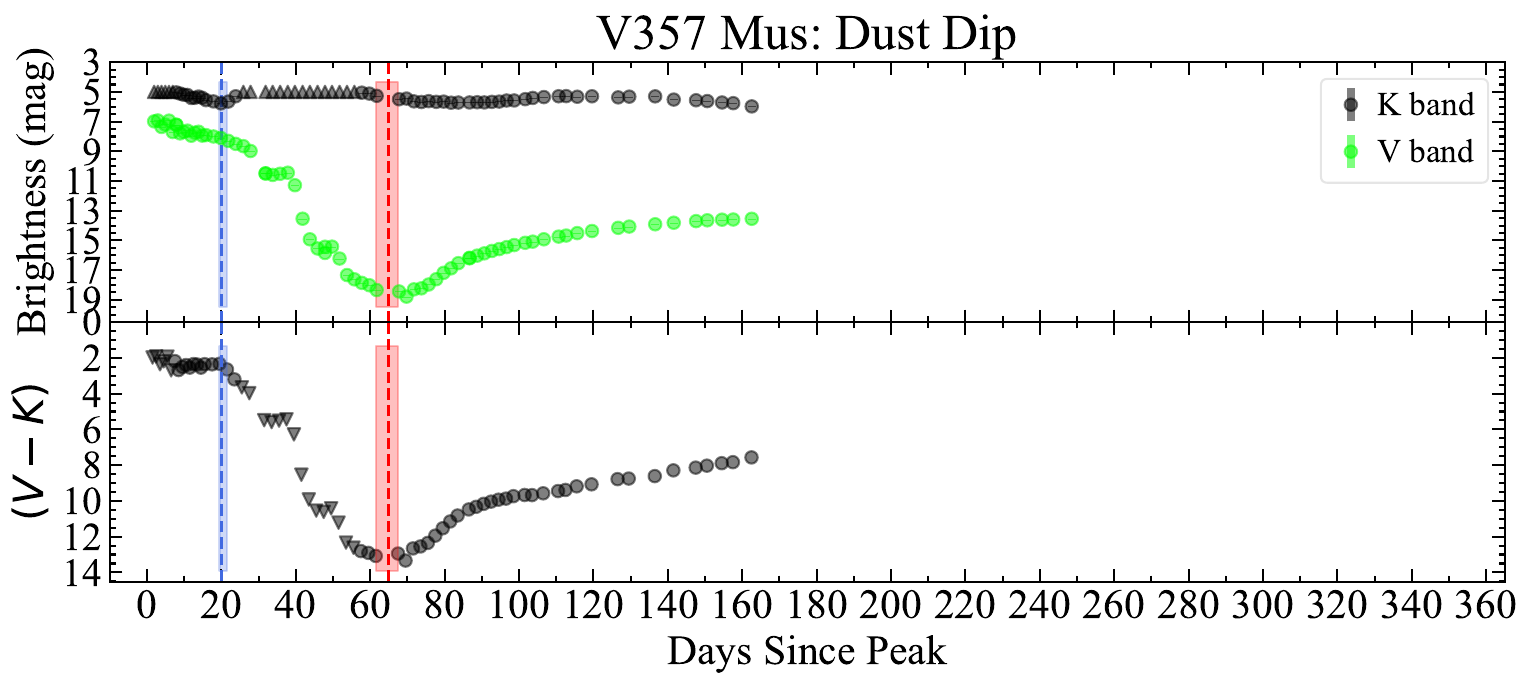}
    \caption{Same as Figure~\ref{fig:app_v475sct} but for nova V357~Mus, which we classify as `Dust Dip'.}
    \label{fig:app_v357mus}
\end{figure}

\begin{figure}
    \centering
    \includegraphics[width=1\linewidth]{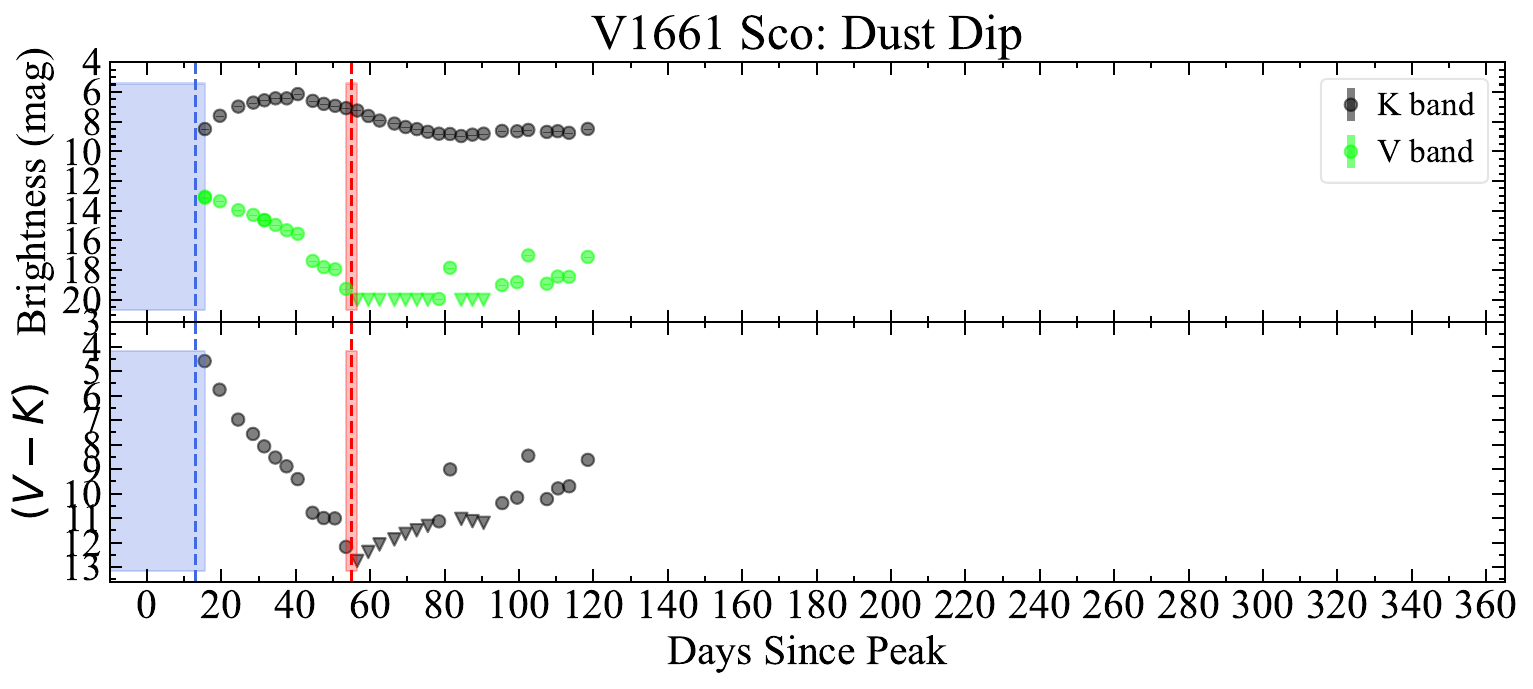}
    \caption{Same as Figure~\ref{fig:app_v475sct} but for nova V1661~Sco, which we classify as `Dust Dip'.}
    \label{fig:app_v1661sco}
\end{figure}

\begin{figure}
    \centering
    \includegraphics[width=1\linewidth]{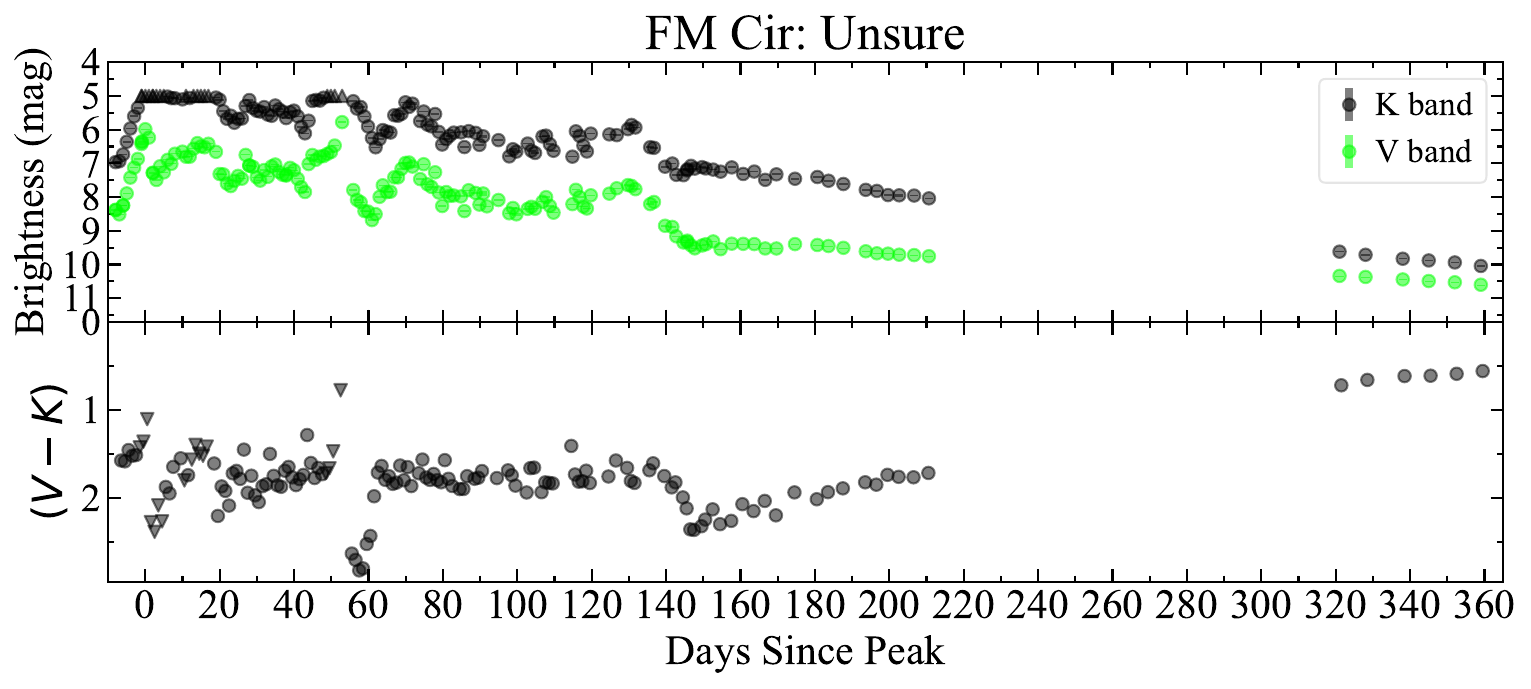}
    \caption{Same as Figure~\ref{fig:app_v475sct} but for nova FM~Cir, which we classify as `Unsure'.}
    \label{fig:app_fmcir}
\end{figure}

\begin{figure}
    \centering
    \includegraphics[width=1\linewidth]{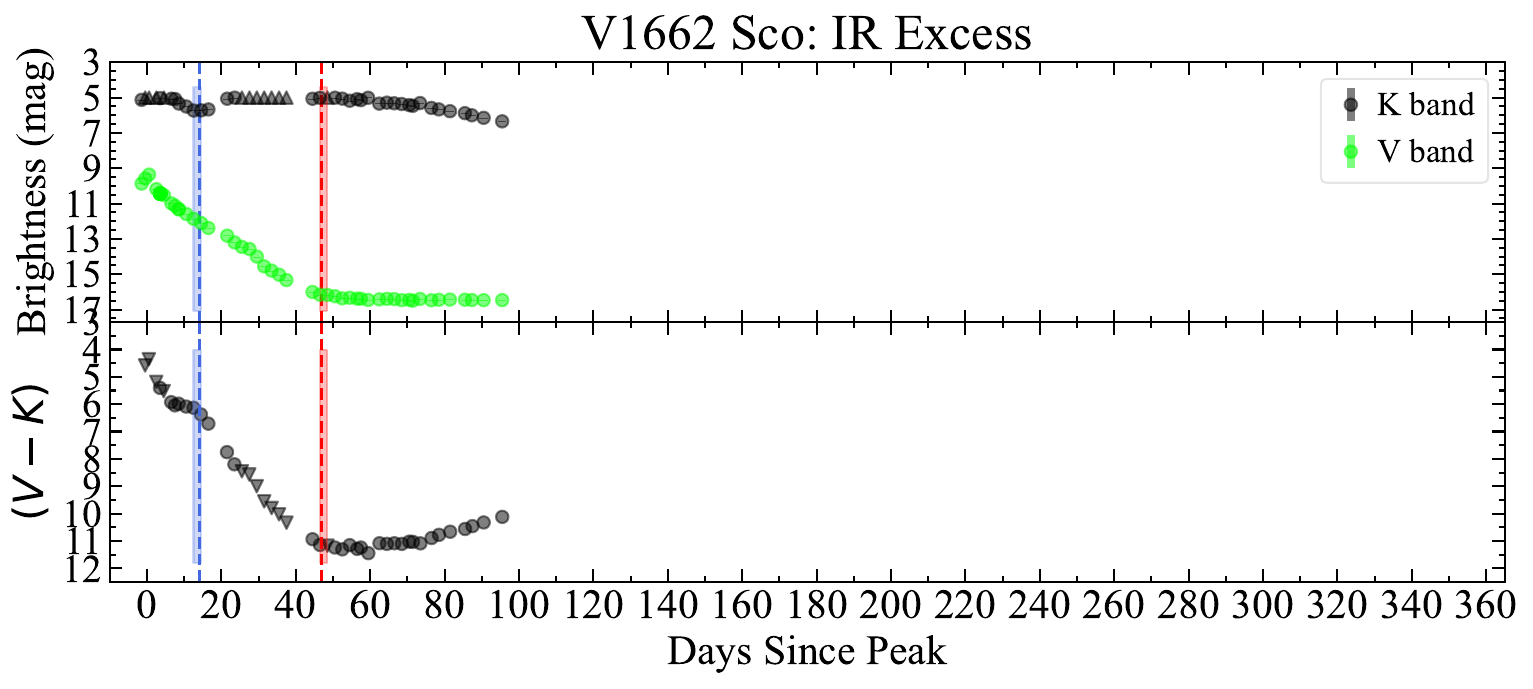}
    \caption{Same as Figure~\ref{fig:app_v475sct} but for nova V1662~Sco, which we classify as `IR Excess'.}
    \label{fig:app_v1662sco}
\end{figure}

\begin{figure}
    \centering
    \includegraphics[width=1\linewidth]{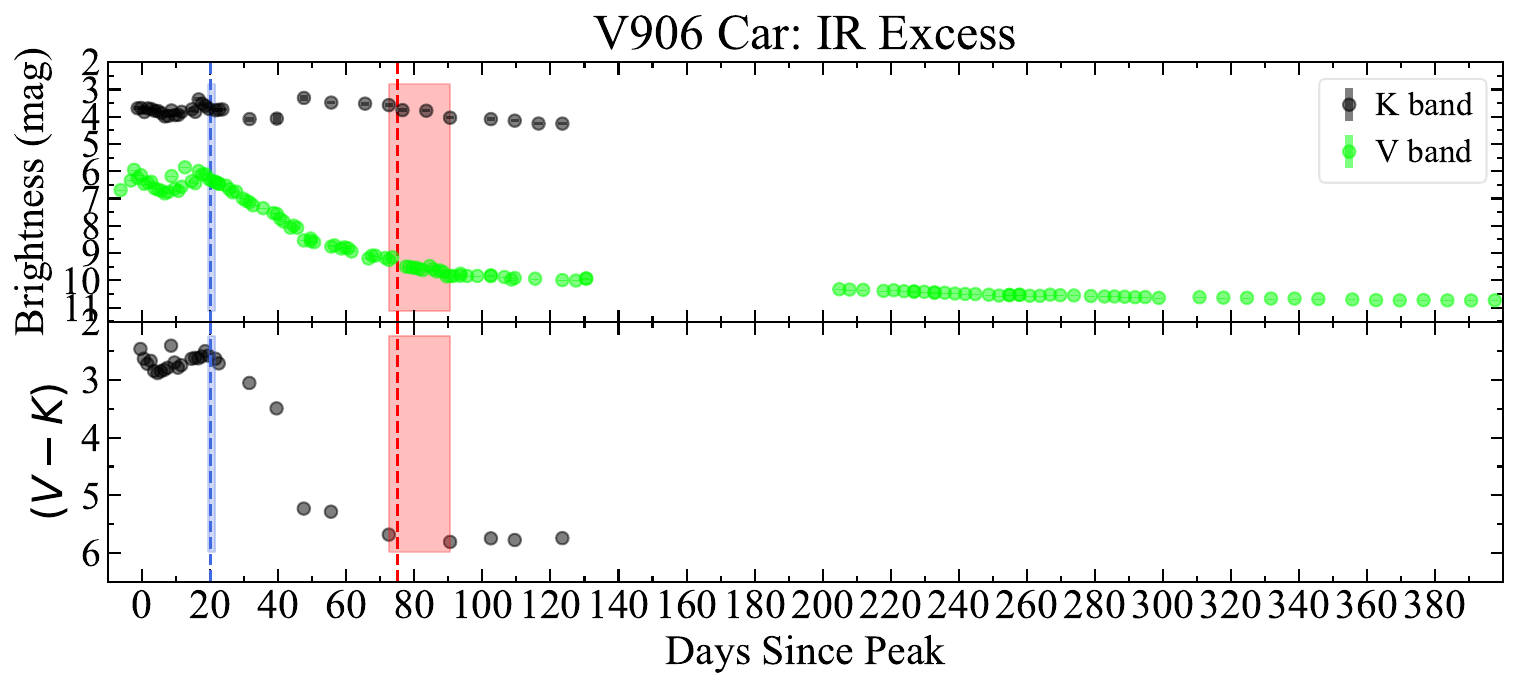}
    \caption{Same as Figure~\ref{fig:app_v475sct} but for nova V906~Car, which we classify as `IR Excess'.}
    \label{fig:app_v906car}
\end{figure}

\clearpage

\section{Evolution of Optical--IR Colours around Light Curve Maximum}\label{sec:colorchange}

\begin{figure}
    \centering
    \includegraphics[width=1\linewidth]{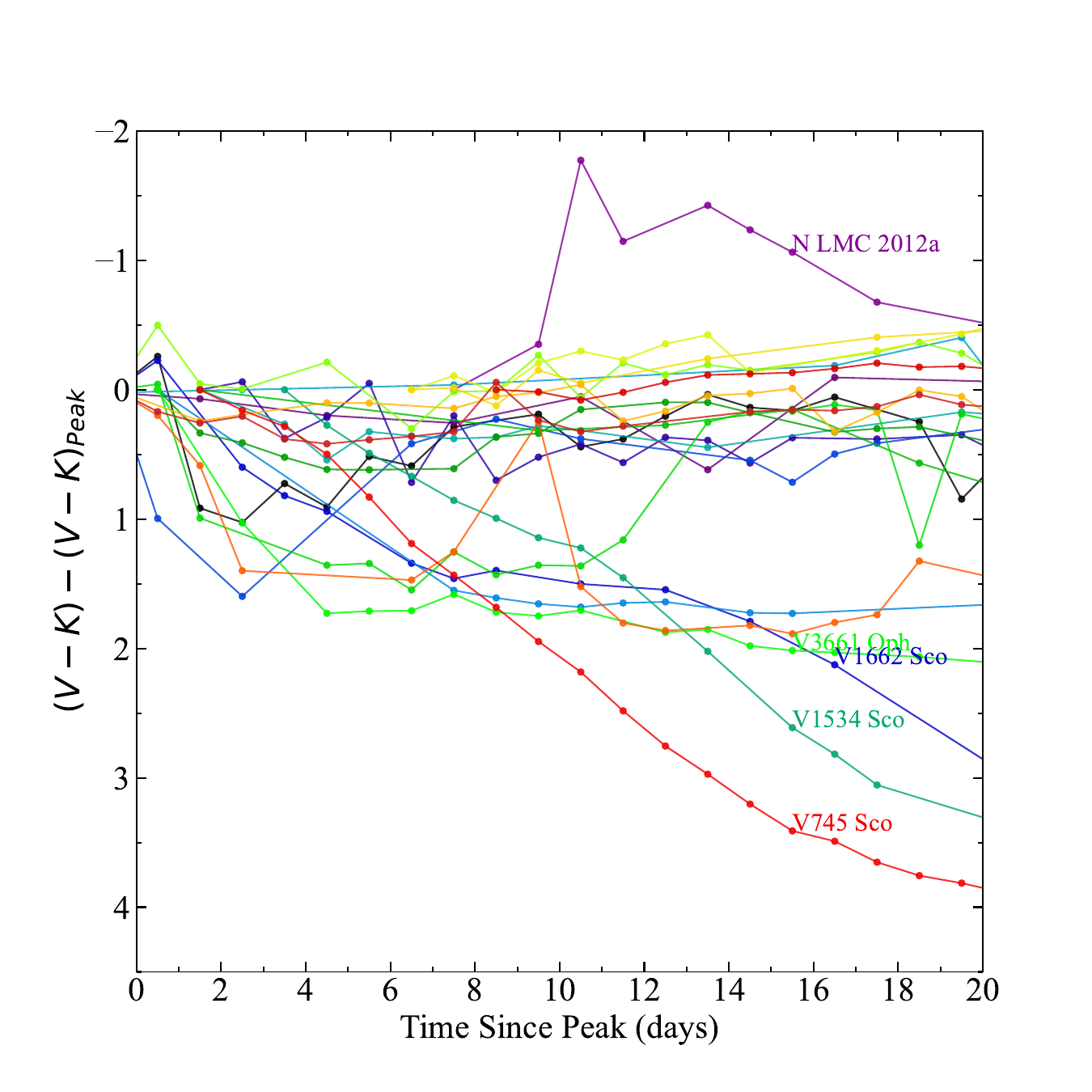}
    \caption{The evolution of $V-K$ colour for 20 days following $V$-band peak. Includes light curves with at least 2 measurements between $t_{peak}$ and 10 days after peak, and at least 2 measurements between 10 and 20 days after peak.}
    \label{fig:vk_evolution}
\end{figure}

\begin{figure}
    \centering
    \includegraphics[width=1\linewidth]{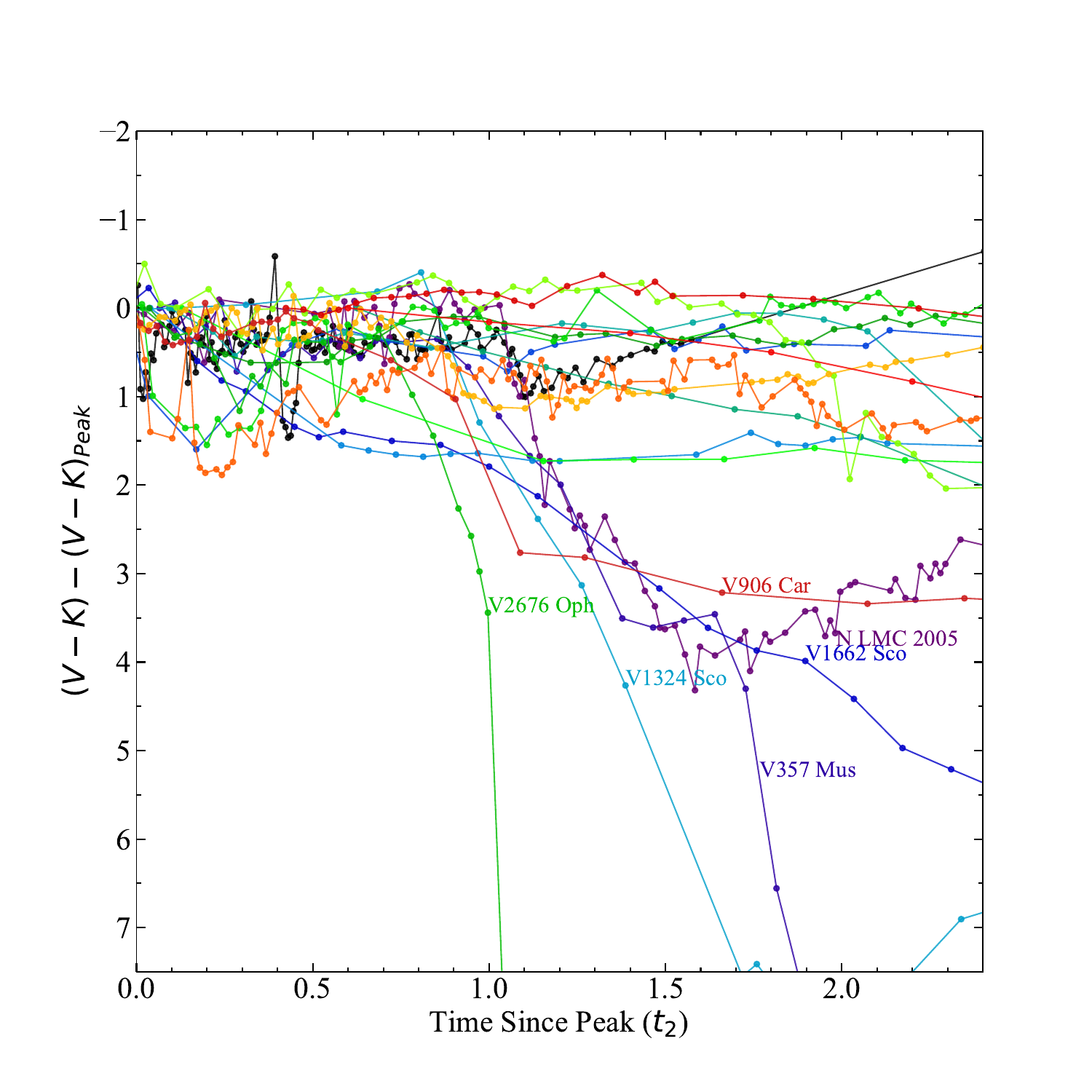}
    \caption{The evolution of ($V-K$) colour as a function of time in units of $t_2$. Again, light curves with at least 2 measurements between $t_{peak}$ and 10 days after peak, and at least 2 measurements between 10 and 20 days after peak, are included.}
    \label{fig:vk_evolution_long}
\end{figure}


In Section \ref{sec_diagnostic}, we suggest using the colour change from nova peak as a diagnostic of dust formation. In practice, the cadence of photometry can be limited, and $V$ and/or $K$ may not be measured right at $V$-band peak. Here we assess how much ($V-K$) changes around peak, in order to recommend a reasonable time window over which $(V-K)_{peak}$ can be measured.

We track the evolution of the ($V-K$) colours around the $V$-band light curve peak, to see how the colours vary with time. The colour evolution for the first 20 days of available SMARTS data is shown in Figure~\ref{fig:vk_evolution}, specifically showing the change in ($V-K$) since the nova peak. The majority of novae show relatively flat colour curves over this period of time, with their ($V-K$) reddening slightly from peak but staying within $\sim0.5$\,mag of $(V-K)_{peak}$ out to 20 days. 
For the first $\sim$week of eruption, the colour curves are clustered relatively closely in most novae. Later, the spread in the colour curves increases. The baseline plotted here is too short to capture much of the dust formation in this sample, but it does capture dust formation for those novae that form dust quickly. Three of the novae (V745~Sco, V1534~Sco, and V1662~Sco) redden sufficiently to meet our colour cutoff of $(V-K) - (V-K)_{peak} > 2.35$ on this time-scale. With that said, V745~Sco and V1534~Sco have red giant companions, so in those cases the colour change does not necessarily indicate dust formation. N LMC 2012a shows somewhat uncommon evolution in ($V-K$), growing significantly bluer after the peak. Otherwise, all of the novae in our sample either retain constant colours compared to the peak, or become redder over time.


A different time baseline is displayed in Figure~\ref{fig:vk_evolution_long}, with the time provided in units of $t_2$. While for most novae the time baseline is longer than in Figure~\ref{fig:vk_evolution}, for a few faster novae, 2.4\,$t_2$ (the upper limit on the x-axis) is less than the 20 days displayed in Figure~\ref{fig:vk_evolution}.
Again, we see the colour curves are clustered between $(V-K) - (V-K)_{peak}  \approx 0-1$ for times earlier than $1\,t_2$. At later times, there is far more spread in the colour curves, primarily driven by dust formation. We conclude that a ($V-K$) colour measured before $t_2$ can reasonably be used to estimate the colour at peak, with a typical uncertainty of $\pm0.5$\,mag.

\section{Intrinsic Optical--IR Colours of Novae at Light Curve Maximum}\label{sec_intrinsic}

If a nova is not observed at peak, but later optical/IR photometry is obtained and an estimate of the foreground extinction can be made, it is still possible to use the diagnostics suggested in Section~\ref{sec_diagnostic} to assess whether the nova formed dust. Here we estimate reddening-corrected colours of novae at peak to enable this use case.

Determining intrinsic colours around peak requires estimates of the interstellar reddening towards each nova. This is only available for a small subset of the novae in our sample. For the purpose of measuring colours, we avoid using extinctions measured using the nova's colour in other bands, to avoid potential biases/circular logic. In the spirit of a recent study of optical colours of novae (Craig et al. 2024, submitted), our $E(B-V)$ measurements mostly consist of spectroscopic measurements using Diffuse Interstellar Bands (DIBs), or 3D dust maps when Gaia DR3 parallaxes are available and have errors less than
30~per~cent of the parallax \citep{GaiaMission, GaiaDR3, Schaefer_2018, Schaefer_2022}. DIB based measurements are available for some of the novae in our sample in Craig et. al. (2024, submitted). For novae in the LMC, we use the 2D dust maps of \cite{Schlafly_etal_2011}, which should be a reasonable extinction estimate towards the LMC.

\begin{figure}
 
    \hspace{-0.7cm}
    \includegraphics[width=1.15\linewidth]{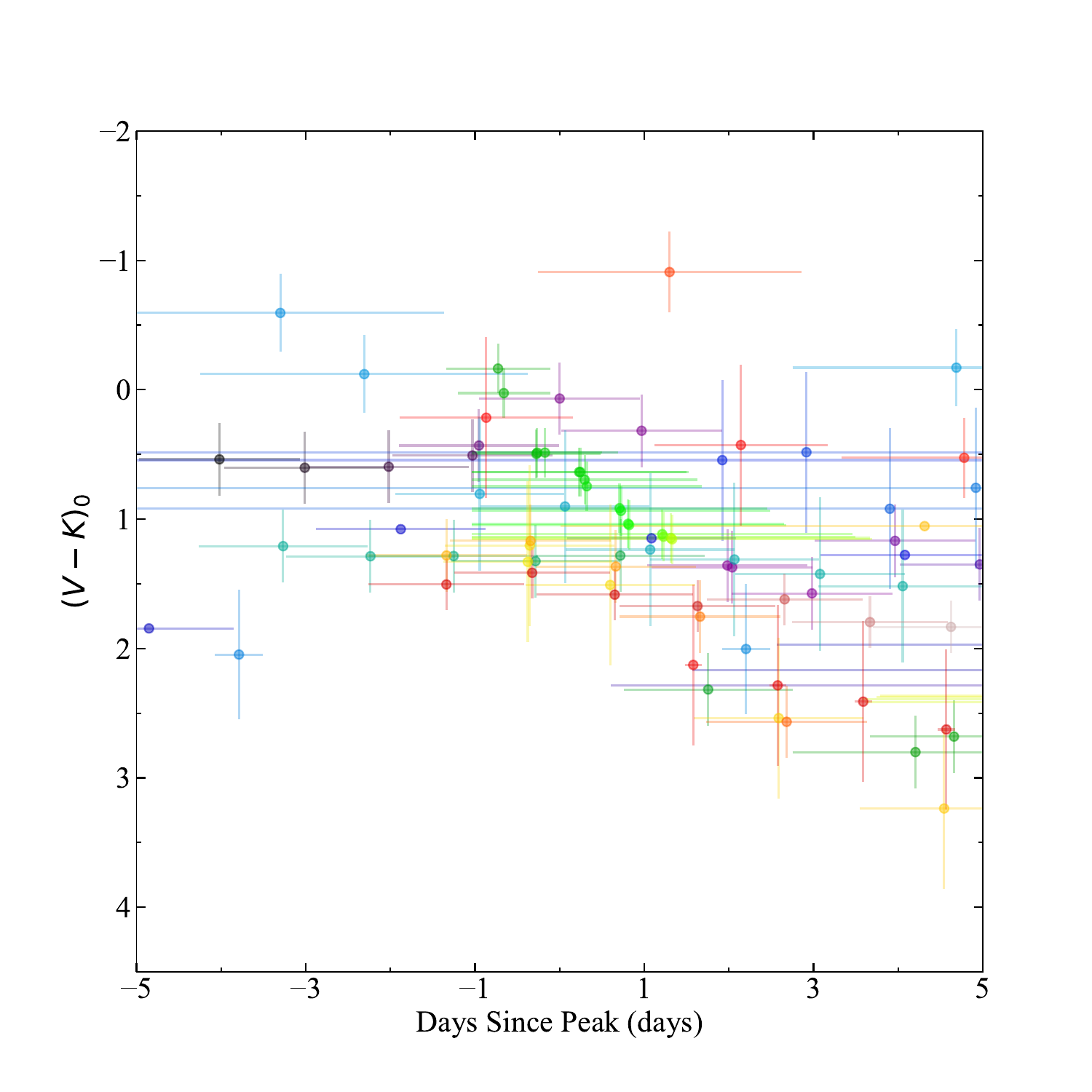}
    \caption{$(V-K)_0$ colour around peak, corrected for extinction, as a function of time from peak. Vertical error bars are dominated by errors in extinction. Horizontal error bars include uncertainty in $t_{\rm peak}$ and the time difference between the $V$-and $K$-band measurements.}
    \label{fig:vk_peak}
\end{figure}

Figure~\ref{fig:vk_peak} plots the reddening corrected $(V-K)_0$ colours around $V$-band peak. 
Based on the sample that we have extinctions for, the average colour at peak is $<(V-K)>_{\rm 0,peak} = 1.1$ based on a sample of $9$ novae. The $(V-K)$ colours show significantly more variability compared to the colours in optical bands, with a standard deviation of $0.8$\,mag around peak. The colours near peak in $(B-V)_0$, for comparison, typically vary by $\sim 0.3$ magnitudes, and even less around $t_2$ \citep[][Craig et al. 2024, submitted]{Van_den_Bergh_Younger_1987,Schaefer_2022}. 
At later times, there is a great deal of spread in $(V-K)_0$, likely caused by dust formation in some novae causing large amounts of reddening (some novae can get redder by more than 5 mag). The variability at peak is unlikely to be caused by dust formation in the nova, and likely is a combination of intrinsic variations in the $(V-K)_0$ colours of novae and the uncertainty in determining the interstellar reddening towards these novae.

The $(V-J)_0$ and $(V-H)_0$ colours have also been calculated at the optical peak for this sample. As with the $(V-K)_0$ colours, this is across our sample with extinction measurements and non-saturated SMARTS photometry near the peak. We find that $<(V-J)>_{0,peak} = 0.8$ based on $10$ novae, with a standard deviation of $0.6$. The standard deviation here is large compared to optical colours, but small compared to $(V-K)_0$. The distribution for $(V-H)_0$ is similar, with  $<V-H>_{0,peak} = 0.7$ and a standard deviation of $0.6$, based on $9$ novae.

\end{document}